\documentclass[floatfix,amssymb,amsmath,prd,twocolumn,superscriptaddress,nofootinbib]{revtex4-1}

\usepackage{graphicx}
\usepackage{bm}
\usepackage{amssymb,amsmath}
\usepackage{mathrsfs}
\usepackage{latexsym}
\usepackage{color}
\usepackage[normalem]{ulem} 
\usepackage{dcolumn}
\usepackage[colorlinks=true,citecolor=blue,urlcolor=blue]{hyperref}
\usepackage[usenames,dvipsnames]{xcolor}
\usepackage{multirow}
\usepackage{tabularx}
\usepackage{pgf} %hyperlinks in figures

\allowdisplaybreaks

\newcommand{\CITA}{\affiliation{Canadian Institute for Theoretical Astrophysics, 60 St. George Street, Toronto, Ontario, M5S 3H8, Canada}}
\newcommand{\GATech}{\affiliation{Center for Relativistic Astrophysics, School of Physics, Georgia Institute of Technology, Atlanta, Georgia 30332, USA}} 

\newcommand{\GSI}{\affiliation{GSI Helmholtzzentrum f\"ur Schwerionenforschung, Planckstra{\ss}e 1, 64291 Darmstadt, Germany}}

\newcommand{\HITS}{\affiliation{Heidelberg Institute for Theoretical Studies, Schloss-Wolfsbrunnenweg 35, 69118 Heidelberg, Germany}}

\newcommand{\UdeS}{\affiliation{D\'{e}partement de physique, Universit\'{e} de Sherbrooke, 2500 boulevard de l'Universit\'{e}, Sherbrooke, Qu\'{e}bec J1K 2R1, Canada}}

\newcommand{\CCA}{\affiliation{Center for Computational Astrophysics, Flatiron Institute, 162 5th Ave, New York, NY 10010}}

\definecolor{azgreen}{rgb}{0.03,0.47,0.19}
\definecolor{chmagenta}{rgb}{0.54, 0.17, 0.88}

\newcommand{\cn}[1]{{\textcolor{blue}{$^{\text{\sf{[citation needed]}}}~$}}}

\DeclareMathOperator{\fpeak}{f_{{\tiny peak}}}

\DeclareMathOperator{\fsub}{f_{{\tiny sub}}}

\begin{document}

\title{Observing the post-merger signal of GW170817-like events with improved gravitational-wave detectors}

\author{Andoni Torres-Rivas}
\thanks{CITA Summer Fellow}
\CITA \UdeS 

\author{Katerina Chatziioannou}
\CITA \CCA

\author{Andreas Bauswein}
\GSI 
\HITS

\author{James Alexander Clark}
\GATech

\date{\today}

\begin{abstract}
The recent detection of a neutron star binary through gravitational waves, GW170817, has offered another source of information about the properties of cold supranuclear matter. Information from the signal emitted before the neutron stars merged has been used to study the equation of state of these bodies, however, any complementary information included in the signal emitted after the merger has been lost in the detector noise. In this paper we investigate the prospects of studying GW170817-like post-merger signals with future gravitational-wave detectors. We first compute the expected properties of the possible GW170817 post-merger signal using information from premerger analyses. We then quantify the required improvement in detector sensitivity in order to extract key features of the post-merger signal. We find that if we observe a signal of similar strength to GW170817 when the aLIGO detectors have been improved by $\sim 2-3$ times over their design sensitivity in the kHz regime, we will be able to extract the dominant frequency component of the post-merger. With further improvements and next-generation detectors we will also be able to extract subdominant frequencies. We conclude that post-merger signals could be brought within our reach in the coming years given planned detector upgrades, such as A+, Voyager, and the next-generation detectors.  
\end{abstract}

%\pacs{4.00000}

\maketitle
  
%%%%%%%%%%%%%%%%%%%%%%%%%%%%%%%%%%%%%
\section{Introduction}

To this day several compact binary mergers have been detected with gravitational waves (GWs)~\cite{2016PhRvL.116f1102A,2016PhRvL.116x1103A,2017PhRvL.118v1101A,Abbott:2017oio,TheLIGOScientific:2017qsa,Abbott:2017gyy} by the LIGO~\cite{2015CQGra..32g4001T} and Virgo~\cite{TheVirgo:2014hva} detectors, with one being consistent with having been emitted from a  neutron star binary (BNS), GW170817~\cite{TheLIGOScientific:2017qsa}. The coalescing NSs in events such as GW170817 are natural laboratories with which to study the physics of cold nuclear matter at densities several times above the nuclear saturation density~\cite{Ozel:2016oaf,Lattimer:2015nhk,Oertel:2016bki}, conditions that are challenging to access with terrestrial experiments. 

NS coalescences are characterized by two distinct phases: the \emph{premerger} and the \emph{post-merger}. During the premerger phase, the two NSs orbit each other, radiating away orbital energy~\cite{lrr-2014-2}. The frequency of the resulting GW signal increases over time reaching approximately a few kHz when the bodies merge. The post-merger phase is characterized by a highly deformed post-merger remnant. Depending on its properties, the remnant might collapse directly into a black hole, survive for some time emitting a content-rich signal featuring a number of frequency components, or even survive indefinitely, see~\cite{lrr-2012-8,Baiotti:2016qnr} for reviews.

Both the late premerger and the post-merger signal carry information about the properties of NS matter, usually parametrized through the equation of state (EoS), a relation between the NS interior pressure, density, and temperature. Tidal interactions between the two NSs cause the late premerger phase to accelerate compared to point-particle dynamics~\cite{2008PhRvD..77b1502F,2010PhRvD..81l3016H} and  can be used to place constraints on the EoS~\cite{2009PhRvD..79l4033R,2013PhRvD..88d4042R,PhysRevLett.111.071101,2014PhRvD..89j3012W,2015PhRvD..91d3002L,2015arXiv150305405A,Chatziioannou:2015uea,Hotokezaka:2016bzh,Chatziioannou:2018vzf,Carney:2018sdv}. Moreover, the frequency content of the post-merger signal depends sensitively on the structure -and hence the EoS- of the stellar remnant and offers information that is complementary to the premerger~\cite{1994PhRvD..50.6247Z,1996A&A...311..532R,2005PhRvL..94t1101S,2005PhRvD..71h4021S,shibata:06bns,2007PhRvL..99l1102O,2011MNRAS.418..427S,2011PhRvD..83l4008H,2012PhRvL.108a1101B,bauswein:12,2013PhRvL.111m1101B,PhysRevD.78.084033,2011PhRvL.107e1102S,hotokezaka:13,2014PhRvL.113i1104T,2014arXiv1412.3240T,2015arXiv150401764B,bauswein:15,Foucart2016,Lehner:2016lxy,Kawamura2016,Maione2017,Bauswein:2018bma,Most2018a}. 

Indeed, premerger data from GW170817 have been used to measure the tidal parameters and radii of the coalescing NSs~\cite{TheLIGOScientific:2017qsa,2041-8205-850-2-L19,2041-8205-850-2-L34,Zhou:2017pha,2041-8205-852-2-L25,
PhysRevLett.120.172702,Nandi:2017rhy,PhysRevD.97.021501,
PhysRevLett.120.172703,Raithel:2018ncd,Most:2018hfd,Radice2018,De:2018uhw,TheLIGOScientific:2017qsa,Abbott:2018wiz,LigoEoS2018}, and to place constraints on their EoS, yielding results in agreement with terrestrial experiments~\cite{Tsang:2018kqj}. However, the post-merger emission of GW170817 remained undetected~\cite{TheLIGOScientific:2017qsa,Abbott:2018wiz,Abbott:2017dke}. Even though GW170817 was the loudest GW event observed to date~\cite{TheLIGOScientific:2017qsa}, its post-merger emission remained buried in detector noise resulting in our inability to determine if the merger remnant collapsed promptly to a black hole and obscuring possible further information about the EoS.

The properties of the post-merger remnant, including its EoS, are encoded in the post-merger signal through its frequency content, see for example Fig. 1 of~\cite{bauswein:15}. In particular, the dominant frequency component of the signal, appearing as a pronounced peak in the GW spectrum at a frequency $\fpeak$, carries information about the stellar structure of the remnant~\cite{bauswein:12}. 

A number of studies using numerical simulations of merging NSs have found empirical relations between various peaks in the post-merger spectrum and stellar properties, such as the NS radius~\cite{bauswein:12,2014PhRvL.113i1104T,bauswein:15,2014arXiv1412.3240T,Clark:2015zxa,bauswein:july15} and tidal deformability~\cite{2015arXiv150401764B}, which are uniquely linked to the EoS.

From a data analysis perspective, though, the post-merger signal is particularly challenging to detect and analyze. Uncertainty in and sparsity of numerical simulations mean that fully analytic, physically parametrized waveform templates which are consistent with the premerger signal are currently unavailable, reducing the feasibility of matched-filtering. Generic analyses that target signals of unknown morphology might be less efficient than matched-filtering, but they have been shown to be able to extract the main features of post-merger signals, such as its main frequency components~\cite{Clark:2015zxa,2014PhRvD..90f2004C,Chatziioannou:2017ixj}. In particular, Ref.~\cite{Chatziioannou:2017ixj} showed that the morphology-independent algorithm {\tt BayesWave}~\cite{Cornish:2014kda,Littenberg:2014oda} can provide a measurement of $\fpeak$ to within a few dozens of Hz and the radius to a few hundred meters for signals of an SNR $\sim 5$ with no prior knowledge of the signal properties.

Despite the non detection of a GW post-merger signal from GW170817, Ref.~\cite{Abbott:2018wiz} used {\tt BayesWave} to place upper bounds on the energy emitted by the merger remnant~\cite{Chatziioannou:2017ixj}. It was estimated that improvements of $\sim 3-15$ in amplitude sensitivity are required before analyses can extract information from the post-merger signal of a GW170817-like event. The desired improvements can be achieved in two ways: by improving the detectors' sensitivity and by improving the efficiency of our data analysis tools. 

In this paper we focus on the former. The network of advanced GW detectors is expected to expand in number and improve in sensitivity over the next years. As an outcome, dozens of BNSs such as GW170817 will be detected per year once the detectors reach their design sensitivity~\cite{Aasi:2013wya}. Moreover, third generation detectors are at the planning stages~\cite{2010CQGra..27h4007P,2011CQGra..28i4013H,ISwhitePaper,Miao:2017qot}. With these expected advances in mind, we calculate the improvement compared to the aLIGO~\cite{2015CQGra..32g4001T} design sensitivity required in order to extract features of the post-merger frequency spectrum of a GW170817-like event. We find that improvements of $\sim 2-3$ times the currently planned design sensitivity are necessary to measure the dominant frequency component of the signal. This corresponds to a strain sensitivities around $3\times10^{-24} \sqrt{1/\text{Hz}}$ at $2000$Hz. Such improvements are achievable with planned upgrades to existing facilities~\cite{Miller2015,ISwhitePaper}. Moreover, an improvement of $\sim 4-5$ times the aLIGO design sensitivity is required in order to observe sub-dominant features of post-merger signal. 

This paper is structured as follows. In Sec.~\ref{sec:premergeranalysis} we discuss the results of 
the premerger analysis on GW170817 and how they can be used to infer the post-merger properties of the signal.
In Sec.~\ref{sec:analysis} we describe the details of our analysis, while in Sec.~\ref{sec:results} we present our results.
We conclude in Sec.~\ref{sec:conclusions}.

%%%%%%%%%%%%%%%%%%%%%%%%%%%%%%%%%%%%%
\section{Constraints from premerger}
\label{sec:premergeranalysis}

Observations of the premerger signal from GW170817 can be used to inform our expectation for the properties of the undetected post-merger signal and select appropriate simulations to study the performance of a variety of detector sensitivities. In this section we describe the premerger information we use and what it implies for the potential post-merger signal from GW170817.

%---------------------------------------
\subsection{premerger analysis}

In the premerger phase, the GW signal emitted from the merger of two NSs differs from that of coalescing black holes due the effects of matter. Specifically, the tidal field of each star induces a quadrupole moment in its companion. The dimensionless tidal deformability parameter $\Lambda$ is proportional to the ratio of the induced quadrupole moment to the tidal field and it quantifies how easily a star is deformed and impacts the GW phase evolution~\cite{2008PhRvD..77b1502F,Vines:2011ud}.

Reference~\cite{LigoEoS2018} used two methods to measure the tidal parameter and the radius of each NS in GW170817. The first makes use of an EoS-insensitive relation between the tidal parameters of the two stars given the ratio of their masses~\cite{Yagi:2015pkc,Chatziioannou:2018vzf}. The second utilized an efficient spectral parametrization of the EoS itself in order to model the stellar structure directly~\cite{2015PhRvD..91d3002L,PhysRevD.82.103011,Carney:2018sdv}. Both analyses yield consistent results when applied on the GW data, yielding a measurement of the NS radius to within $\sim 3.6$km at the $90\%$ level~\cite{LigoEoS2018}. Moreover, the second analysis has the flexibility of imposing that the EoS supports masses of at least $1.97M_{\odot}$, motivated by pulsar observations~\cite{Antoniadis26042013}. 

We use the publicly available posterior samples\footnote{ \href{https://dcc.ligo.org/LIGO-P1800115/public}{dcc.ligo.org/LIGO-P1800115/public}} produced in the analysis of~\cite{LigoEoS2018} to estimate the expected properties of the post-merger signal for GW170817. In particular, we use four sets of posterior samples:

\begin{enumerate}
    \item masses, radii, and tidal parameters from the `EoS-insensitive' analysis,
    \item masses, radii, and tidal parameters from the `parametrized EoS' analysis without a maximum mass constraint,
    \item masses, radii, and tidal parameters from the `parametrized EoS' analysis with a maximum mass constraint, and
    \item EoS pressure--rest-mass density posterior from the `parametrized EoS' analysis with a maximum mass constraint.
\end{enumerate}

%---------------------------------------
\subsection{Expected post-merger properties}
We use the posterior samples for the masses, radii and tidal parameters to estimate the expected $\fpeak$ for the GW170817 post-merger signal. We use three such EoS-insensitive relations:

\begin{enumerate}
    \item A relation between $\fpeak$ and $R_{1.6}$, the radius of a $1.6 M_{\odot}$ NS~\cite{bauswein:july15},
    \item A relation between $\fpeak$ and the tidal parameter $\kappa_2^T$ which characterizes the binary tidal interactions during the late-inspiral~\cite{2015arXiv150401764B}, 
    \item A relation between $\fpeak$ and $f_c$, the contact frequency~\cite{Lehner:2016lxy},
\end{enumerate}
where

\begin{align}
\kappa_2^T &\equiv 3 \left( \frac{q^4}{(1+q)^5}\Lambda_1 + \frac{q}{(1+q)^5}\Lambda_2 \right),\\
f_c &\equiv \frac{1}{\pi M} \left( \frac{R_1+R_2}{M}\right)^{-3/2}.
\end{align}
In the above equations $m_i$, $R_i$, $\Lambda_i$ are the mass, radius and dimensionless tidal deformability for each binary component $i\in \{1,2\}$ respectively, $q\equiv m_2/m_1 <1$ is the mass ratio of the binary, $M\equiv m_1+m_2$ is its total mass, and we use units where $G=c=1$.

\begin{figure}[!htbp]
\includegraphics[width=\columnwidth,clip=true]{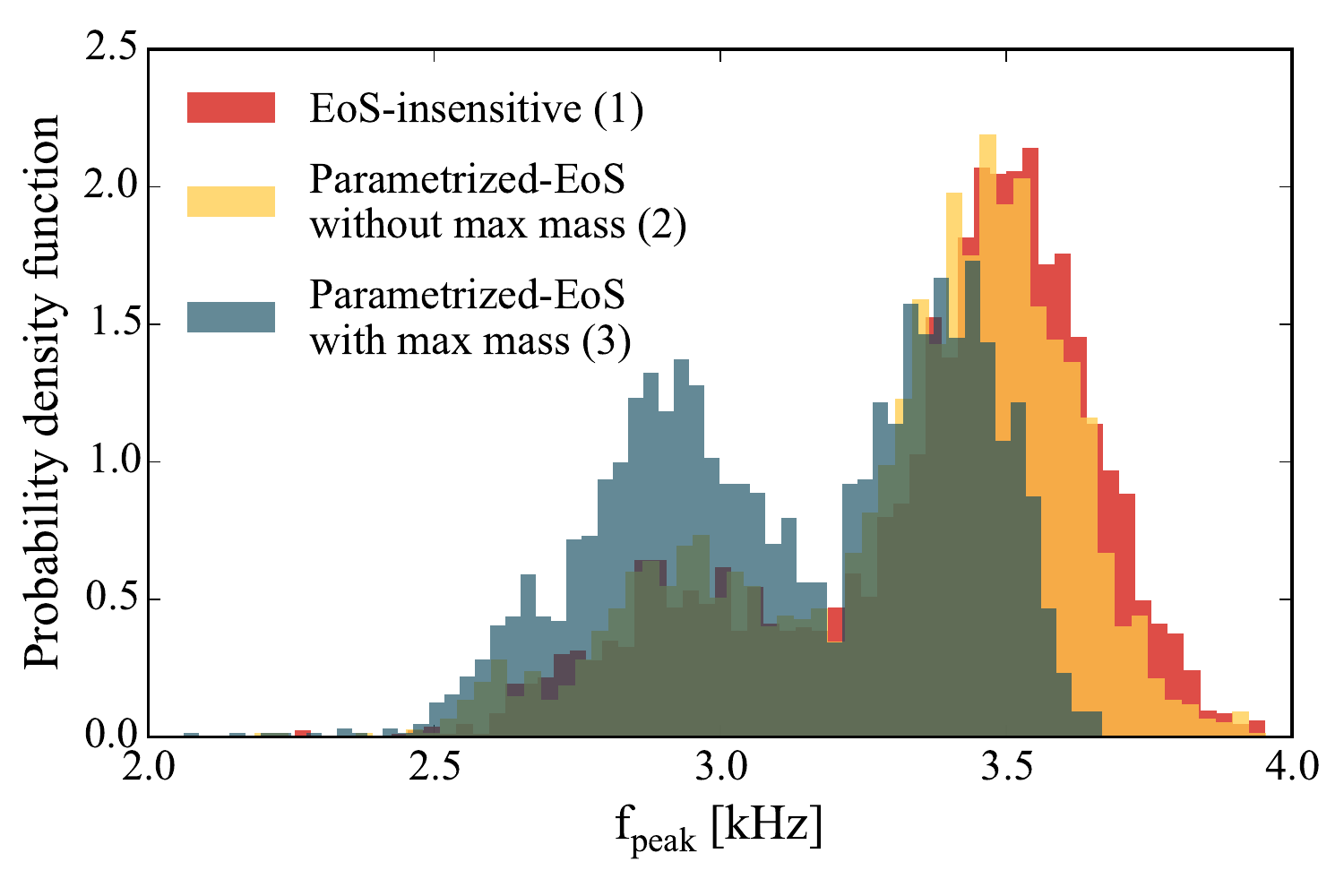}
\includegraphics[width=\columnwidth,clip=true]{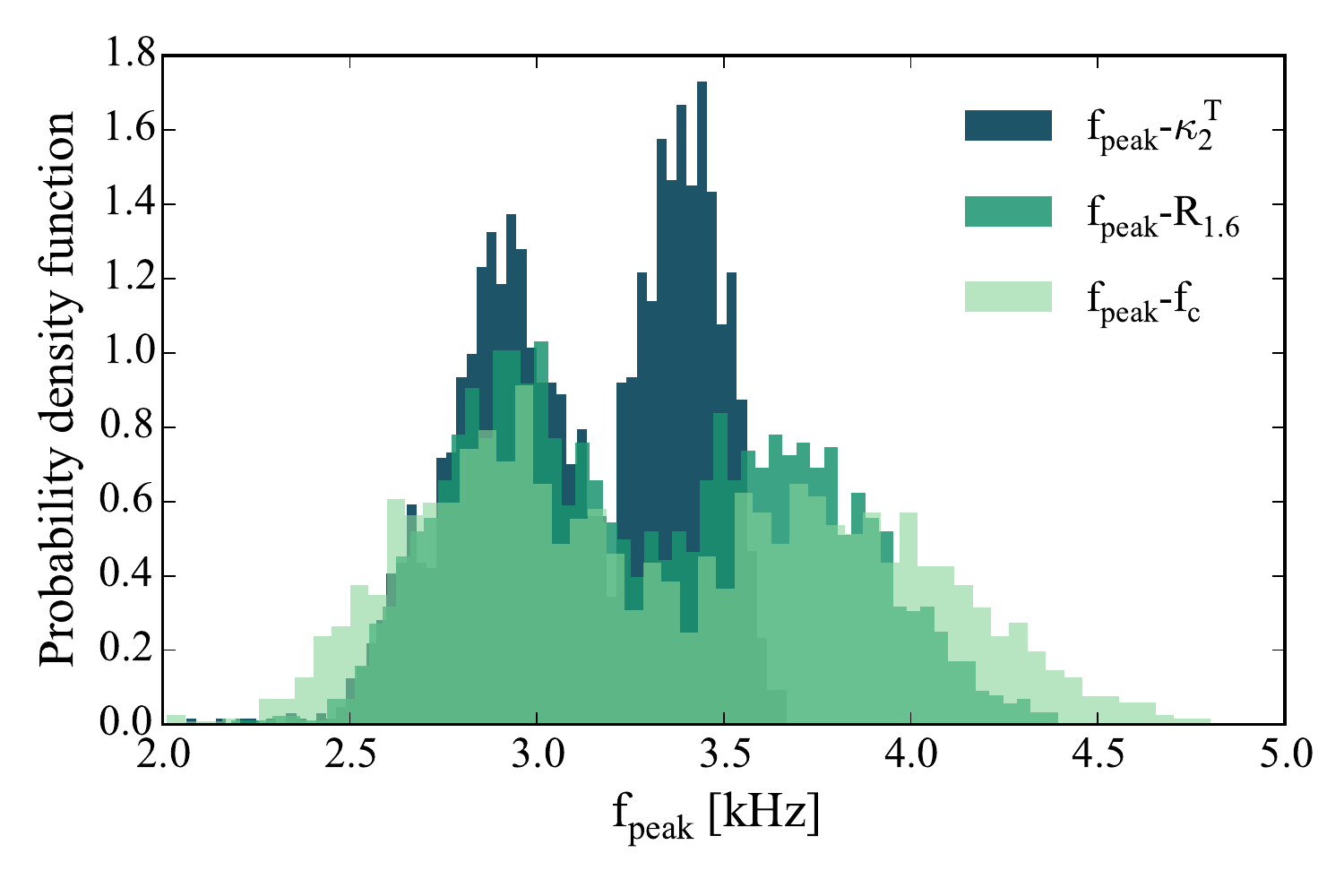}
\caption{ \label{fig:fpeakposteriors} Probability density for $\fpeak$, inferred from the premerger data from GW170817. The top panel uses the first three samples sets from~\cite{LigoEoS2018} described in Sec.~\ref{sec:premergeranalysis} and the $\fpeak-\kappa_2^T$ EoS-insensitive relation. The bottom panel uses the `parametrized EoS' analysis with a maximum mass samples from~\cite{LigoEoS2018} (third set in Sec.~\ref{sec:premergeranalysis}) and three EoS-insensitive relations. The bimodal structure of the posterior distributions is an outcome of the bimodality of the radius and tidal parameter posterior already observed in~\cite{LigoEoS2018}. }
\end{figure}

We use the above relations to derive posterior samples for $\fpeak$, given samples from the mass, radii, and $\Lambda$ posteriors which were obtained from studying the premerger phase of GW170817. Figure~\ref{fig:fpeakposteriors} shows the inferred posterior for $\fpeak$. The top panel uses the first $3$ sets of posterior samples from Sec.~\ref{sec:premergeranalysis} and the $\fpeak-\kappa_2^T$ EoS-insensitive relation described above. The bottom panel uses the results from the `parametrized EoS' analysis with a maximum mass requirement and computes $\fpeak$ with the three EoS-insensitive relations described above. For the first relation, $\fpeak-R_{1.6}$, we use the radius of the heaviest of the coalescing stars, rather than the radius of a $1.6M_{\odot}$ star.
Given the large statistical uncertainty in the radius measurement from the premerger phase, we expect the error in the radius from this mass approximation to be negligible.
Moreover, we neglect systematic uncertainties in the three EoS-insensitive relations, since they are expected to be smaller than the statistical errors from the premerger observations of GW170817.

We find that $\fpeak$ is expected to be approximately in $[2.5,4]$kHz. 
The results of~\cite{LigoEoS2018} disfavor large NS radii and stiff EoSs, which translates to a more compact post-merger remnant that emits GWs at relatively higher frequencies. Unsurprisingly, we obtain the tightest $\fpeak$ measurement from the posterior samples obtained after imposing that the EoS supports a maximum mass of at least $1.97M_{\odot}$. Additionally, the requirement that the EoS supports such a large maximum mass results in a stiffer EoS at high densities, translating to slightly larger radii and lower $\fpeak$ values. We verify that the two analyses that do not impose a maximum mass on the EoS lead to consistent results, as was originally noted in~\cite{LigoEoS2018}. 

Finally, we find that the three EoS-insensitive relations under study give broadly consistent results (bottom panel), though the $\fpeak-\kappa_2^T$ relation with  lead to a tighter $\fpeak$ estimate than the relations with $\fpeak (R)$. The largest disagreement between the EoS-insensitive relations happens at large values of $\fpeak$, or smaller radii and soft EoSs. The $\fpeak(R)$ relations become steeper at higher frequencies and thus larger deviations in $\fpeak$ are to be expected. Another possible reason is that the radius posterior for GW170817 includes values that are outside the calibration region of these relations. Therefore it is not surprising that they disagree in that region. Note that none of these relations informs about the occurrence of a prompt collapse of the remnant and thus they also predict $\fpeak$ values for systems where no strong post-merger GW emission is expected.

\begin{figure}[!htbp]
\includegraphics[width=\columnwidth,clip=true]{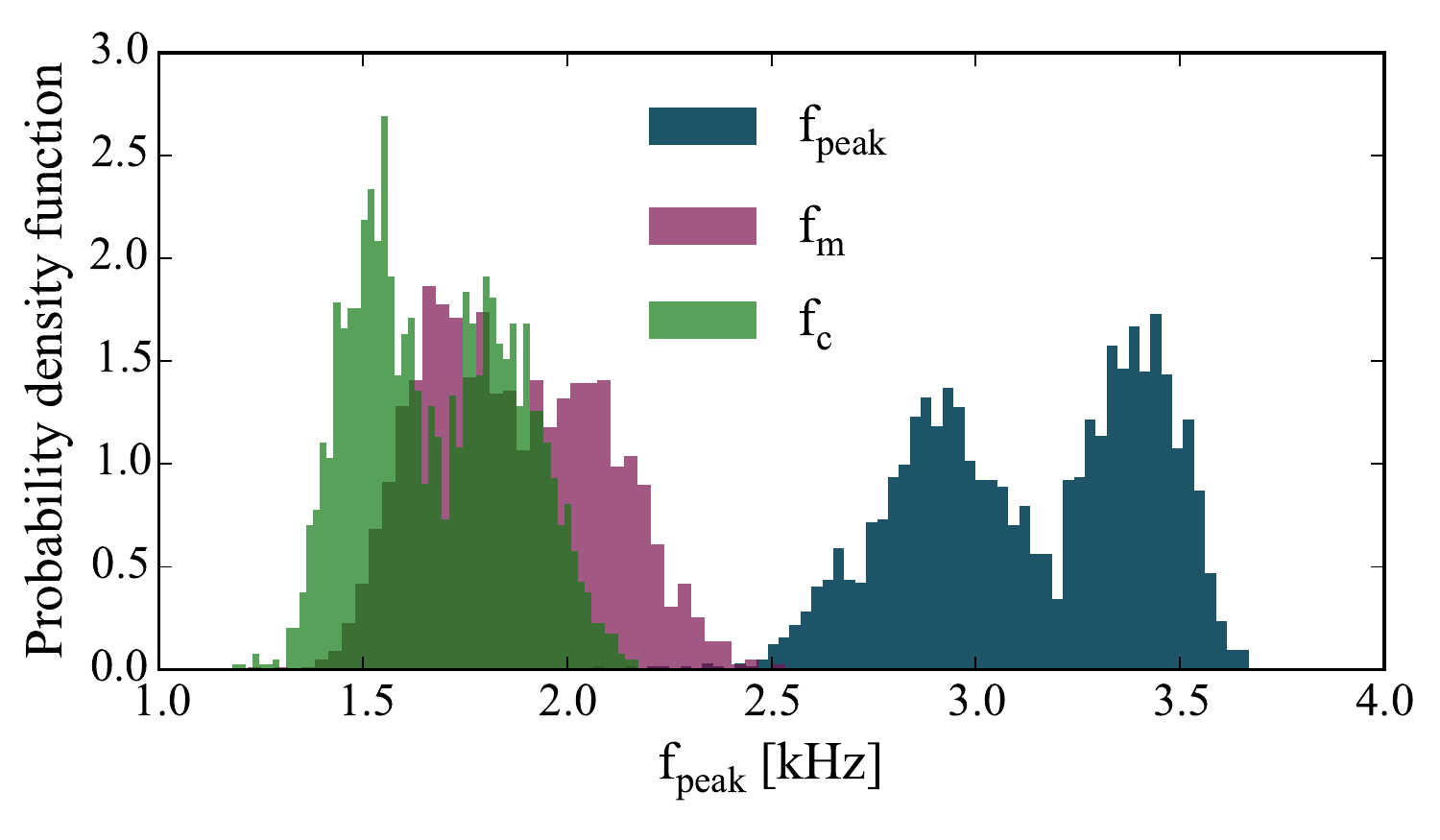}
\caption{Probability density for the contact frequency, the merger frequency, and the dominant post-merger frequency of GW170817, computed from the premerger results of~\cite{LigoEoS2018}.}
\label{fig:frequency_hists}
\end{figure}

Fig.~\ref{fig:frequency_hists} shows the posteriors on the frequency at the latter stages of the GW170817 coalescence inferred from premerger data.  We show the posterior for the contact frequency $f_c$, the merger frequency $f_m$, and the dominant post-merger frequency $\fpeak$. The first is a Newtonian estimate of the GW frequency at which the two NSs touch\footnote{For a discussion on its definition and meaning within the context of relativistic systems see~\cite{PhysRevD.81.084016,Baiotti:2011am,Bernuzzi:2012ci}.}, the second is the GW frequency when the GW signal reaches its maximum amplitude, while the third describes a property of the post-merger remnant. For this plot we use the `parametrized EoS' posteriors with a maximum mass constraint. To compute $\fpeak$ we use the $\fpeak-\kappa_2^T$ relation, while to compute $f_m$ we use the EoS-insensitive relation between the merger frequency and $\kappa_2^T$ proposed in~\cite{PhysRevLett.112.201101} and updated in~\cite{TDietrich2018}. Our results suggest that an analysis of the post-merger signal starting at $1024$Hz~\cite{Chatziioannou:2017ixj,Abbott:2018wiz} would include the very late inspiral, the merger, as well as the post-merger stages of the signal.

%----------------------------------------------------------------------------------------------------
\subsection{Simulated post-merger signals}
\label{sec:simulations}
\begin{figure}[!htbp]
\centering
\hspace*{-0.49cm}
\includegraphics[width=1.02\columnwidth,clip=true]{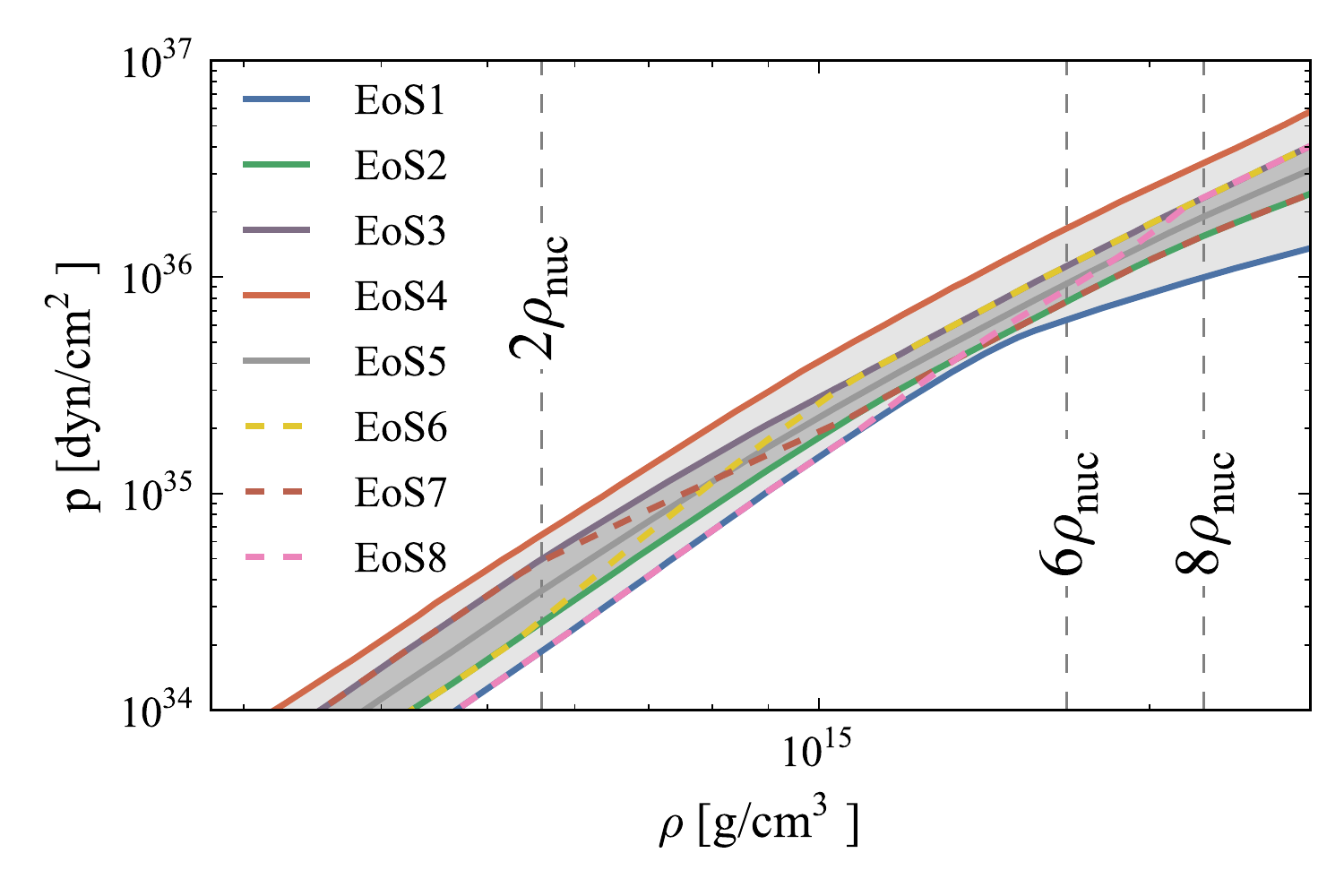}\\
\includegraphics[width=\columnwidth,clip=true]{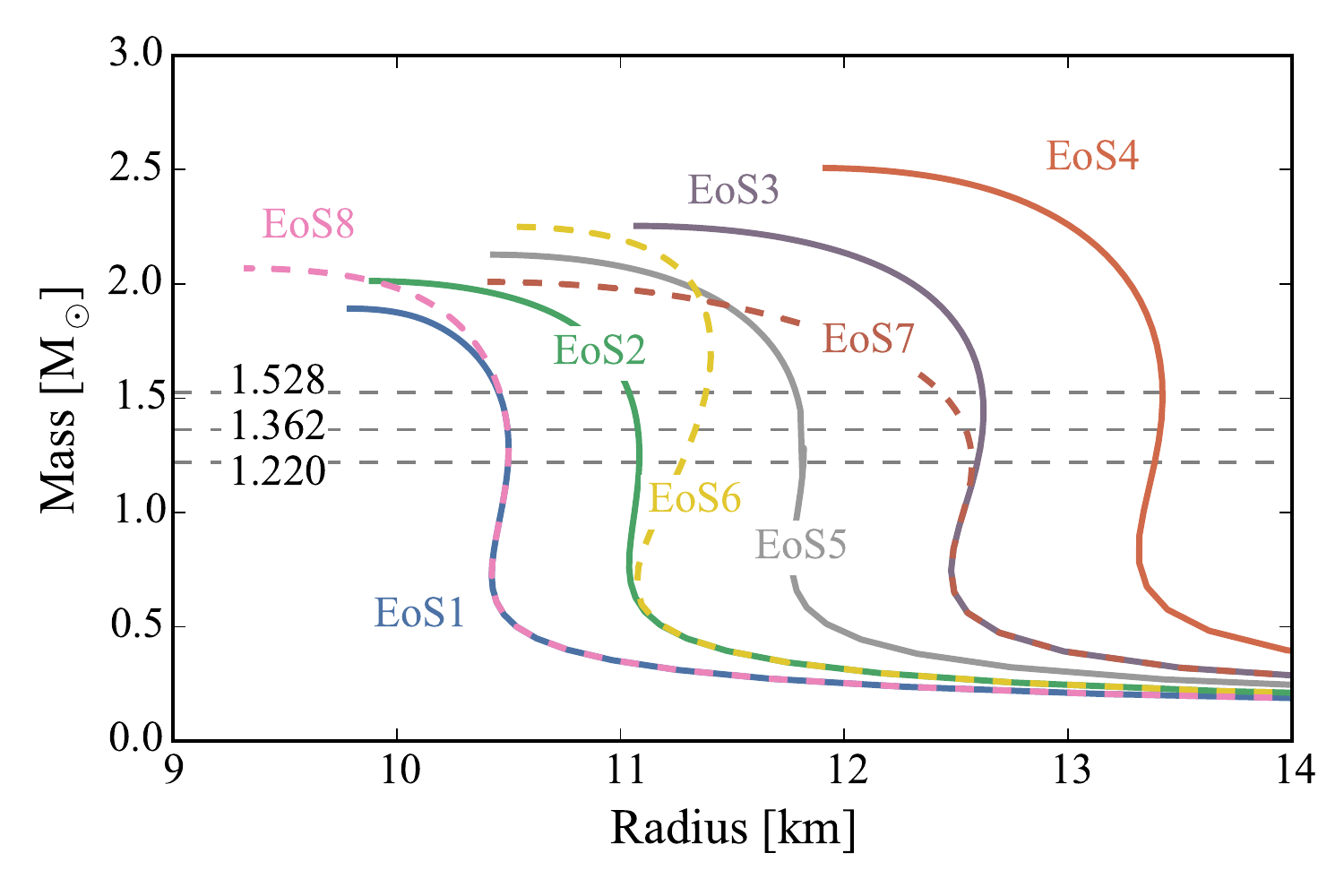}
\caption{Pressure--rest-mass density (top panel) and mass-radius (bottom panel) for the $8$ EoS models that we study. In the top panel we include for illustration purposes the pressure--rest-mass posterior (50\% and 90\% credible intervals) computed in~\cite{LigoEoS2018}. The dashed vertical lines denote the transition points for EoS6, EoS7, and EoS8. The horizontal lines in the bottom panel denote the NS masses in our simulations.} 
\label{fig:EoS-simulated}
\end{figure}

The analysis of the premerger data from GW170817 yielded a posterior for the pressure as a function of the rest-mass density~\cite{LigoEoS2018}.
We use this posterior to construct $8$ EoS models that are consistent with the GW170817 data. These EoS models are then used to generate BNS simulations from which we extract the expected GW emission and probe the efficacy of potential future GW detector instrumentation.   Our EoS models are given  by combining the various published pressure credible levels for different values of the density, expressed as a multiple of $\rho_{nuc} = 2.8\times10^{14}$gcm$^{-3}$:

\begin{itemize}
\item EoS1:  $5^{th}$ percentile of the pressure posterior for all densities,
\item EoS2:  $25^{th}$ percentile of the pressure posterior for all densities,
\item EoS3:  $75^{th}$ percentile of the pressure posterior for all densities, 
\item EoS4:  $95^{th}$ percentile of the pressure posterior for all densities, 
\item EoS5:  midpoint in the logarithm of the pressure of the $25^{th}$ and the $50^{th}$ percentiles of the pressure posterior for all densities, 
\item EoS6:  $25^{th}$ percentile of the pressure posterior until $2\rho_{nuc}$, $75^{th}$ percentile above $4\rho_{nuc}$, and a linear-in-logP transition in-between,
\item EoS7:  $75^{th}$ percentile of the pressure posterior until $4\rho_{nuc}$, $25^{th}$ percentile above $8\rho_{nuc}$, and a linear-in-logP transition in-between,
\item EoS8:  $5^{th}$ percentile of the pressure posterior until $4\rho_{nuc}$, $75^{th}$ percentile above $8\rho_{nuc}$, and a linear-in-logP transition in-between,
\end{itemize}
Note that none of these is formally a sample in the EoS posterior. They are, however, indicative of the allowed pressure range for the EoS of GW170817. In particular, EoS1 does not support a $1.97M_{\odot}$ star, however we choose to use it here as an example of the soft end of the EoSs allowed. Figure~\ref{fig:EoS-simulated} shows the pressure--rest-mass density (top panel) and mass-radius (bottom panel) relation for each EoS.

The resulting merger simulations are conducted with relativistic smooth particle hydrodynamics code~\cite{2002PhRvD..65j3005O,2007A&A...467..395O,2010PhRvD..82h4043B}, which has been used before for EoS surveys~\cite{2012PhRvD..86f3001B,2013ApJ...773...78B,bauswein:15}. The implementation adopts the conformal flatness condition to solve the Einstein field equations~\cite{1980grg..conf...23I,1996PhRvD..54.1317W}. Since the EoSs of our sample do not provide the temperature dependence of the pressure and the energy density, we employ a common approximate treatment of thermal effects, which allows to simulate BNS mergers based on barotropic EoSs. Within this scheme one has to specify a coefficient $\Gamma_\mathrm{th}$, which determines the strength of the thermal pressure. We choose $\Gamma_\mathrm{th}=1.75$ because this choice simulates fairly well the behavior of available temperature-dependent microphysical EoS models~\cite{2010PhRvD..82h4043B}. We assume initially nonspinning NSs on circular orbits and simulate the inspiral through the last few orbits. For each EoS we simulated two sets of binaries with the same chirp mass ${\cal{M}}=1.186M_{\odot}$ and different values of the mass ratio $q=\{1,0.8\}$, corresponding to component masses $m_1=m_2=1.362M_\odot$ and  $m_2=1.22M_\odot, \, m_1=1.525M_\odot$ respectively. These configurations were chosen to be consistent with GW170817~\cite{Abbott:2018wiz}. 

\begin{figure}[!htbp]
\centering
\includegraphics[width=\columnwidth,clip=true]{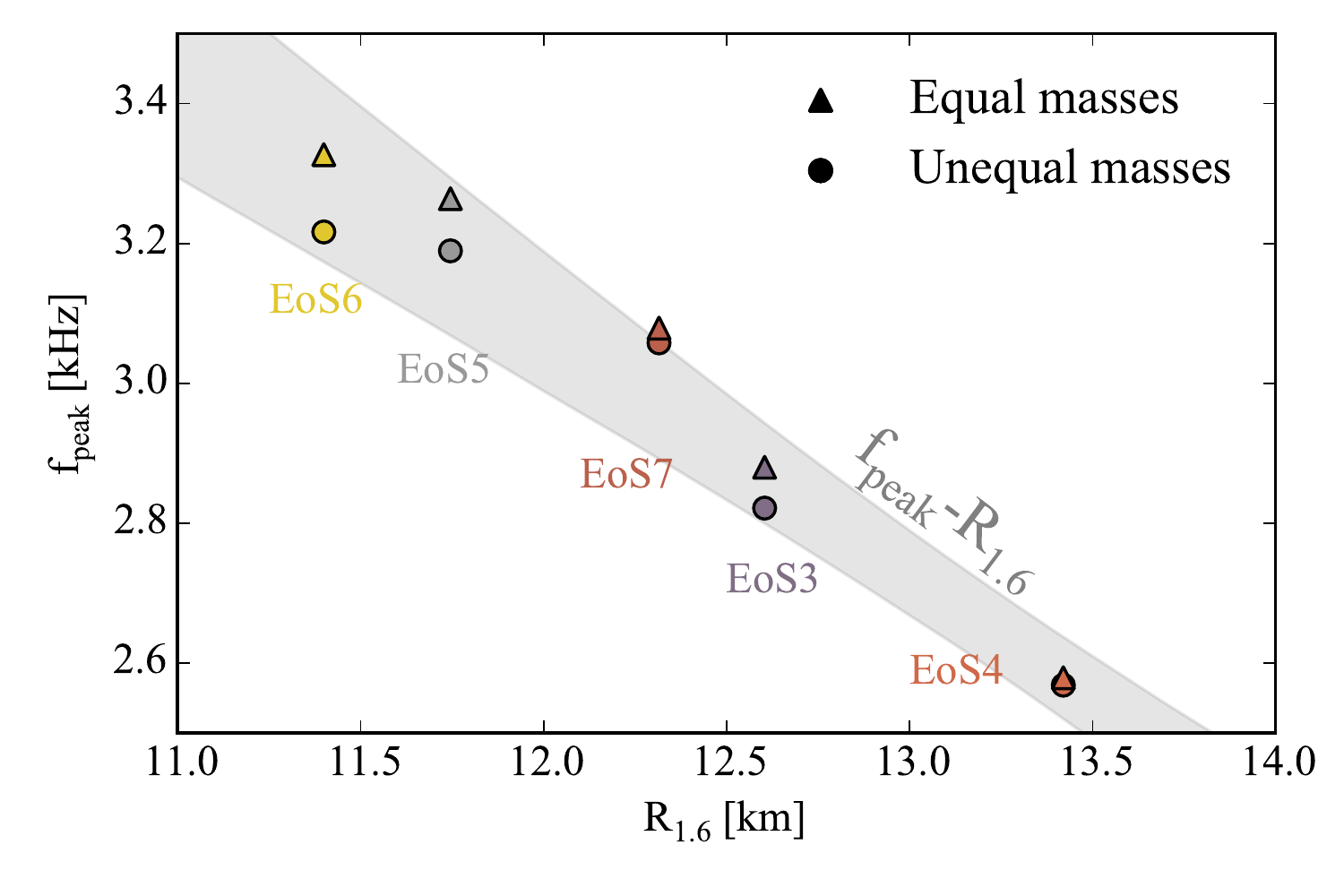}
\caption{Main frequency content of the simulated signals. We show $\fpeak$ as a function of the radius of a nonrotating $1.6M_{\odot}$ star for each EoS with points. The grey band shows the 90\% credible interval of the expected EoS-insensitive relation between $\fpeak$ and $R_{1.6}$ as computed in~\cite{Chatziioannou:2017ixj}.} 
\label{fig:FpeakvsR16}
\end{figure}

Of the total of $16$ simulated mergers, $5$ resulted in the merger remnant collapsing to a BH immediately (EoS1 and EoS8 for both mass ratios, and EoS2 for $q=0.8$). In Fig.~\ref{fig:FpeakvsR16} we characterize the dominant post-merger frequency of the simulations that resulted in an NS remnant. We show the dominant post-merger frequency $\fpeak$ as a function of the radius of a nonrotating $1.6M_{\odot}$ star for each EoS. We also plot the $\fpeak-R_{1.6}$ fit obtained in~\cite{Chatziioannou:2017ixj}, confirming that our simulations follow the empirical (EoS-insensitive) relation.
Regarding subdominant peaks in the spectrum, we note that they can be generated by different physical mechanisms and that the strength of these different peaks can vary with the binary masses and the EoS~\cite{2014PhRvL.113i1104T,2011MNRAS.418..427S,bauswein:15,bauswein:july15,2014arXiv1412.3240T,Clark:2015zxa,Rezzolla2016,Maione2017} (see \cite{bauswein:15,bauswein:july15} for a unified picture of the postmerger dynamics and GW emission). In what follows and for the purposes on this study we do not distinguish between the different origin of subdominant features and define $\fsub$ as the frequency of the second highest peak with a frequency at least 400Hz below $\fpeak$.

\section{Analysis setup}
\label{sec:analysis}

The $16$ simulated mergers described in Sec.~\ref{sec:simulations} are used to simulate the signal waveforms observed by a network of GW detectors~\cite{Schmidt:2017btt}, assuming the known sky location of the GW170817 host galaxy and a distance of $40$Mpc~\cite{Abbott:2018wiz}.
In this section we describe the set up of our analysis of these waveforms, namely the detector configurations we assume and the morphology-independent reconstruction algorithm {\tt BayesWave}. 

%----------------------------------------------------------------------------------------------------
\subsection{Detector configurations}

The simulated signals are projected onto networks of second- and third- generation detectors and analyzed with the noise-weighting appropriate for each instrument.  We note that these \emph{signal injections} do not contain a specific noise-realization: such analysis of noise-free injections has previously been shown to be equivalent to averaging over many noise realizations~\cite{Nissanke:2009kt}. Second generation ground-based detectors are observational facilities currently operational or under construction. Two LIGO detectors in Hanford (H) and Livingston (L) and VIRGO (V) are operational, while KAGRA~\cite{PhysRevD.88.043007} and LIGO-India~\cite{LIGOINDIA} are under construction. These detectors are expected to reach their design sensitivity in the coming years and keep improving towards A+ and Voyager. Eventually the second-generation detectors will be replaced by third-generation ground-based detectors, such as Cosmic Explorer and the Einstein Telescope.

\begin{figure}[!htbp]
\includegraphics[width=\columnwidth,clip=true]{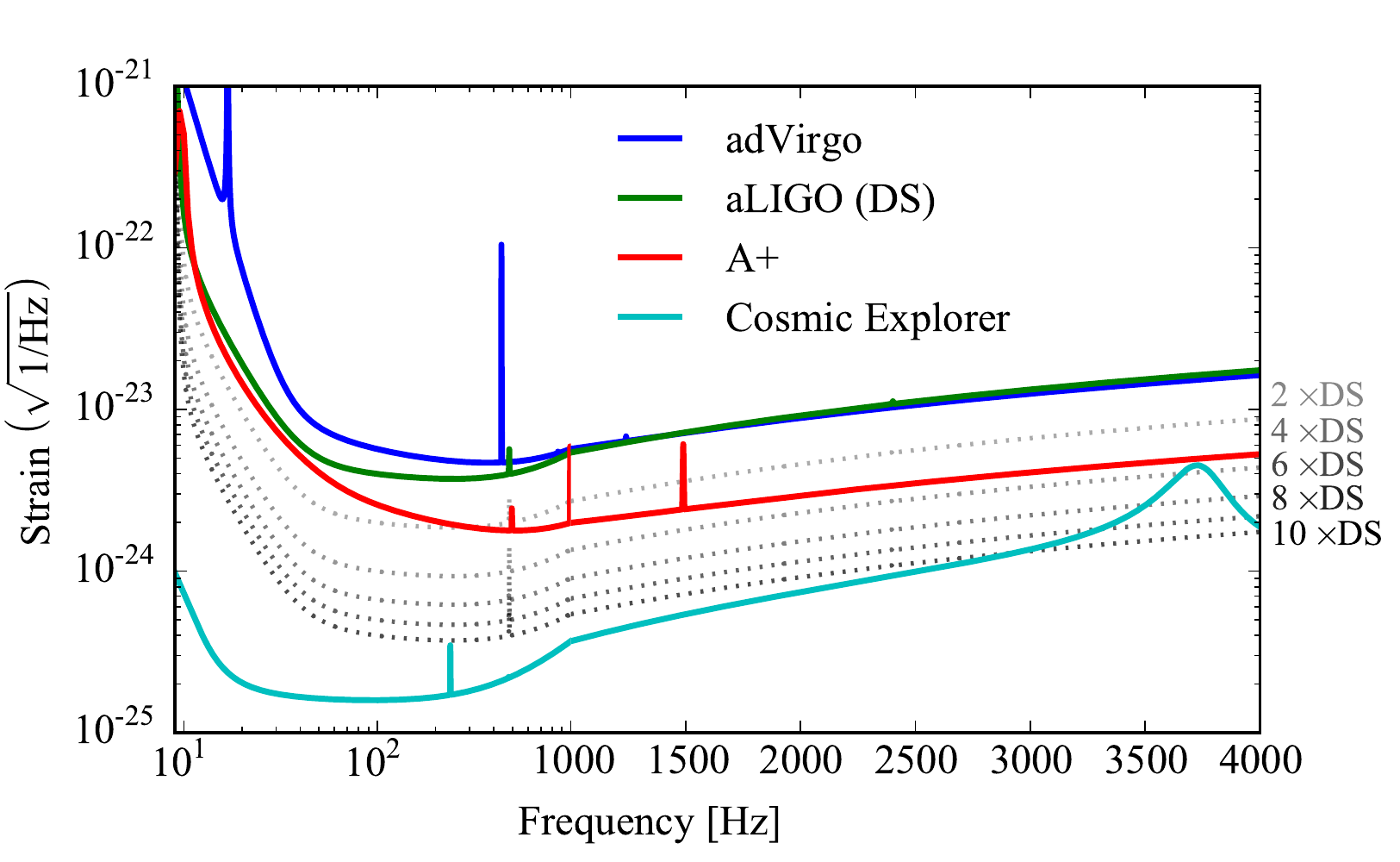}
\caption{Sensitivity curves for the detectors we include in our networks. To illustrate the high-frequency region better, the $x$ axis uses by a logarithmic scale below $1000$Hz and linear above that. Dotted lines represent the aLIGO design sensitivity improved by various constant factors.}
\label{fig:alldesignsensitivities}
\end{figure}

Given the scheduled gradual upgrades and expansion of the network in the coming years, we study networks that are incrementally improved compared to design sensitivity~\cite{AdvLIGO-noise}. In particular we assume a network of three detectors: H, L, V. We keep V at its design sensitivity\footnote{We choose to inject our signals in the known sky location of GW170817. Since this location at that GPS time was very close to a blind spot for V~\cite{TheLIGOScientific:2017qsa}, we do not expect V to contribute significantly to the numbers presented here, even if we chose to increase its sensitivity.} and incrementally increase the sensitivity of H and L by dividing it by a number $Y$; we denote this network as $Y$xDS. Once $Y\geq 7$ we assume a network with only two detectors, L and V in order to make a smooth transition towards third-generation detectors. We also carry out targeted runs using the sensitivity of Cosmic Explorer. 
Figure~\ref{fig:alldesignsensitivities} summarizes the sensitivities we study.

Besides incremental improvements of the whole sensitivity curve, narrow 
band tuning has also been proposed~\cite{AdvLIGO-noise,GWINC}. This design is 
expected to give improved sensitivity at a narrow frequency range. However, in 
order to implement such a design for post-merger studies we would need more 
precise 
knowledge of the approximate location of $\fpeak$ than currently available. 
Moreover, the narrow-band 
tuning may lead to diminished sensitivity across the full spectrum, potentially 
including the high frequencies of interest here. Detailed exploration of the 
capabilities of such a design is the 
subject of ongoing investigations.

%----------------------------------------------------------------------------------------------------
\subsection{Signal reconstruction with minimal assumptions}

The complicated morphology of post-merger signals makes constructing accurate templates challenging. In order to reconstruct the injected signals we instead use a morphology-independent approach. We employ {\tt BayesWave}~\cite{Cornish:2014kda,Littenberg:2014oda}, and carry out a Bayesian analysis where the GW signal is modeled as a sum of sine-Gaussian wavelets where both the parameters and the number of wavelets are marginalized over. {\tt BayesWave} has been shown to accurately reconstruct a range of signal morphologies~\cite{Becsy:2016ofp,Chatziioannou:2017ixj,Pannarale:2018cct} and to facilitate detection of unmodeled signals~\cite{Littenberg:2015kpb,2016PhRvD..93b2002K}.

Reference~\cite{Chatziioannou:2017ixj} applied {\tt BayesWave} to post-merger signals and showed that it can extract various features of the signal, including the dominant frequency component and the energy. We here carry out an analysis similar to~\cite{Chatziioannou:2017ixj}, using $250$ms of data in a frequency band of $[1024,4096]$Hz. We employ the same parameter priors as~\cite{Chatziioannou:2017ixj} and highlight that again we impose a prior on the number of wavelets used to be at least $2$. This is further motivated by Fig.~\ref{fig:frequency_hists} which suggests that the analyzed data contain both the merger and the post-merger phases of the coalescence. We choose to not restrict our analysis bandwidth above 2500Hz, since we are interested in i) also studying the subdominant peaks of the spectrum, and ii) constructing a generic analysis that is applicable to lower mass systems that are expected to have lower values of $\fpeak$ than GW170817.

Once a posterior for the reconstructed signal has been computed, we measure its frequency components $\fpeak$ and $\fsub$. The former is defined as the peak frequency of the post-merger amplitude spectrum, while the latter is the second highest peak with the constraint $\fsub<\fpeak$-400. If a posterior sample for the reconstruction does not have a peak, then a sample is drawn from the $\fpeak$ prior, as explained in~\cite{Chatziioannou:2017ixj}.

%%%%%%%%%%%%%%%%%%%%%%%%%%%%%%%%%%%%%
\section{Results}
\label{sec:results}

Our $16$ simulated signals include $5$ systems where the remnant collapses into a BH immediately and $11$ systems that result in a NS remnant (NSR). We analyze these signals with {\tt BayesWave} and describe here the reconstruction properties of the post-merger signal for the various physical outcomes of the merger. 

%---------------------------------------------------------------
\subsection{Post-merger reconstruction for NS remnants}

We begin by discussing the case of a NSR, which leads to a post-merger GW signal exhibiting a characteristic spectral peak as well as possible subdominant frequency peaks. We first discuss EoS5 in detail, as it represents the midpoint EoS for GW170817. We then turn to the other EoSs that lead to a NSR remnant and determine at which sensitivities we are likely to observe a GW170817-like post-merger signal.

%%%%%%%%%%%%%%%%%%%%%%%%%%%%%%%%%%%%%%%%%%%%%%%%%%%%%%%%%

\begin{figure*}[!htbp]
\centering
\includegraphics[width=0.6\textwidth]{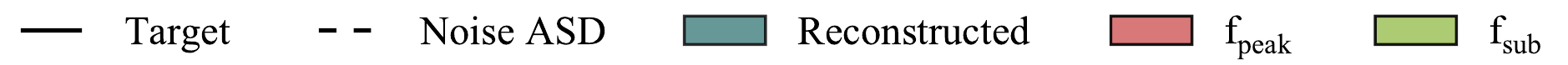}\\
\includegraphics[width=0.32\textwidth]{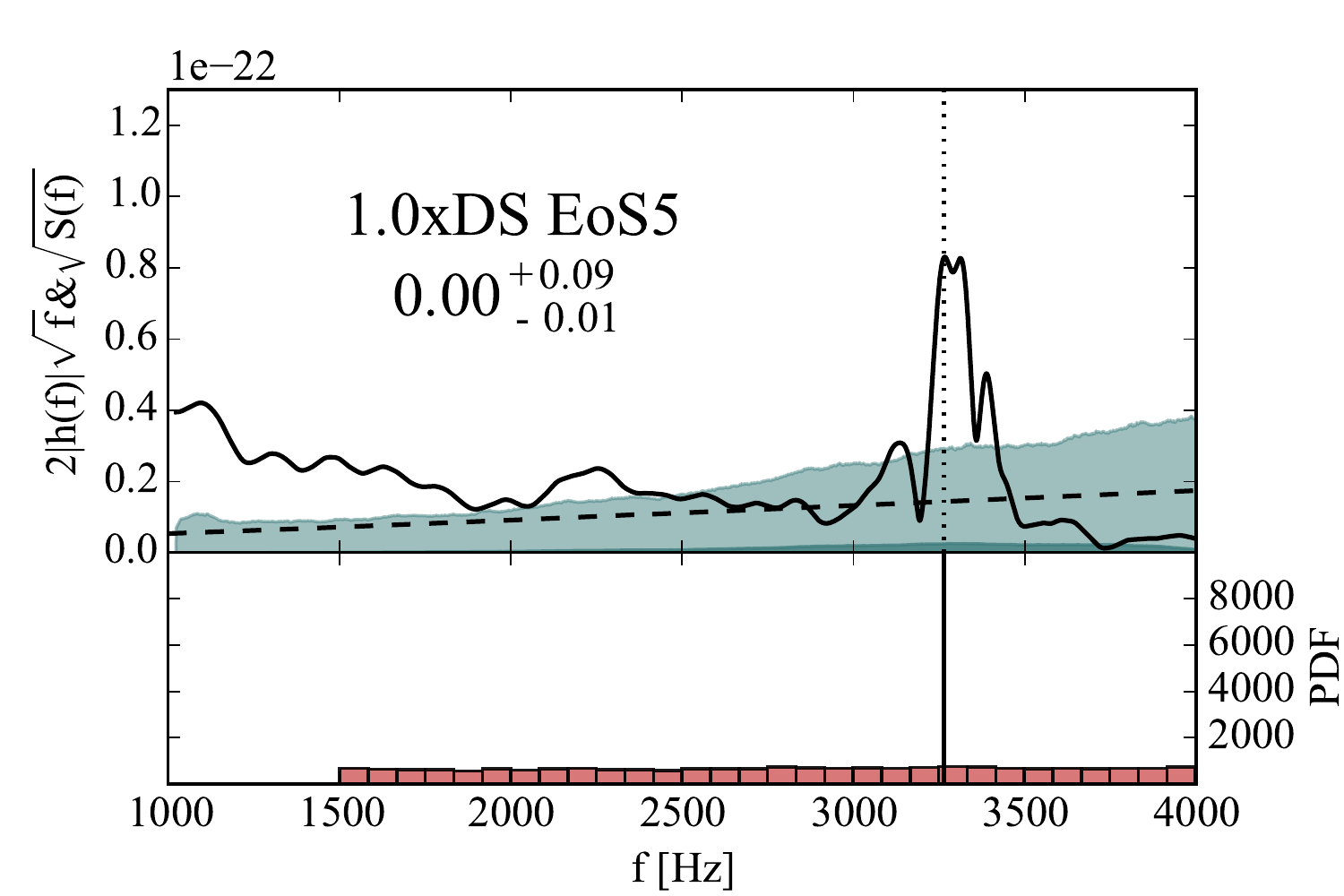}
\includegraphics[width=0.32\textwidth]{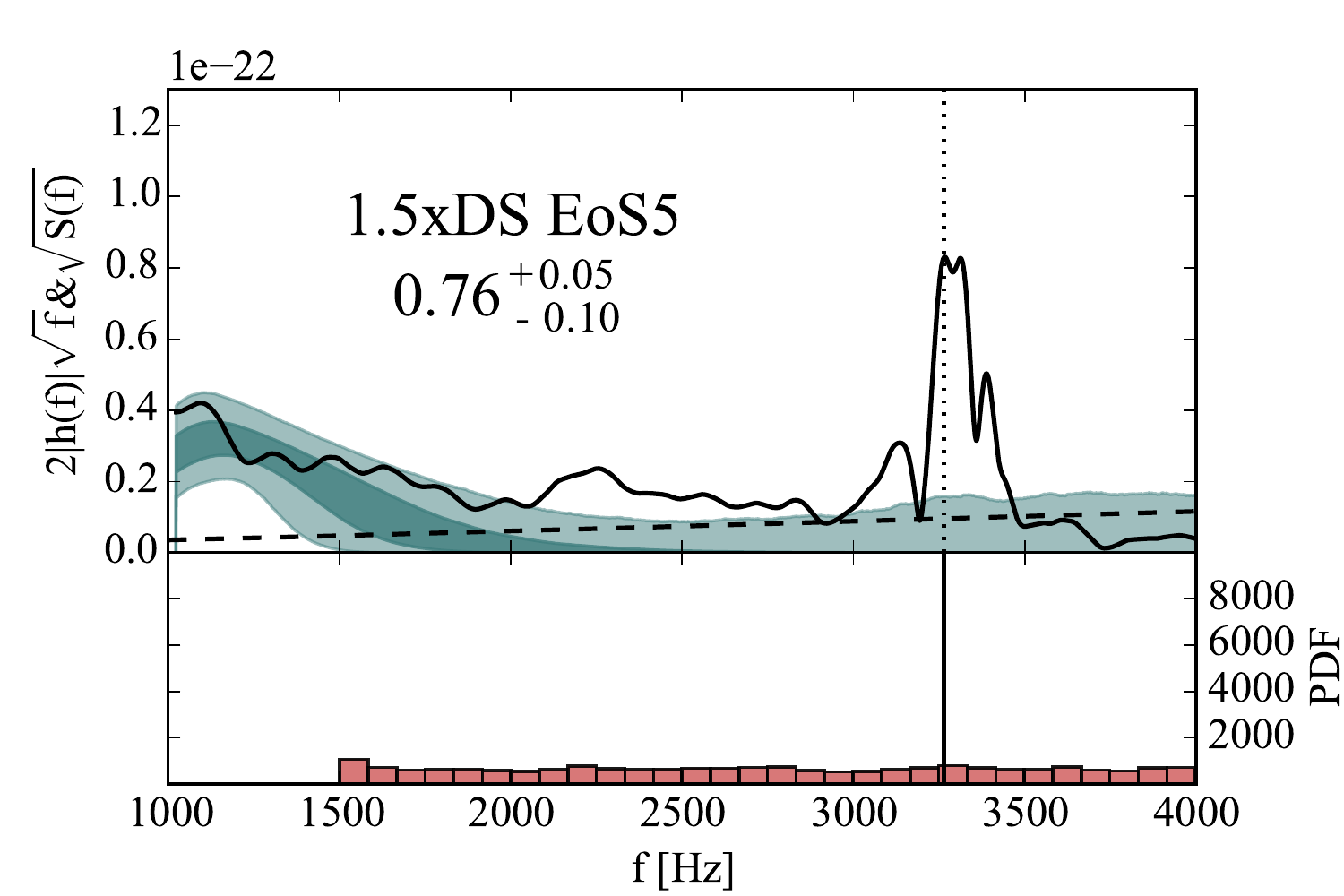}
\includegraphics[width=0.32\textwidth]{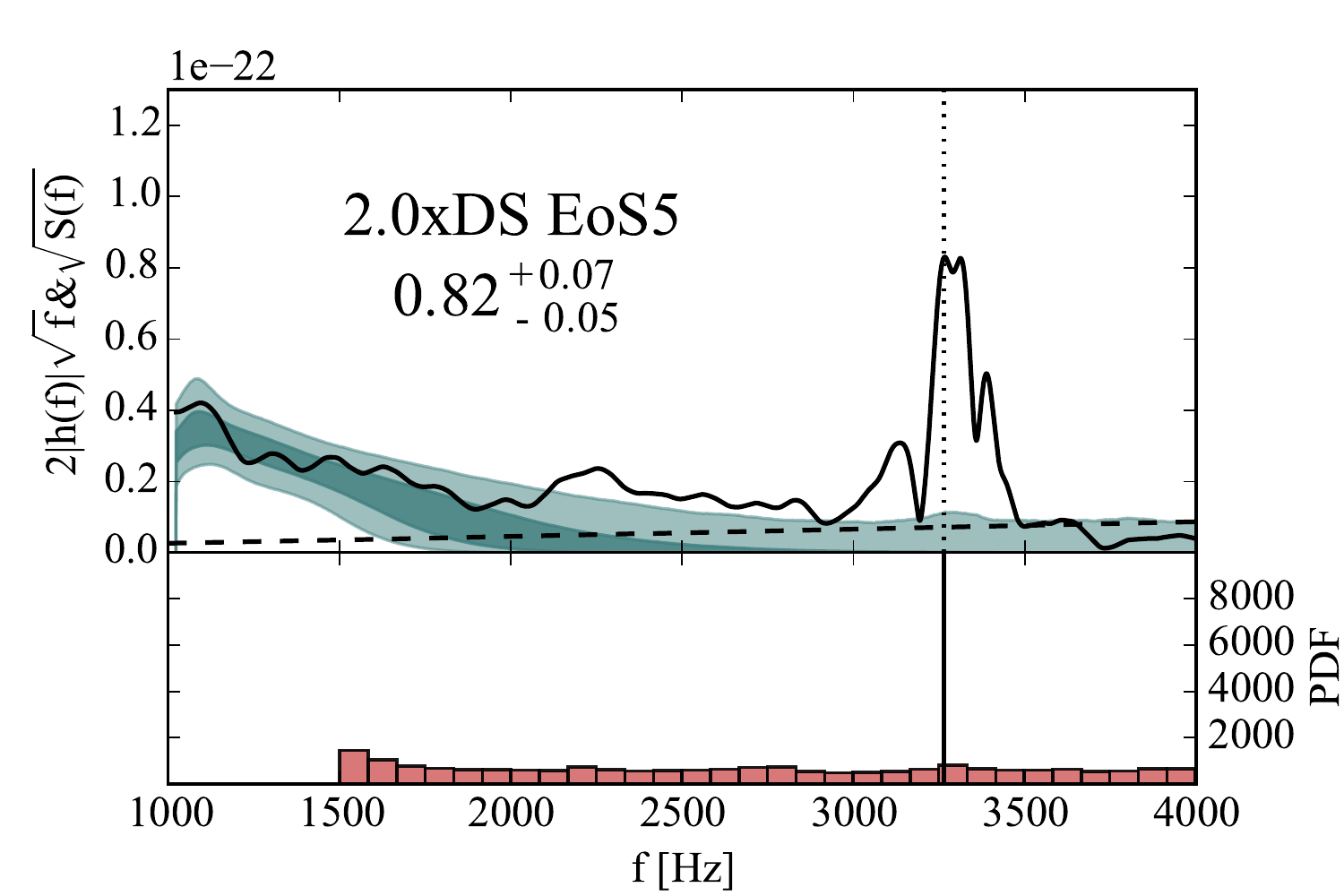}\\
\includegraphics[width=0.32\textwidth]{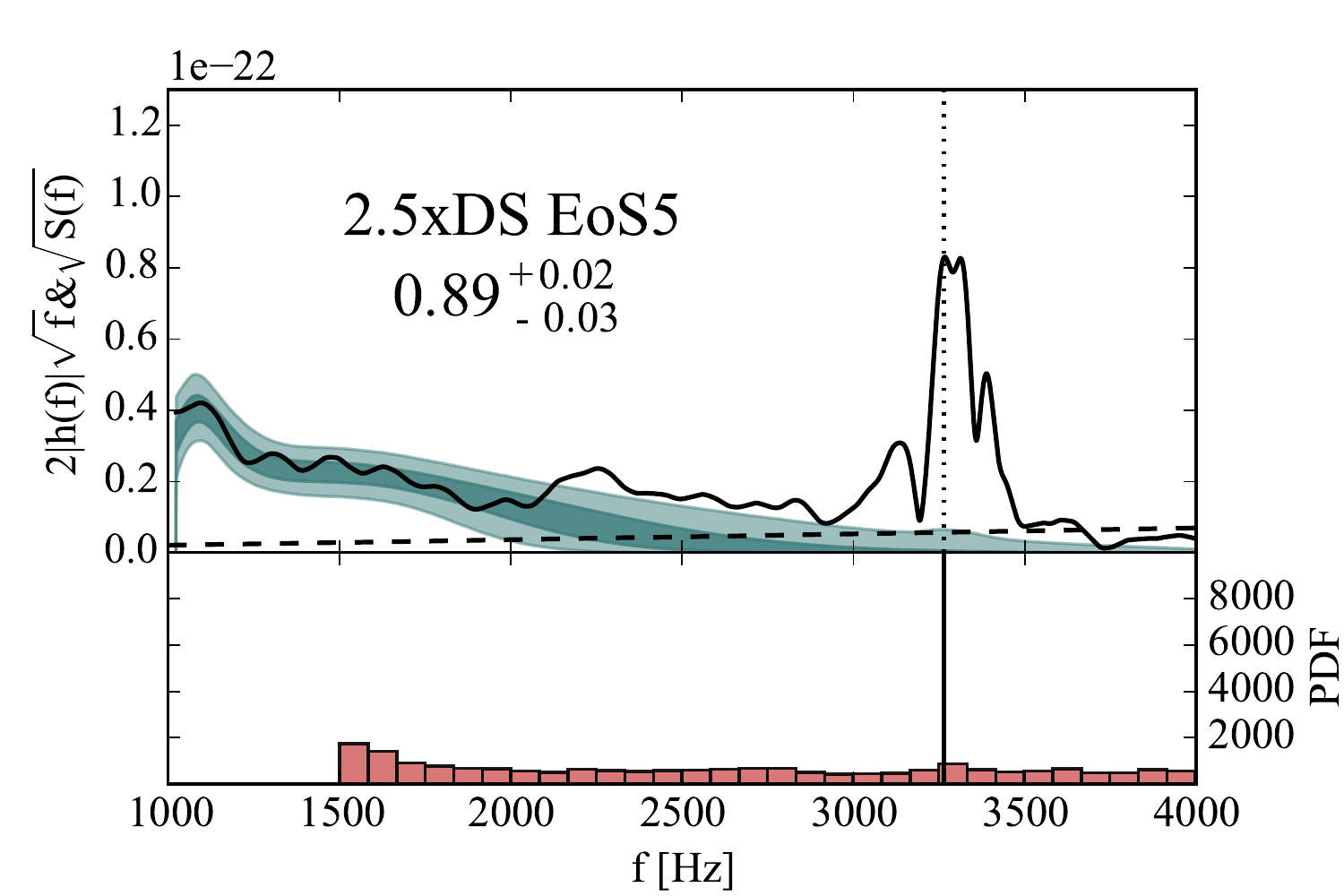}
\includegraphics[width=0.32\textwidth]{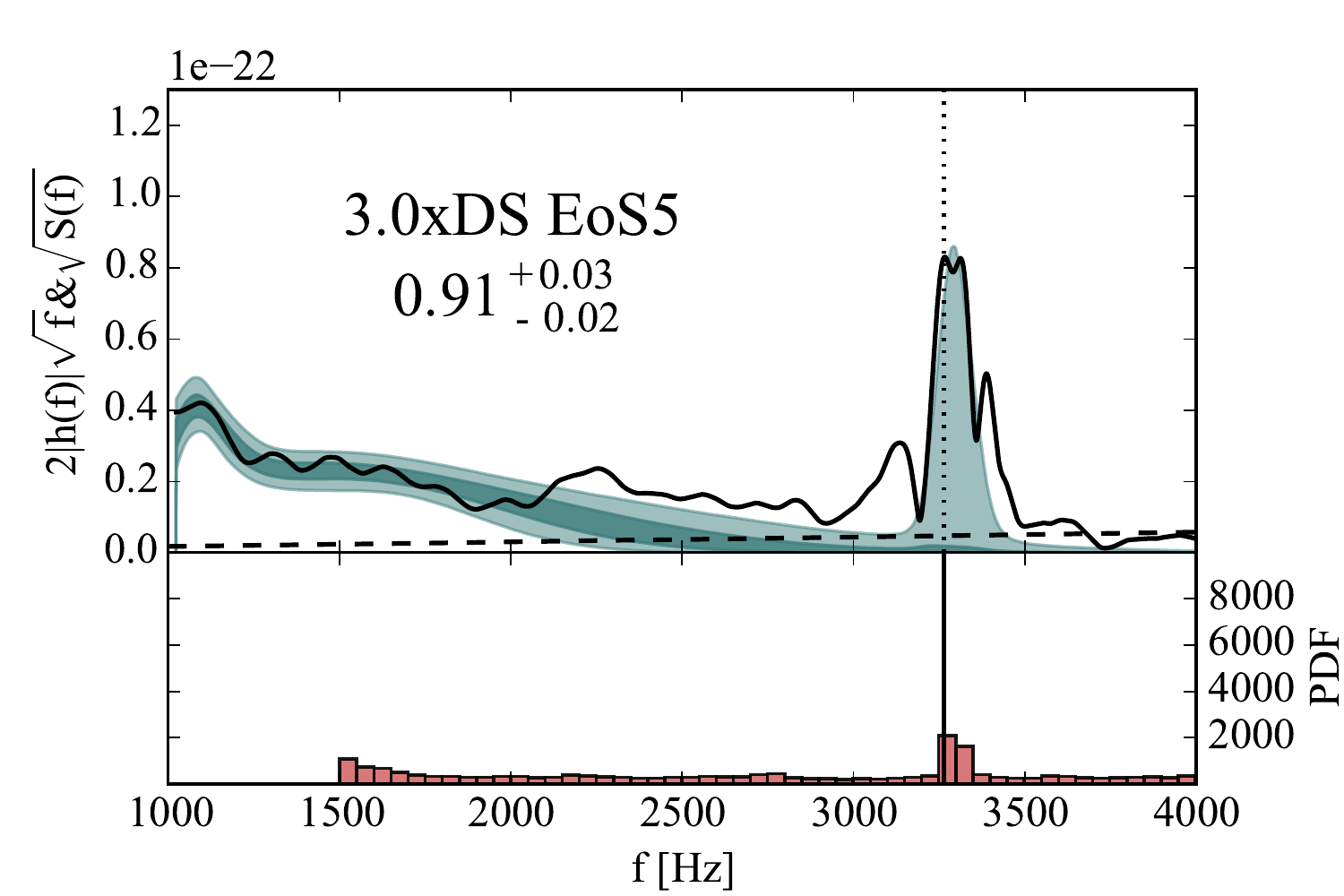}
\includegraphics[width=0.32\textwidth]{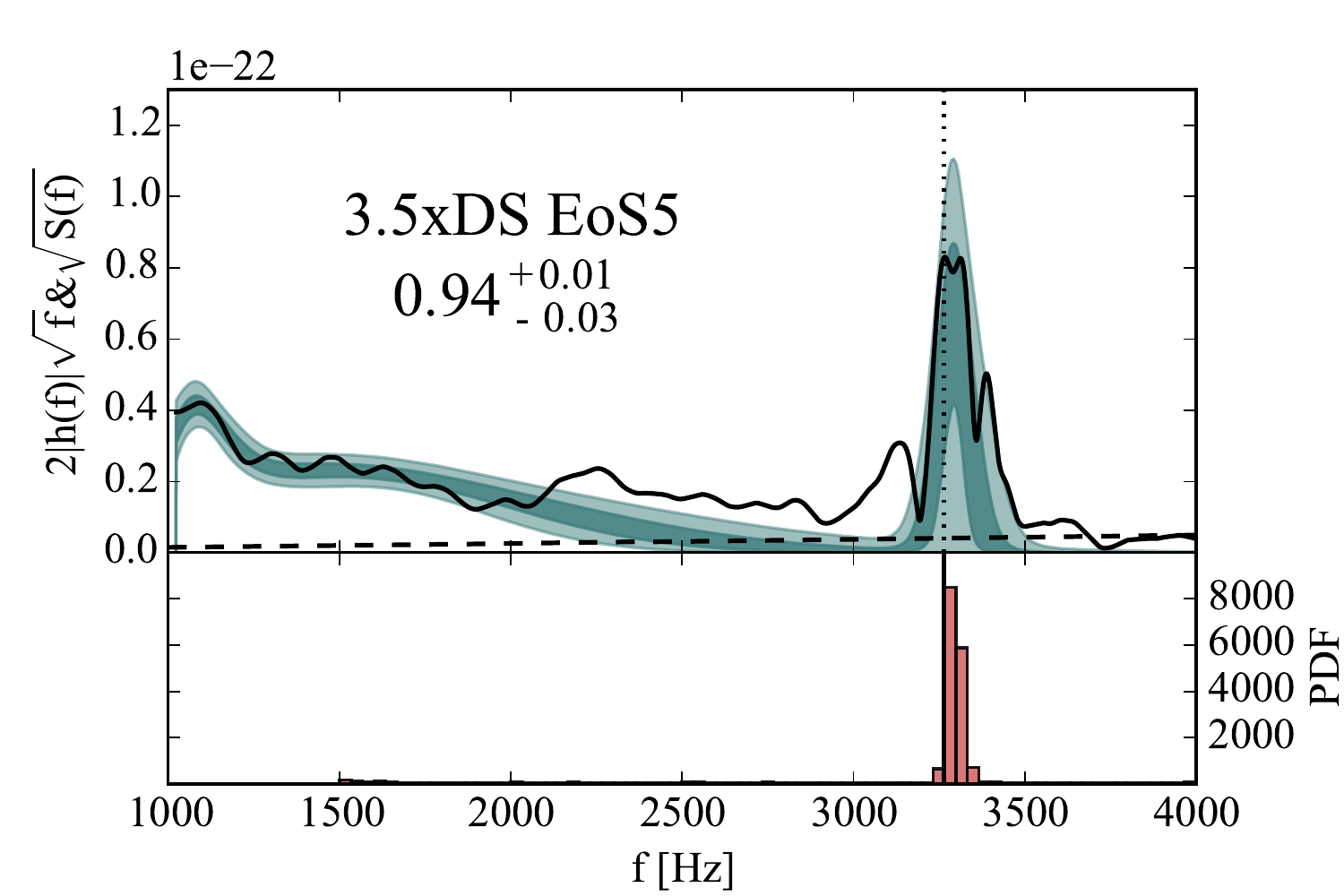}\\
\includegraphics[width=0.32\textwidth]{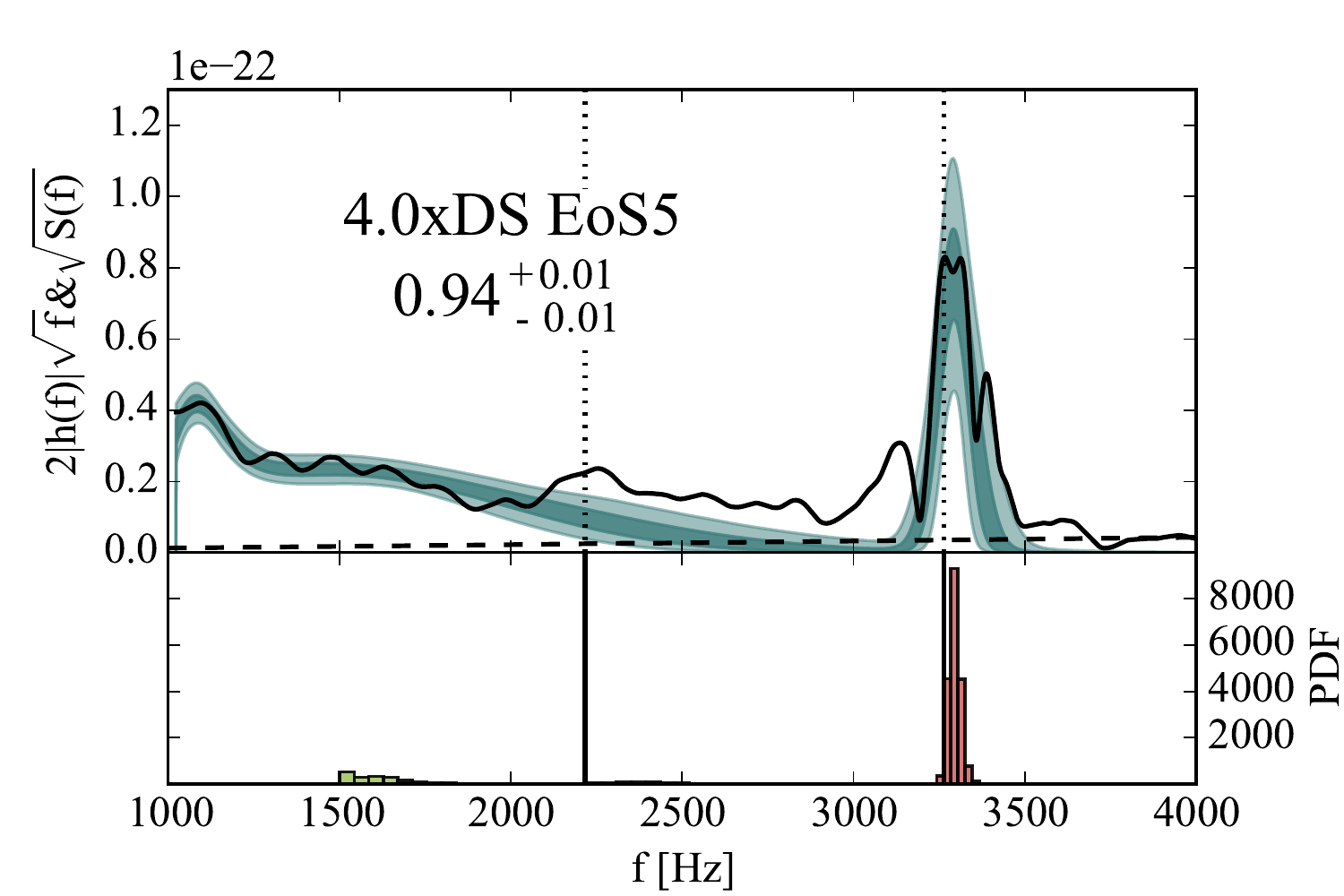}
\includegraphics[width=0.32\textwidth]{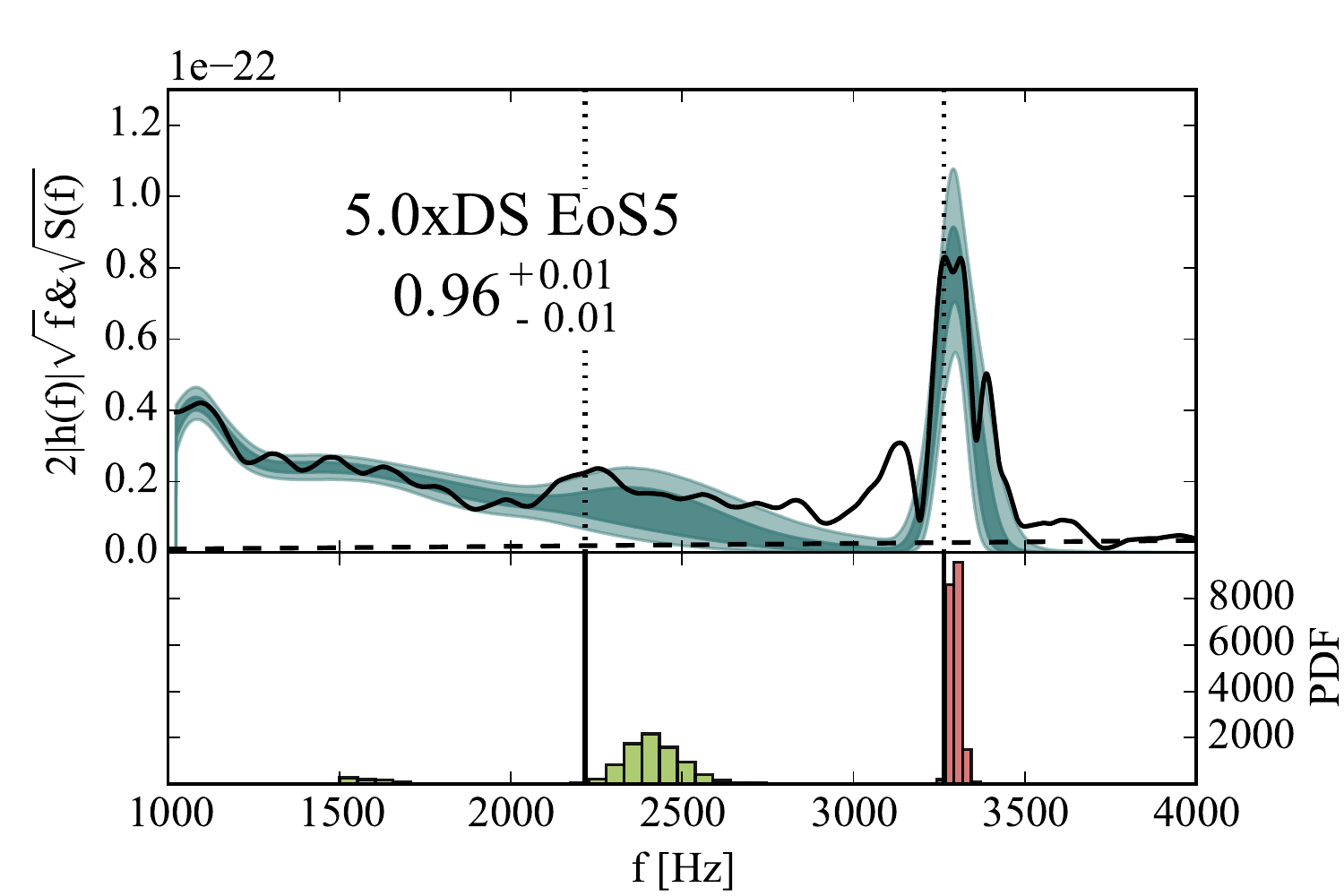}
\includegraphics[width=0.32\textwidth]{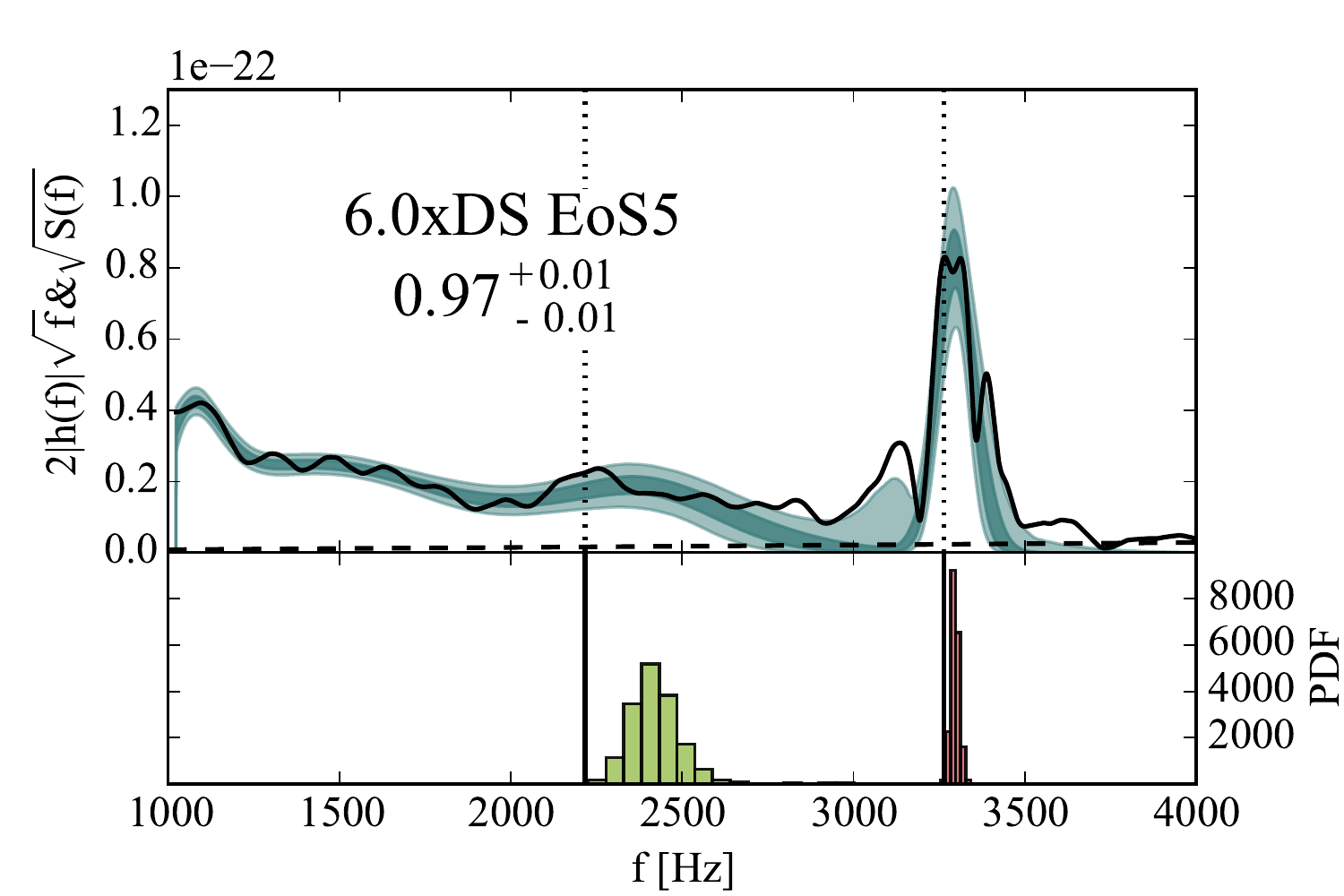}
\caption{Reconstruction of the post-merger signal emitted during the coalescence of an equal-mass binary with EoS5 at various network sensitivities. In each panel the top plot shows the $50\%$ and $90\%$ credible interval for the signal spectrum with dark and light shaded regions respectively. The bottom plot shows the posterior density for $\fpeak$ and $\fsub$ (where applicable). The merger phase is reconstructed at $\sim$1.5xDS, the main post-merger peak is extracted at $\sim$3xDS, while hints of a subdominant peak appear at $\sim$4xDS.}
\label{fig: Prod Runs Eos5}
\end{figure*}

Figure~\ref{fig: Prod Runs Eos5} studies the post-merger signal reconstruction for EoS5 and our equal-mass $q=1$ binary system. The signals are injected in a three detector network including H, L, and V where we gradually improve the sensitivity of the $2$ LIGO detectors, while keeping V at its design sensitivity. Each panel shows the $50\%$ (dark shade) and the $90\%$ (light shade) credible interval of the reconstructed spectral amplitude (top) and the $\fpeak$ and $\fsub$ (where applicable) posterior densities (bottom). Each plot label indicates the sensitivity multiplier (e.g., ``1.0x DS'' indicates this is the aLIGO design sensitivity), the EoS (here, the fifth EoS considered), and the overlap between the injected signal and the inferred waveform at the 90\% level\footnote{The overlap between two waveforms is the noise-weighted inner product, defined in Eq. 6 of~\cite{Chatziioannou:2017ixj} and provides an estimate of how similar two waveforms are both in amplitude and in phase. A high overlap (close to $1$) means that the reconstructed waveform is similar to the injected waveform.}.

The top left panel suggests that a $3$ detector network at its nominal design sensitivity is not sufficient to extract the post-merger signal. In this case our analysis results in upper limits for the spectral amplitude of the signal, similar to what was done for GW170817~\cite{Abbott:2018wiz} and a low overlap. At 1.5xDS (top middle panel) and 2.0xDS (top right panel) we reconstruct the late-inspiral/merger phase at frequencies $\sim(1000,2000)$Hz, achieving overlaps around 80\%; there is only minor evidence for signal power at higher frequencies. Between 2.5xDS (middle left panel) and 3.0xDS (middle middle panel) we start seeing hints of a post-merger spectral peak at around 3300Hz and hence evidence for the presence of a NSR. As the sensitivity increases further, the main spectral peak is reconstructed more accurately. At 4.0xDS (bottom left panel) the reconstructed signal starts exhibiting hints of a subdominant peak at around 2500Hz. Finally, at 5.0xDS (bottom middle panel) and above, both $\fpeak$ and $\fsub$ can be extracted with high confidence, as also reflected in the high overlap value of about 96\%.

Besides the precise measurement of $\fpeak$ and $\fsub$, the bottom plots in Fig.~\ref{fig: Prod Runs Eos5} shows that the inferred posterior distributions do not peak exactly at the target values which correspond to the peak of the injected waveform. This was first noted in~\cite{Chatziioannou:2017ixj} in the context of $\fpeak$ and was attributed to the fact that the post-merger peaks are not symmetric. Something similar is observed here with $\fsub$ and we again argue that this is caused by the shape of the subdominant spectrum peak. Indeed, the reconstructed signal exhibits a broad smooth subdominant peak (for example bottom right plot for 6.0xDS at around $2200-2500$Hz.). The injected signal, on the other hand, exhibits a subdominant peak with more substructure, resulting in a shift between the target and the recovered $\fsub$.

%%%%%%%%%%%%%%%%%%%%%%%%%%%%%%%%%%%%%%%%%%%%%%%%%%%%%%%%%

\begin{figure}[!htbp]
\includegraphics[width=\columnwidth,clip=true]{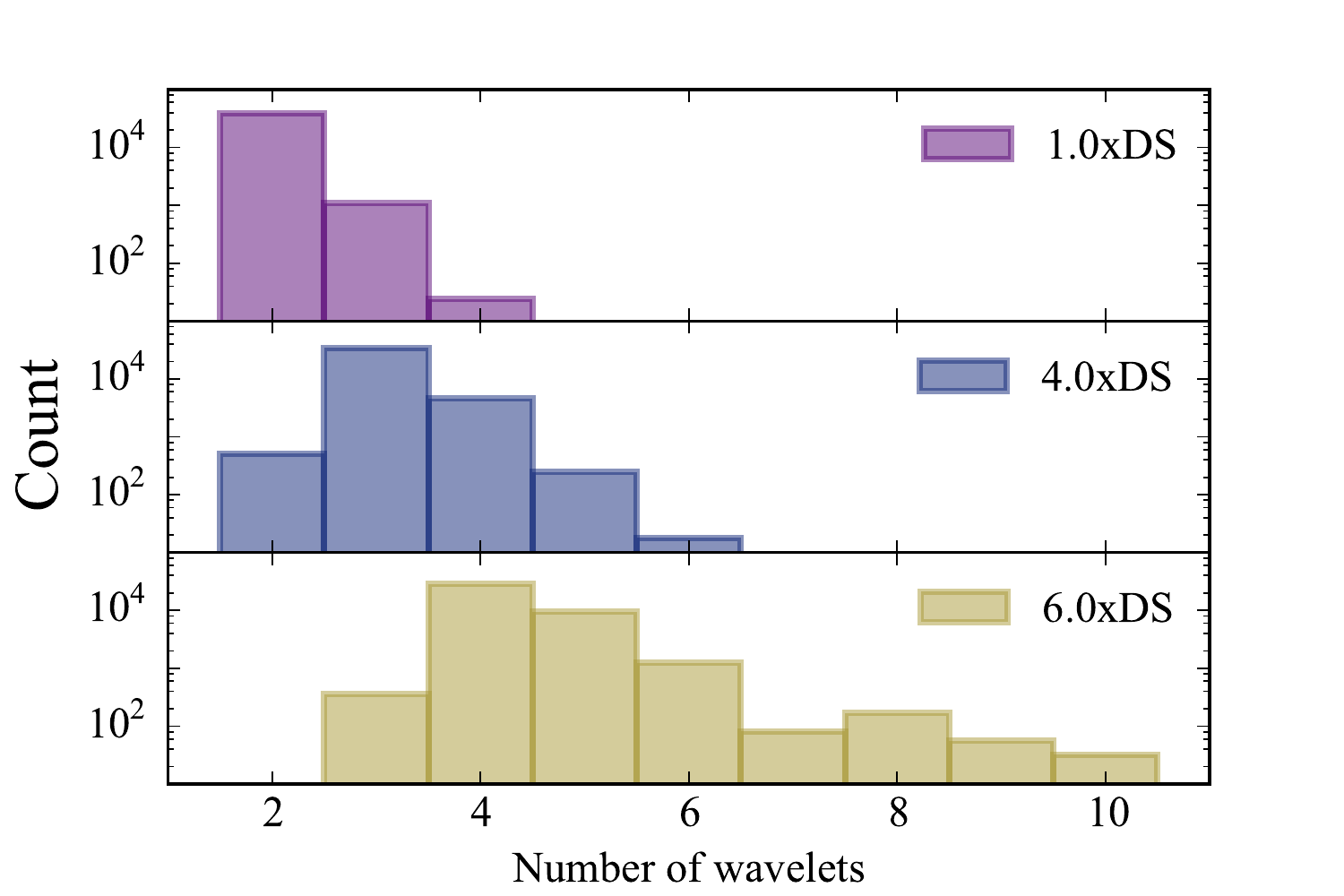}
\caption{Posterior for the number of wavelets used in the reconstruction of selected signals from Fig.~\ref{fig: Prod Runs Eos5}. As the detector sensitivity improves, the reconstruction employs an increasing number of wavelets. This results in a more faithful reconstruction of the injected signal.}
\label{fig: Waves Runs Eos5}
\end{figure}

Overall, we find that, as expected, increasing the detector sensitivity leads to higher quality signal reconstructions: the credible intervals for the dominant and subdominant peaks narrow down and the signal reconstruction includes more subtle details of the injected waveform. {\tt BayesWave} achieves this increasingly detailed reconstruction by utilizing a larger number of wavelets. 
These additional wavelets are used for the reconstruction of the various features of the signal including the merger, main post-merger peak, and subdominant post-merger peaks. Moreover, additional wavelets are needed to capture small changes in the value of these frequency components.
Indeed, Fig.~\ref{fig: Waves Runs Eos5} shows the posterior for the number of wavelets used in the reconstruction of selected signals from Fig.~\ref{fig: Prod Runs Eos5}.

%%%%%%%%%%%%%%%%%%%%%%%%%%%%%%%%%%%%%%%%%%%%%%%%%%%%%%%%%

\begin{figure*}[!htbp]
\centering
\includegraphics[width=0.55\textwidth]{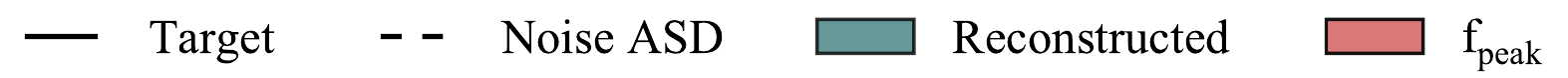}\\
\includegraphics[width=0.32\textwidth]{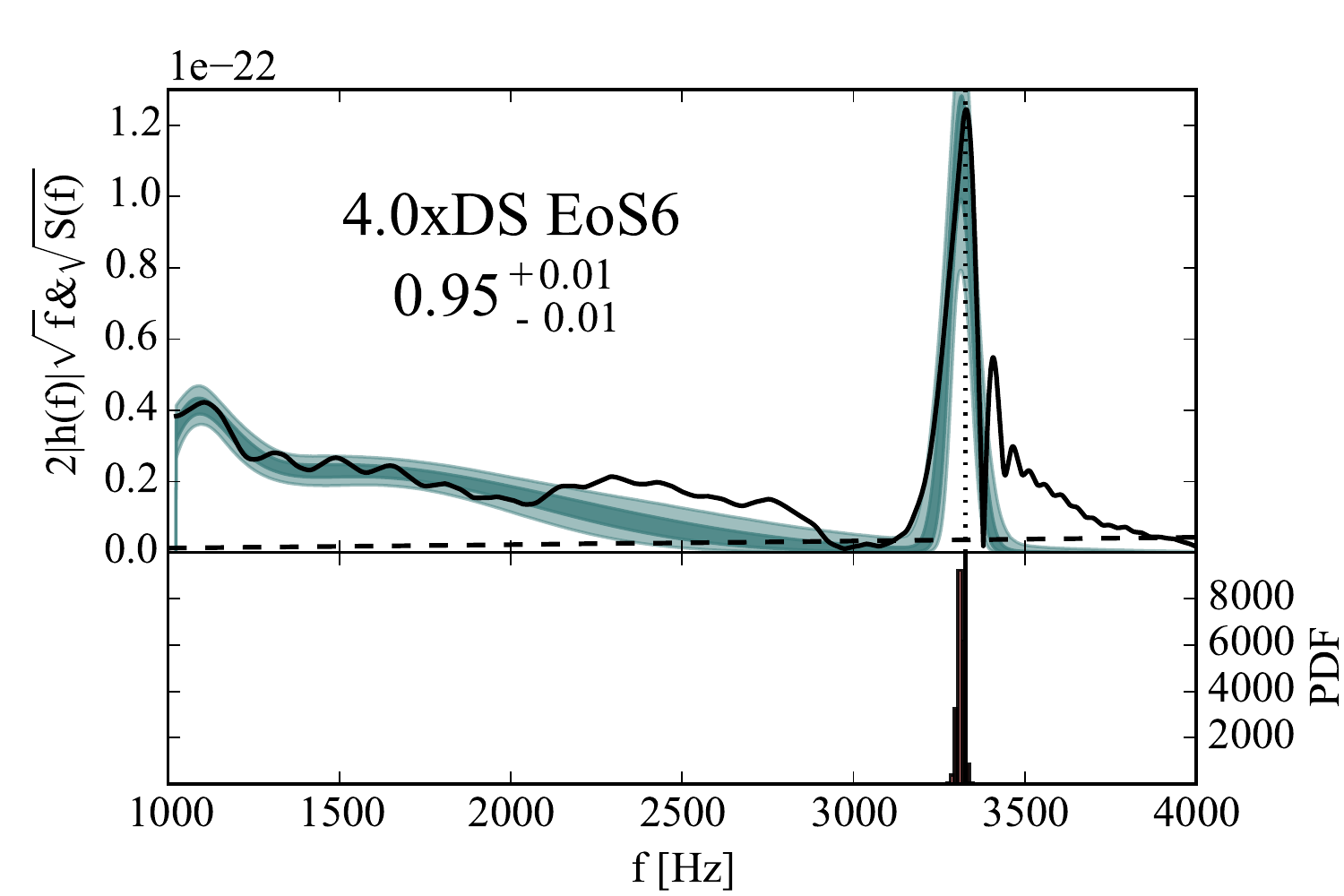}
\includegraphics[width=0.32\textwidth]{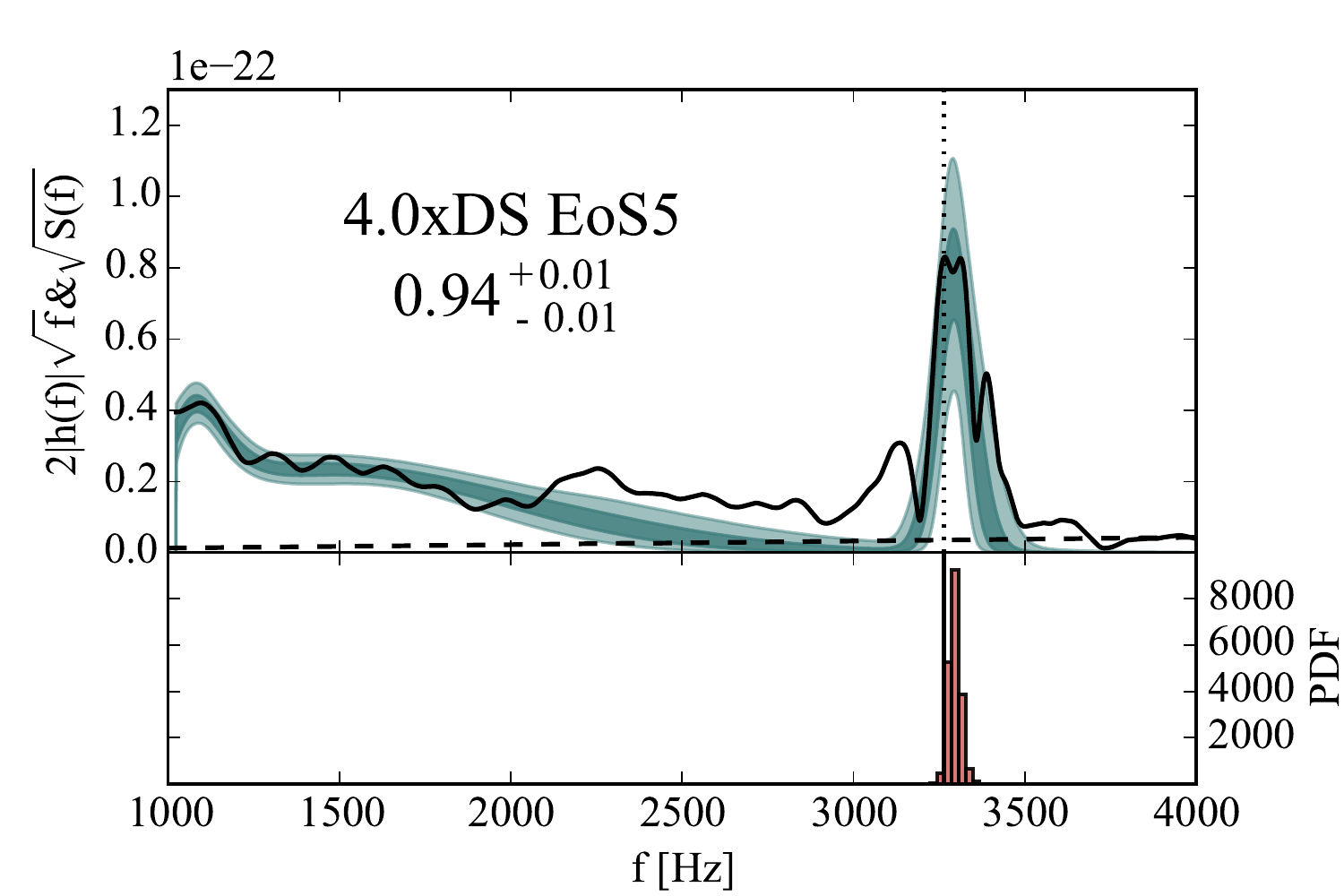}
\includegraphics[width=0.32\textwidth]{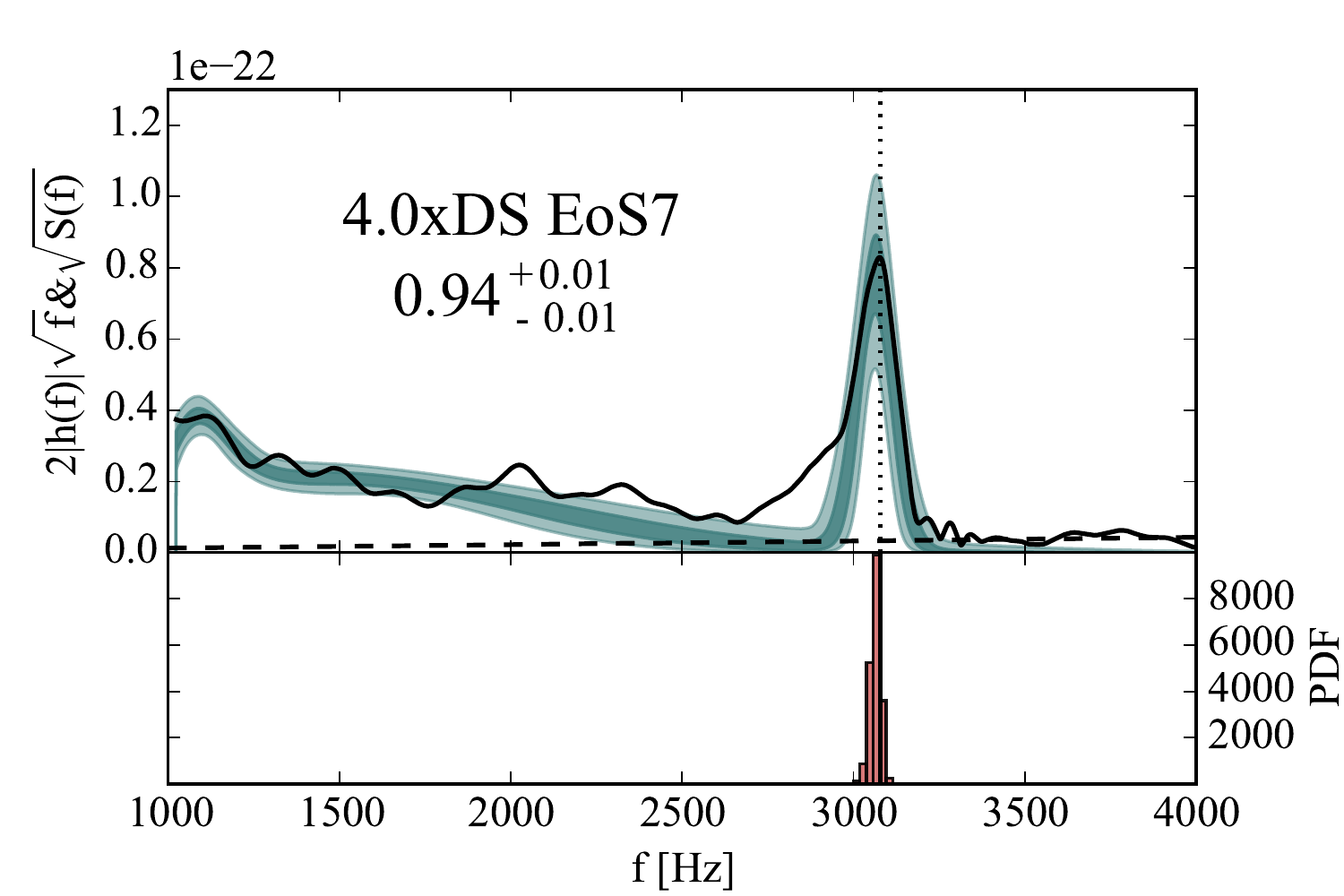}\\
\includegraphics[width=0.32\textwidth]{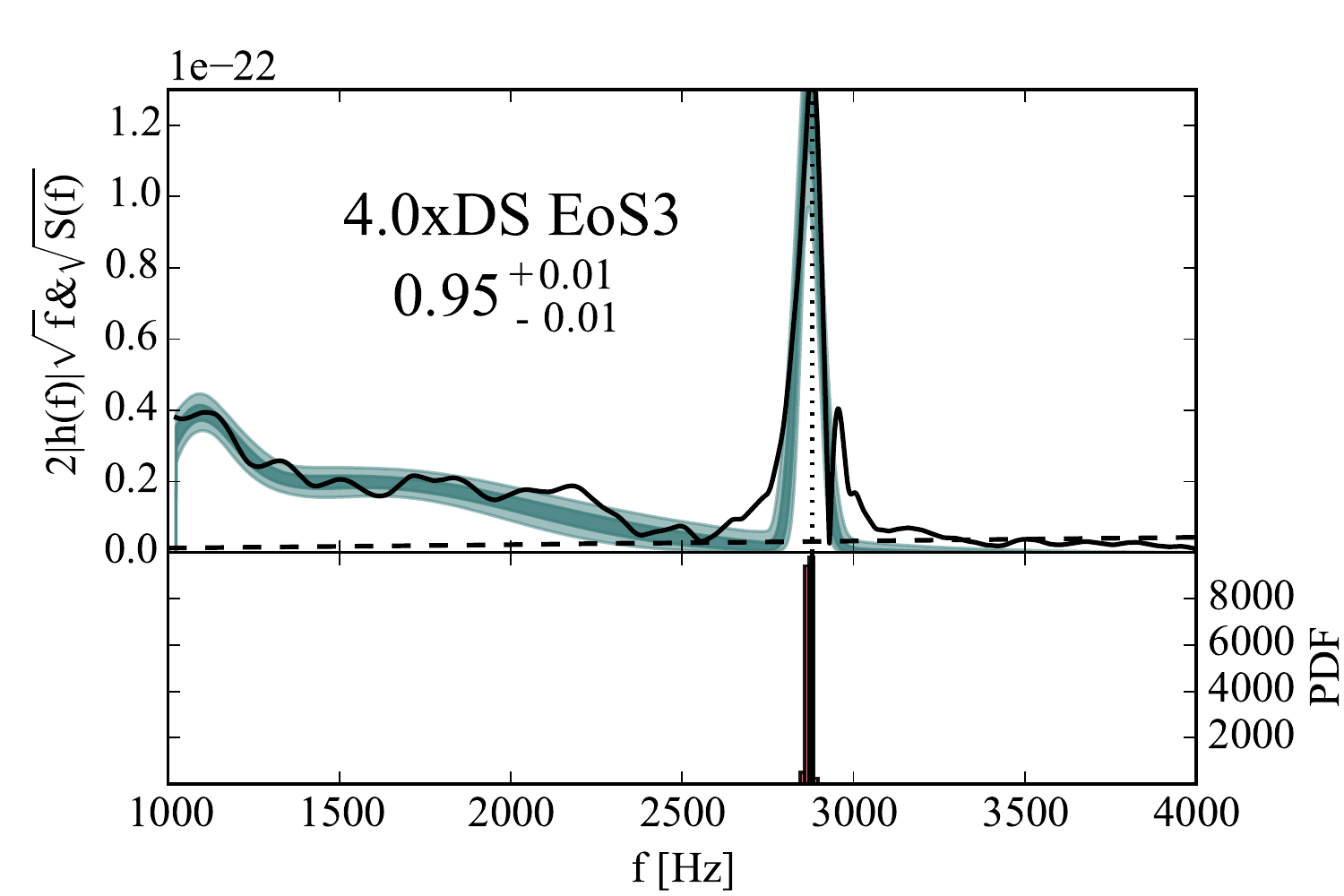}
\includegraphics[width=0.32\textwidth]{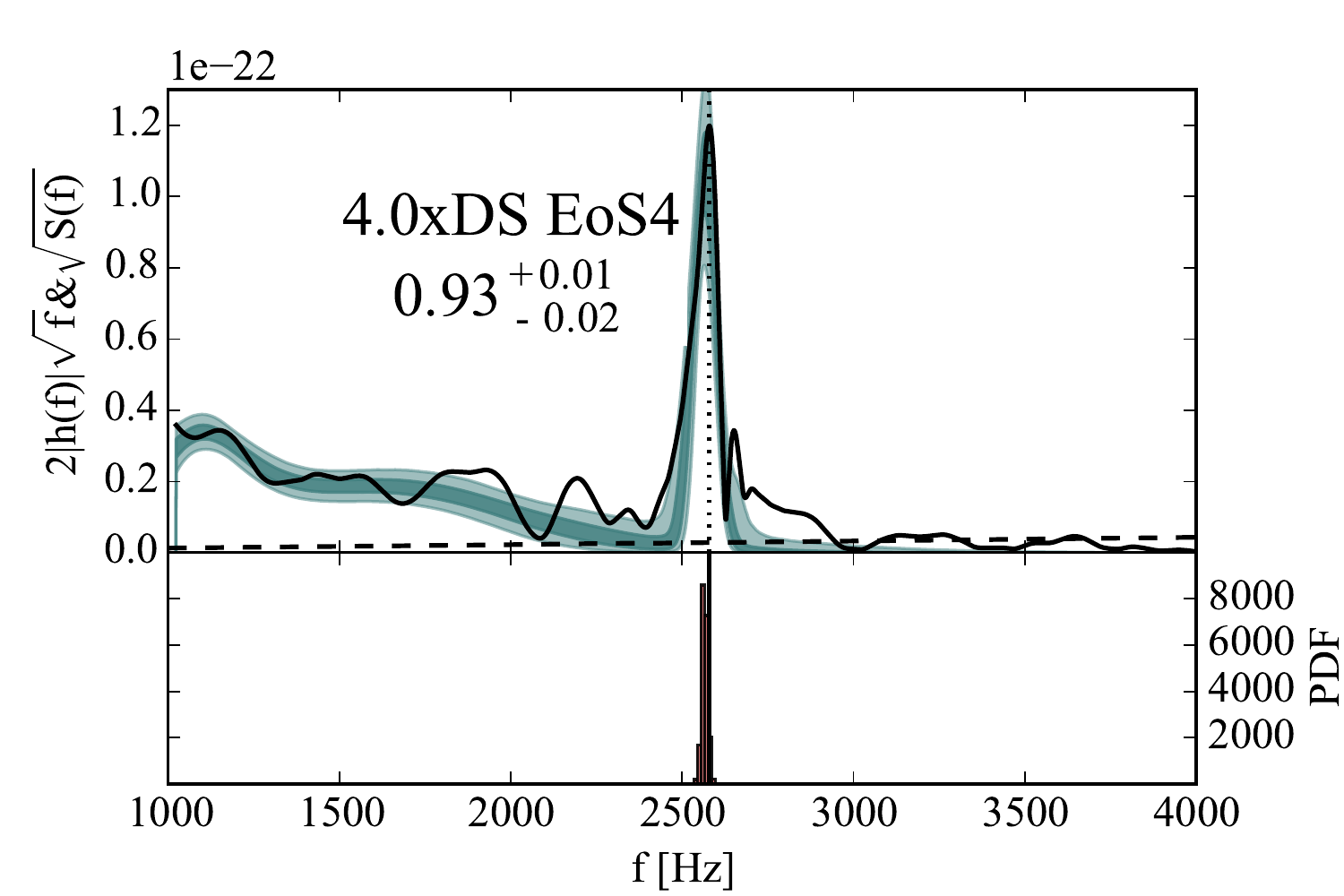}
\caption{Reconstruction of the post-merger signal and posterior density for $\fpeak$ for various EoSs consistent with GW170817 that lead to a NSR and an equal-mass binary. The signals are injected in a network of two LIGO detectors at 4.0xDS and Virgo at is design sensitivity. In each panel the top plot shows the $50\%$ and $90\%$ credible interval for the signal spectrum with dark and light shaded regions respectively. The bottom plot shows the posterior density for $\fpeak$. }
\label{fig: Equal Masses 4xDS EoS}
\end{figure*}

\begin{figure*}[!htbp]
\centering
\includegraphics[width=0.55\textwidth]{legend_nofsub.pdf}\\
\includegraphics[width=0.32\textwidth]{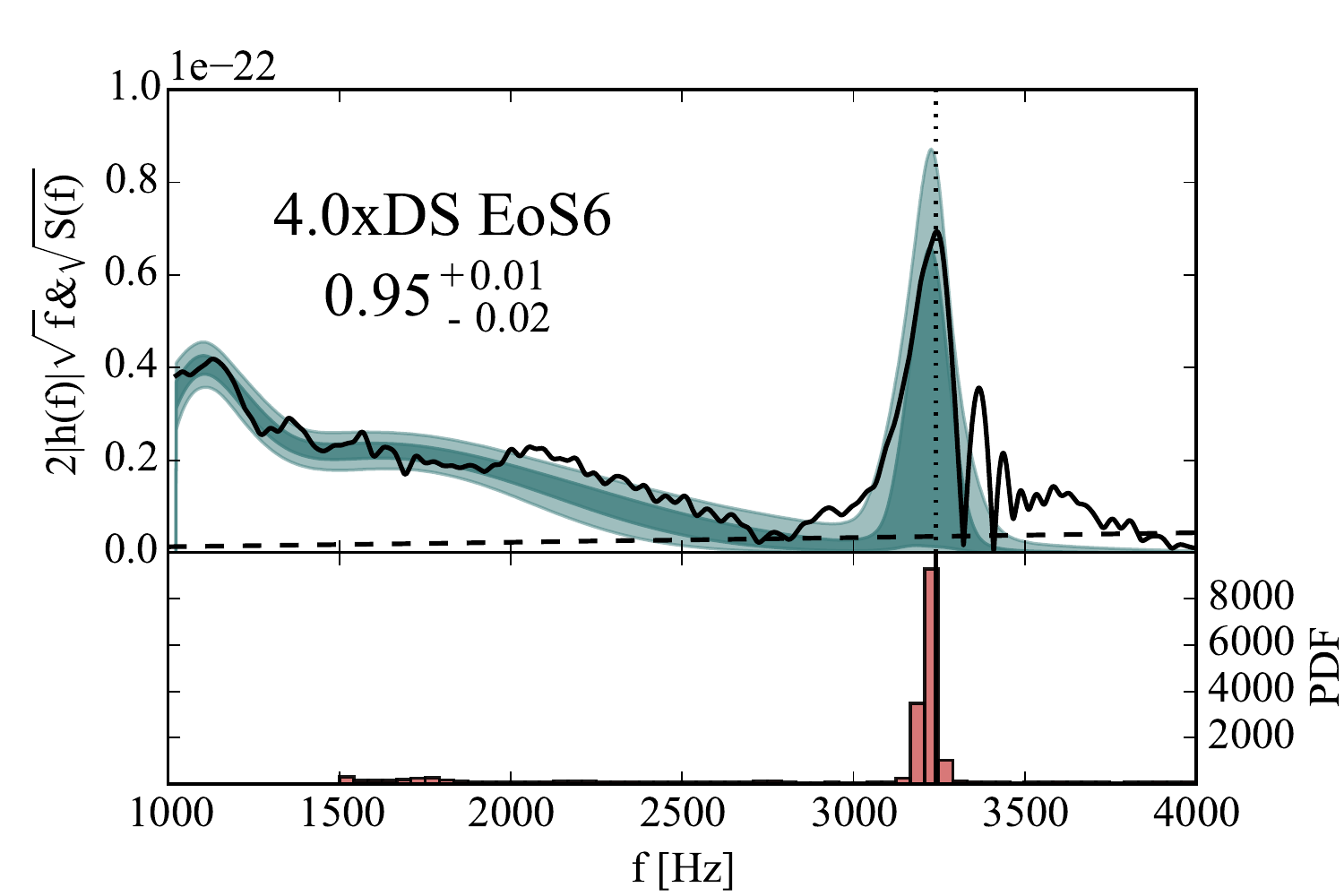}
\includegraphics[width=0.32\textwidth]{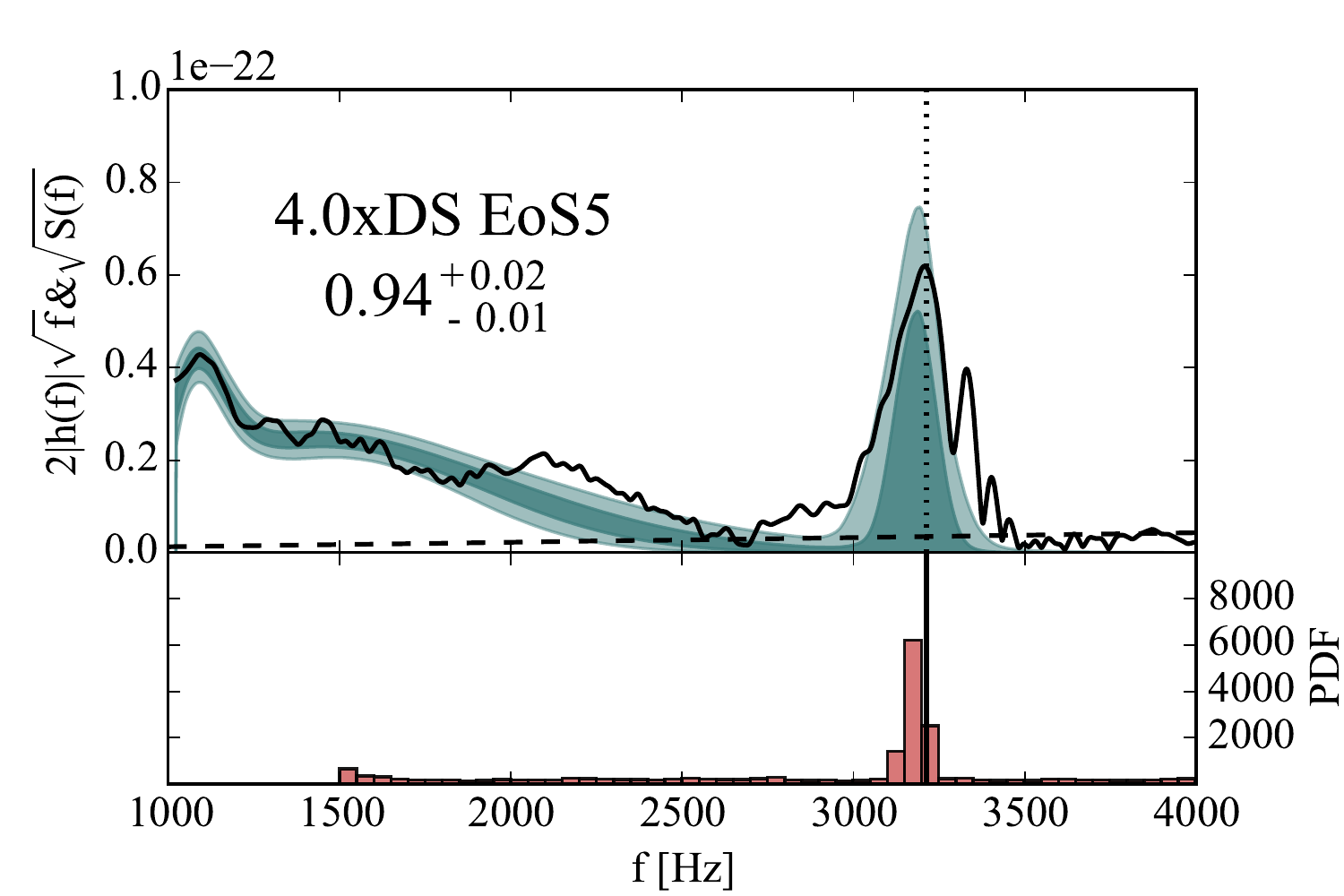}
\includegraphics[width=0.32\textwidth]{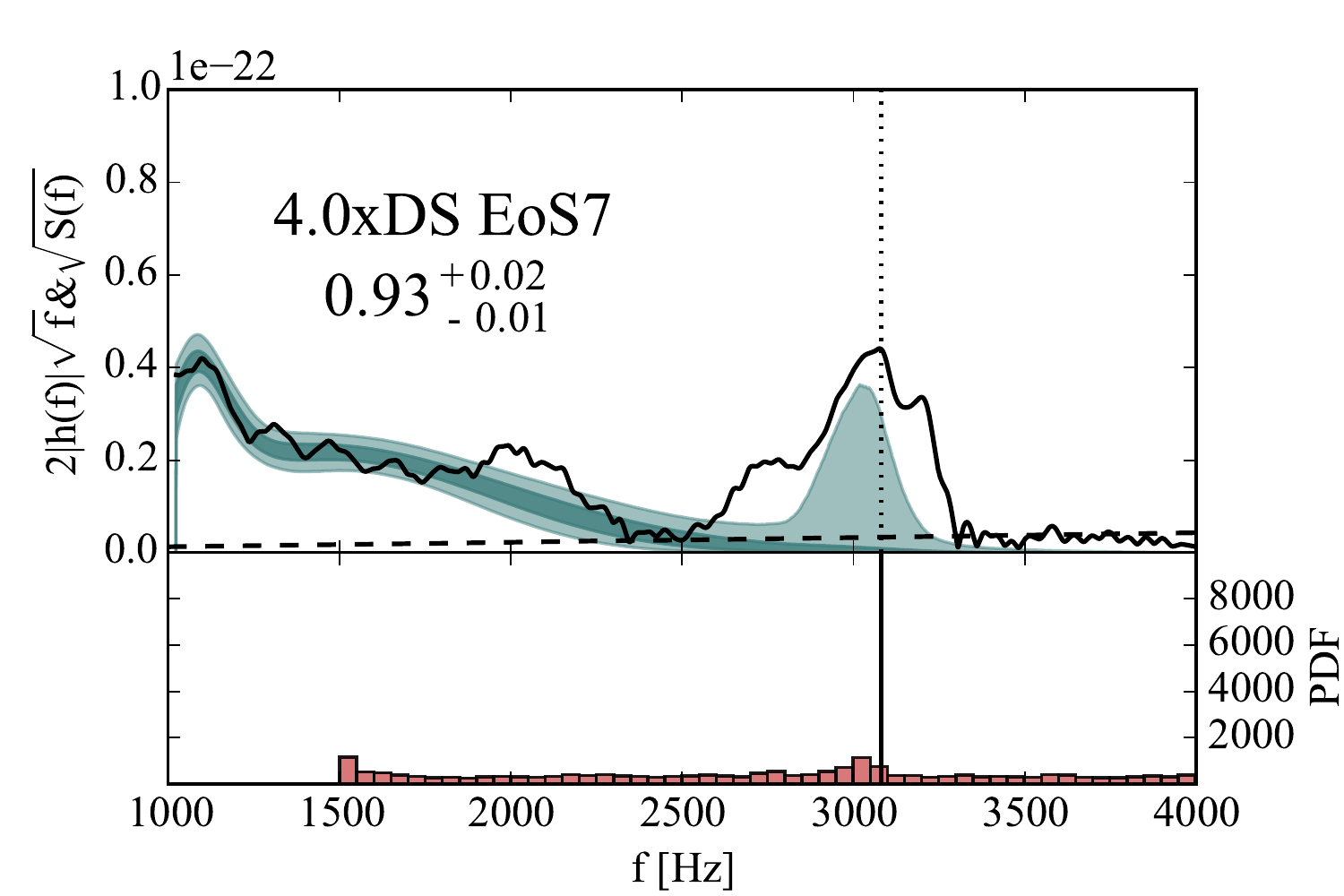}\\
\includegraphics[width=0.32\textwidth]{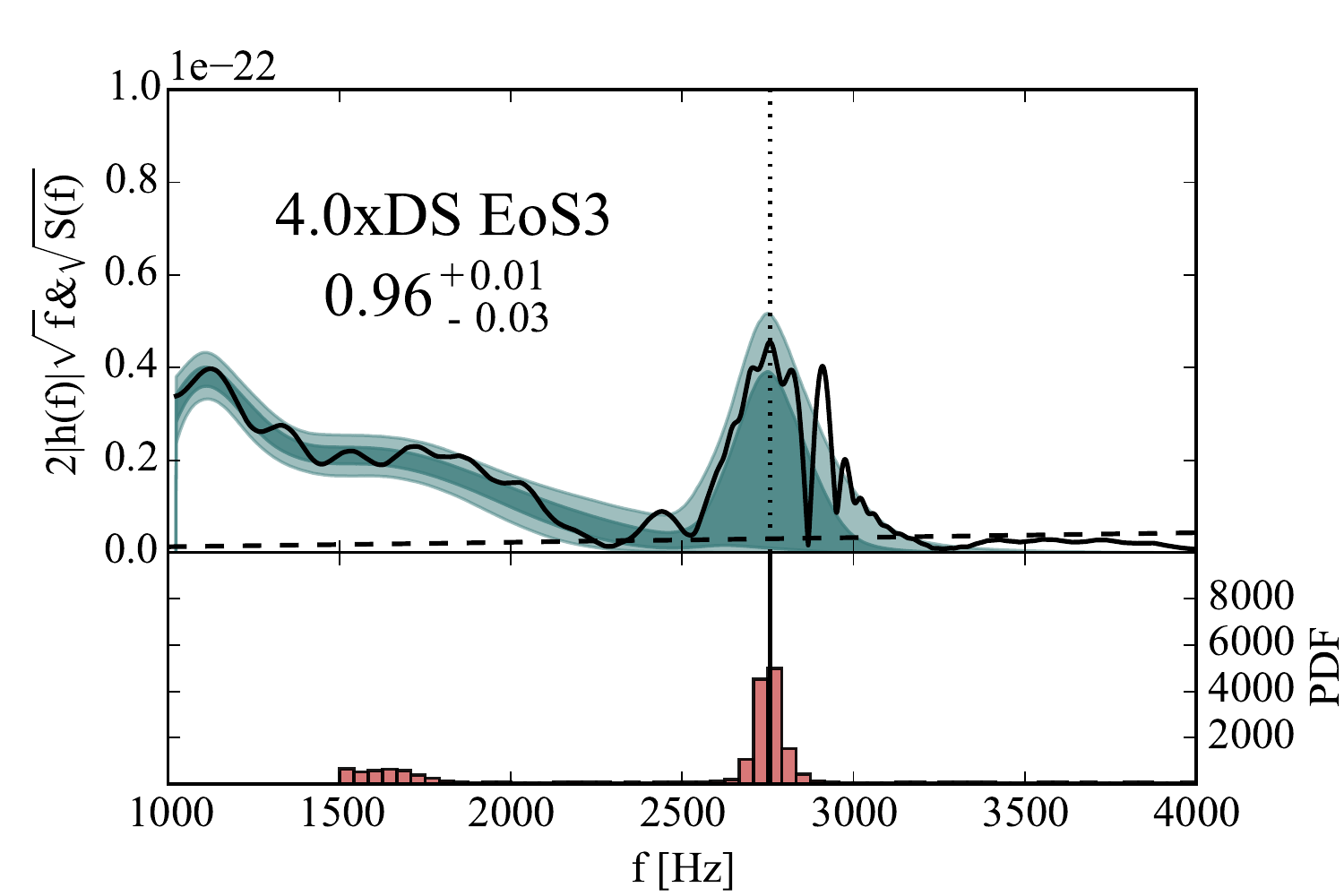}
\includegraphics[width=0.32\textwidth]{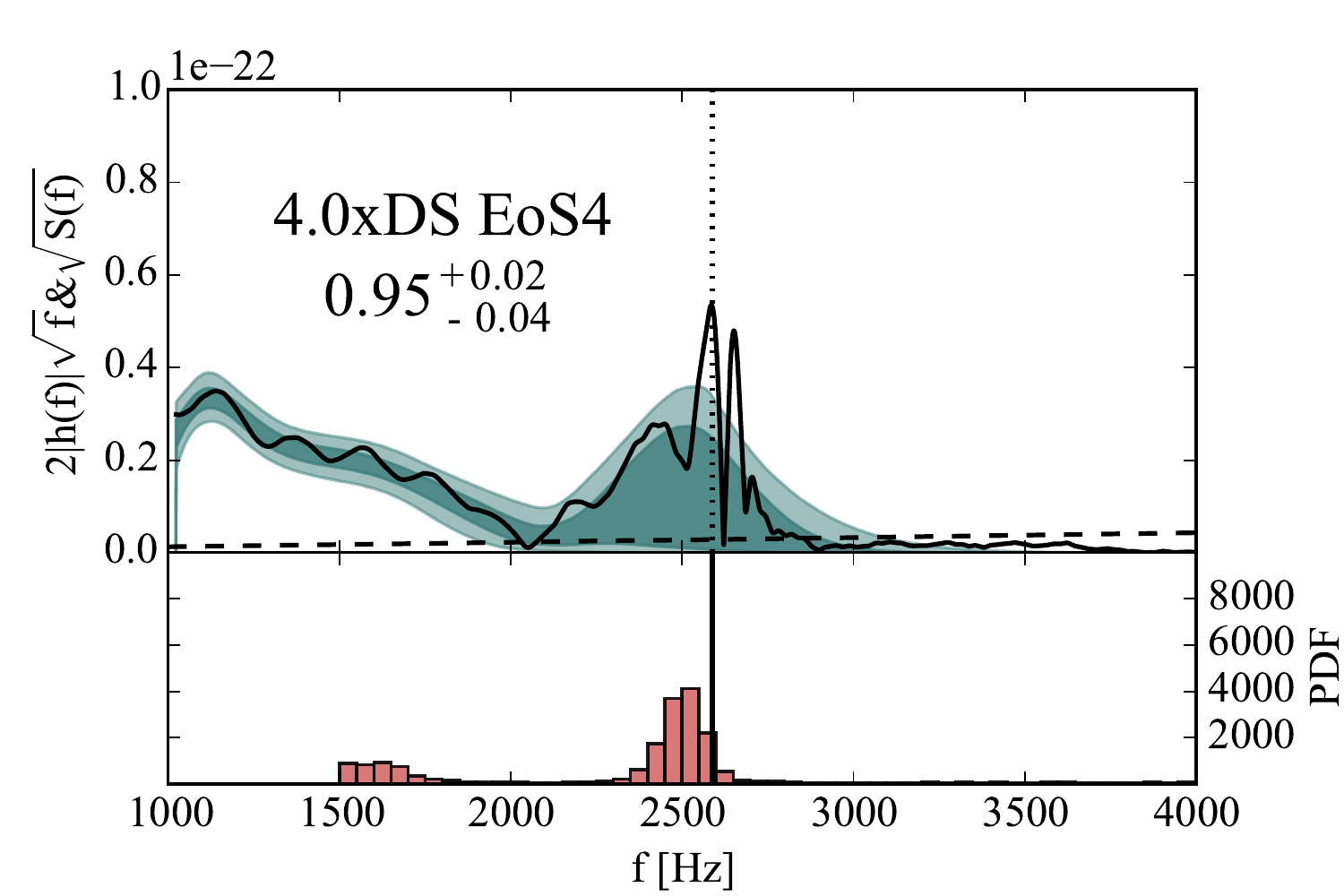}
\caption{Reconstruction of the post-merger signal and posterior density for $\fpeak$ for various EoSs consistent with GW170817 that lead to a SNR and an unequal-mass binary. The signals are injected in a network of two LIGO detectors at 4xDS and Virgo at is design sensitivity. In each panel the top plot shows the $50\%$ and $90\%$ credible interval for the signal spectrum with dark and light shaded regions respectively. The bottom plot shows the posterior density for $\fpeak$.}
\label{fig: Unequal Masses 4xDS EoS}
\end{figure*}

The qualitative results obtained above for EoS5 are representative of EoSs that lead to a NSR. A comparison between the different relevant EoSs studied here is shown in Figs.~\ref{fig: Equal Masses 4xDS EoS} and~\ref{fig: Unequal Masses 4xDS EoS} for equal and unequal masses respectively. We present the reconstructed spectrum and the $\fpeak$ posterior for EoS6, EoS5, EoS7, EoS3, and EoS4 for signals injected in a detector network at 4.0xDS. These plots show the wide range of possible post-merger signals possible for GW170817 assuming that the remnant did not immediately collapse into a BH.

Figures~\ref{fig: Equal Masses 4xDS EoS} and ~\ref{fig: Unequal Masses 4xDS EoS} show that even though the mass ratio does not strongly affect the value $\fpeak$, it affects the general morphology of the post-merger signal. We find that in the equal-mass case the post-merger signal is nicely reconstructed at 4.0xDS and the posterior density for $\fpeak$ peaks at the correct injected values. Moreover, at this sensitivity the reconstructions show hints of the presence of $\fsub$, however improved sensitivities or louder signals will most probably be needed before we can claim the presence of subdominant structure in the spectrum.

In the unequal-mass case, on the other hand, the complicated signal morphology is more difficult to extract. The main post-merger peak has a fairly large width with traces of substructure in some cases. The complicated spectrum leads to a degraded reconstruction, though $\fpeak$ is still extracted at 4.0xDS, with the exception of EoS7. In the future, the quality of the reconstruction could be improved through appropriate priors for {\tt BayesWave}. For example, if an unequal-mass BNS is observed, the prior on the quality factor of the wavelets {\tt BayesWave} uses could be adjusted appropriately so as to favor more wide spectral features.

%%%%%%%%%%%%%%%%%%%%%%%%%%%%%%%%%%%%%%%%%%%%%%%%%%%%%%%%%

\begin{figure}[!htbp]
\centering
\includegraphics[width=\columnwidth]{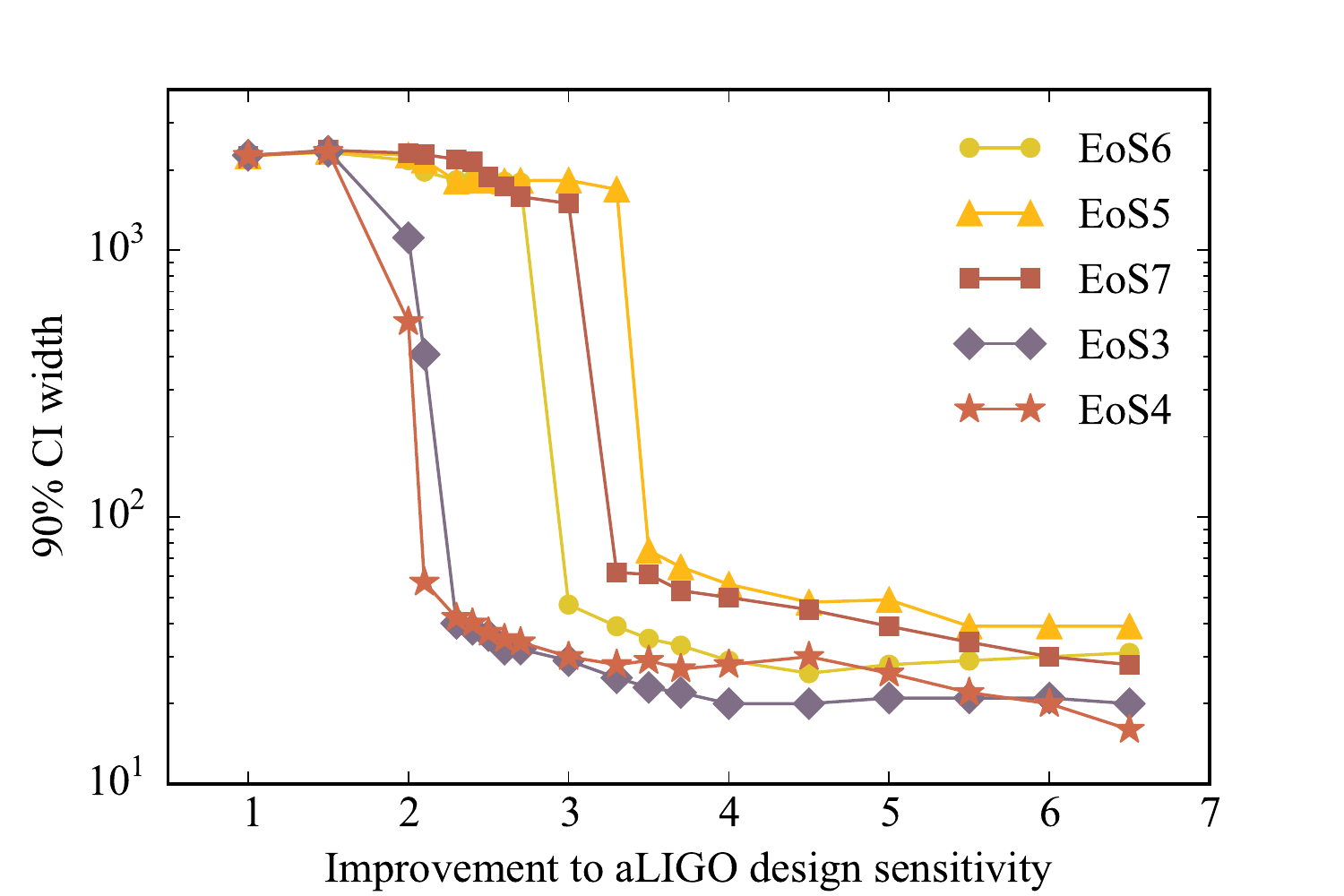}
\includegraphics[width=\columnwidth]{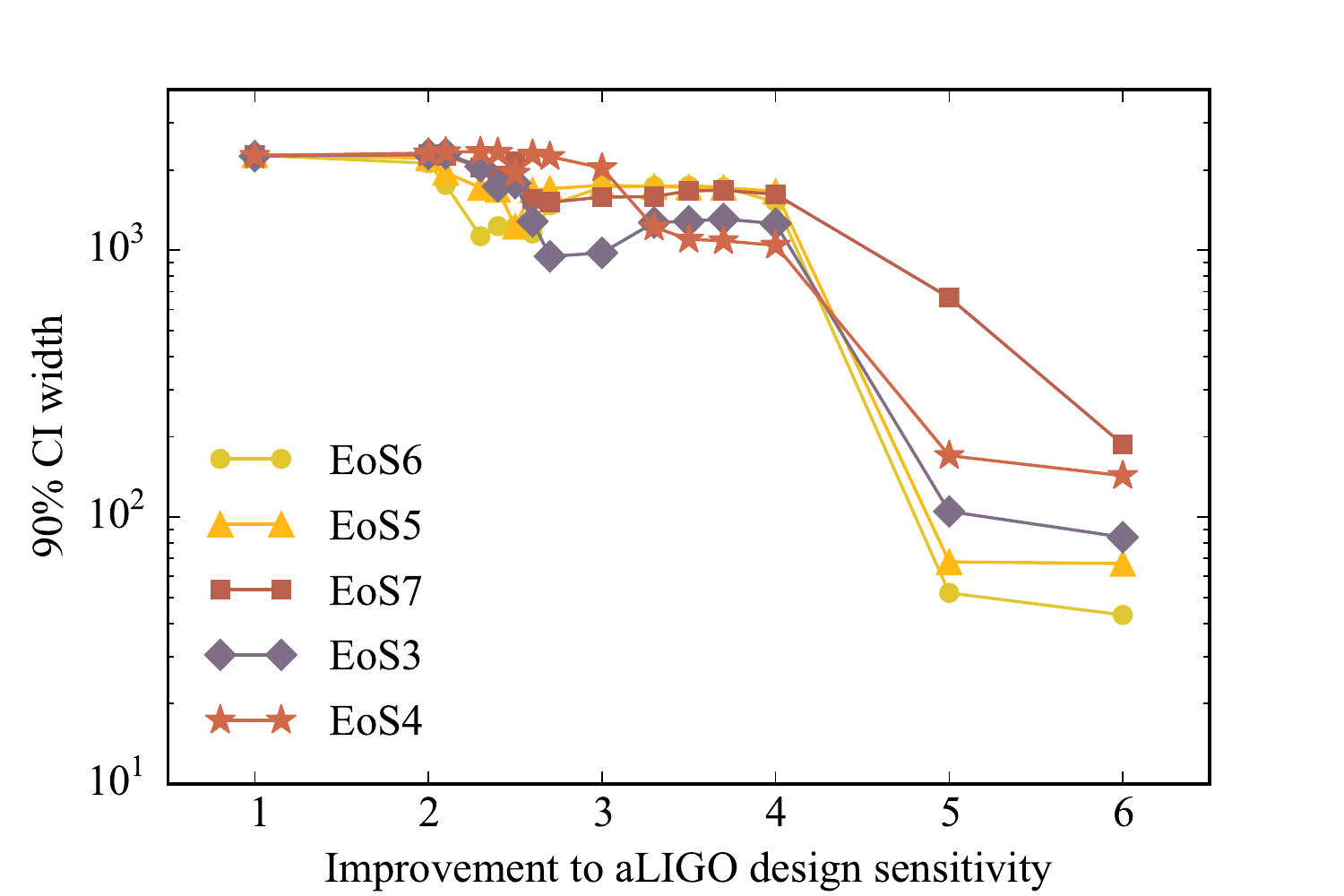}
\caption{Width of the $90\%$ credible interval of the $\fpeak$ posterior for different network sensitivities for equal mass (top) and unequal mass (bottom) systems for the EoSs that result in a NS merger remnant. At low sensitivities the posterior width is equal to the prior width since the signal is not reconstructed. With increasing detector sensitivity we reconstruct the dominant post-merger spectral peak and obtain a measurement of $\fpeak$ to within dozens of Hz. The unequal mass reconstruction and $\fpeak$ measurement is worse than the equal mass one due to the substructure of the spectral peaks, as shown in Fig.~\ref{fig: Unequal Masses 4xDS EoS}. Next generation detectors with sensitivity above 7.0xDS will result in further improvements in the measurement of $\fpeak$, to within tens of Hz for all EoSs, see Sec.~\ref{sec:3g}.}
\label{fig:90c}
\end{figure} 

We summarize the above results in Fig.~\ref{fig:90c}, which shows the width of the $90\%$ credible interval (CI) of the $\fpeak$ posterior for all EoSs as a function of the network sensitivity for equal- (top) and unequal- (bottom) mass systems. At low sensitivity the post-merger spectrum peak is not reconstructed, and the $\fpeak$ posterior is almost equal to its prior, leading to a wide $90\%$ CI of $\gtrapprox 2000$Hz, as also observed in~\cite{Chatziioannou:2017ixj}. Between 2.0xDS and 3.5xDS, depending on the EoS and the mass ratio, the measurement of $\fpeak$ starts vastly improving, resulting in CIs of $\sim100$Hz for the unequal masses scenarios and generally narrower for equal masses. The relatively fast improvement of the $\fpeak$ CI was also observed in~\cite{Chatziioannou:2017ixj} as a function of the signal SNR.

As the sensitivity further improves, the general behavior is for the measurement accuracy of $\fpeak$ to increase. This monotonic reduction of the CI width is obvious in the equal mass case (left panel), however the unequal mass case (right panel) shows a more irregular pattern. We attribute this to the substructure of the broad post-merger peak, see Fig.~\ref{fig: Unequal Masses 4xDS EoS}. In particular as the sensitivity increases, secondary peaks close to the main peak are reconstructed, contributing to the overall uncertainty in the estimation of $\fpeak$. In the future we plan to explore ways to mitigate this, including the already-mentioned priors on the quality factors of the wavelets and $\fpeak$ extraction procedures that take into account the possibility of substructure in the main peak. 

Despite this irregular trend for some simulations, we conclude that the dominant post-merger emission from GW170817 would have been measurable by the $2$ aLIGO detectors, had they been operating at $\sim 2-3\times$ above their design sensitivity, as is expected in the near future. Moreover subdominant features of the post-merger signal can start becoming identifiable at $\sim 4.0$xDS or better.

%-----------------------------------------
\subsection{Third generation detectors}
\label{sec:3g}

The next generation of ground-based gravitational wave detectors is currently in the planning stage, and includes entirely new facilities such as the 10\,km Einstein Telescope and the 40\,km LIGO Cosmic Explorer. To study their capabilities regarding post-merger signals we simulate detector networks with even larger sensitivity that the previous section. In particular we use a 2-detector network of L and V, where V is again assumed to operate at its design sensitivity and L has incrementally increasing sensitivity to match Cosmic Explorer. 

\begin{figure*}[!htbp]
\centering
\includegraphics[width=0.6\textwidth]{legend.pdf}\\
\includegraphics[width=0.45\textwidth]{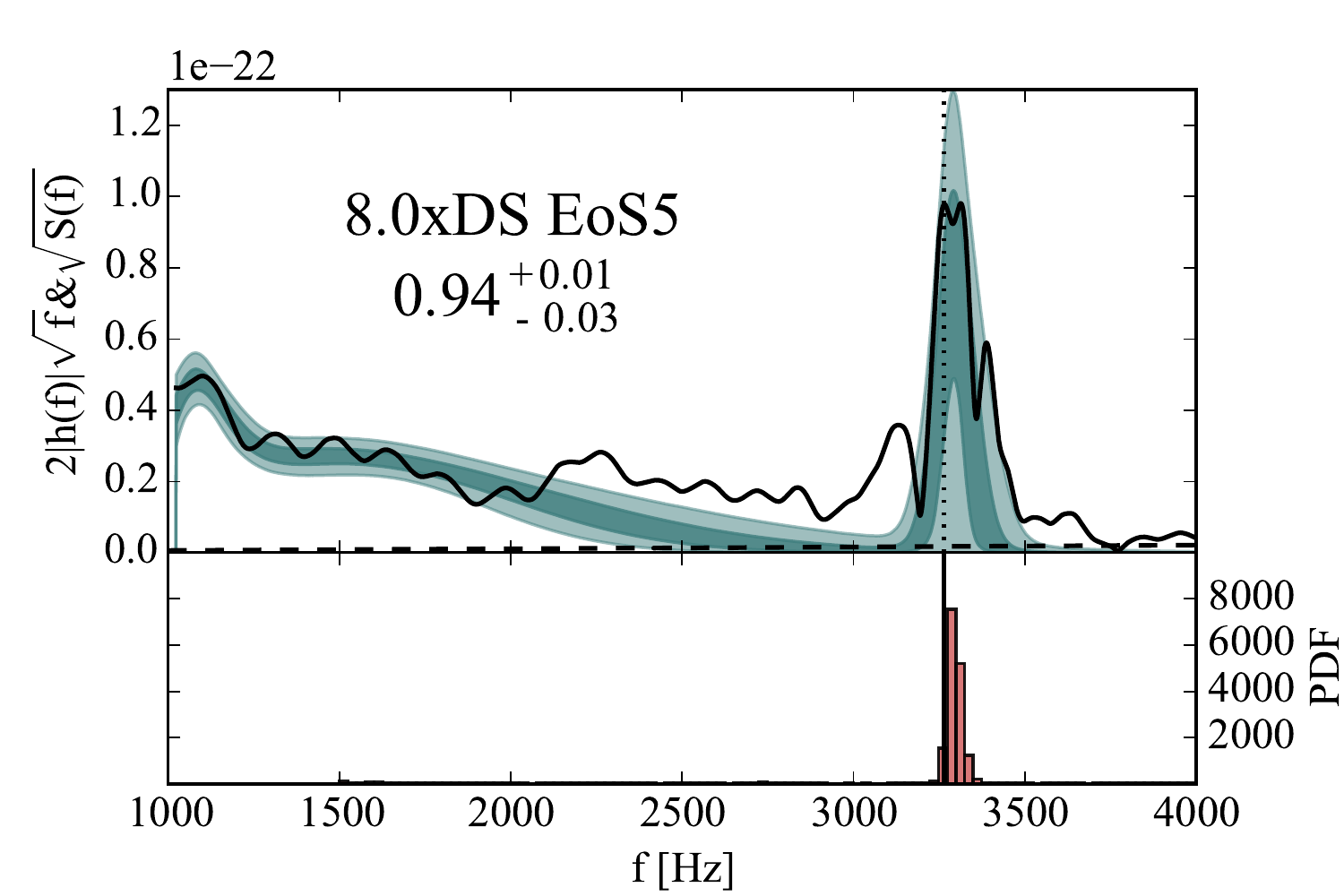}
\includegraphics[width=0.45\textwidth]{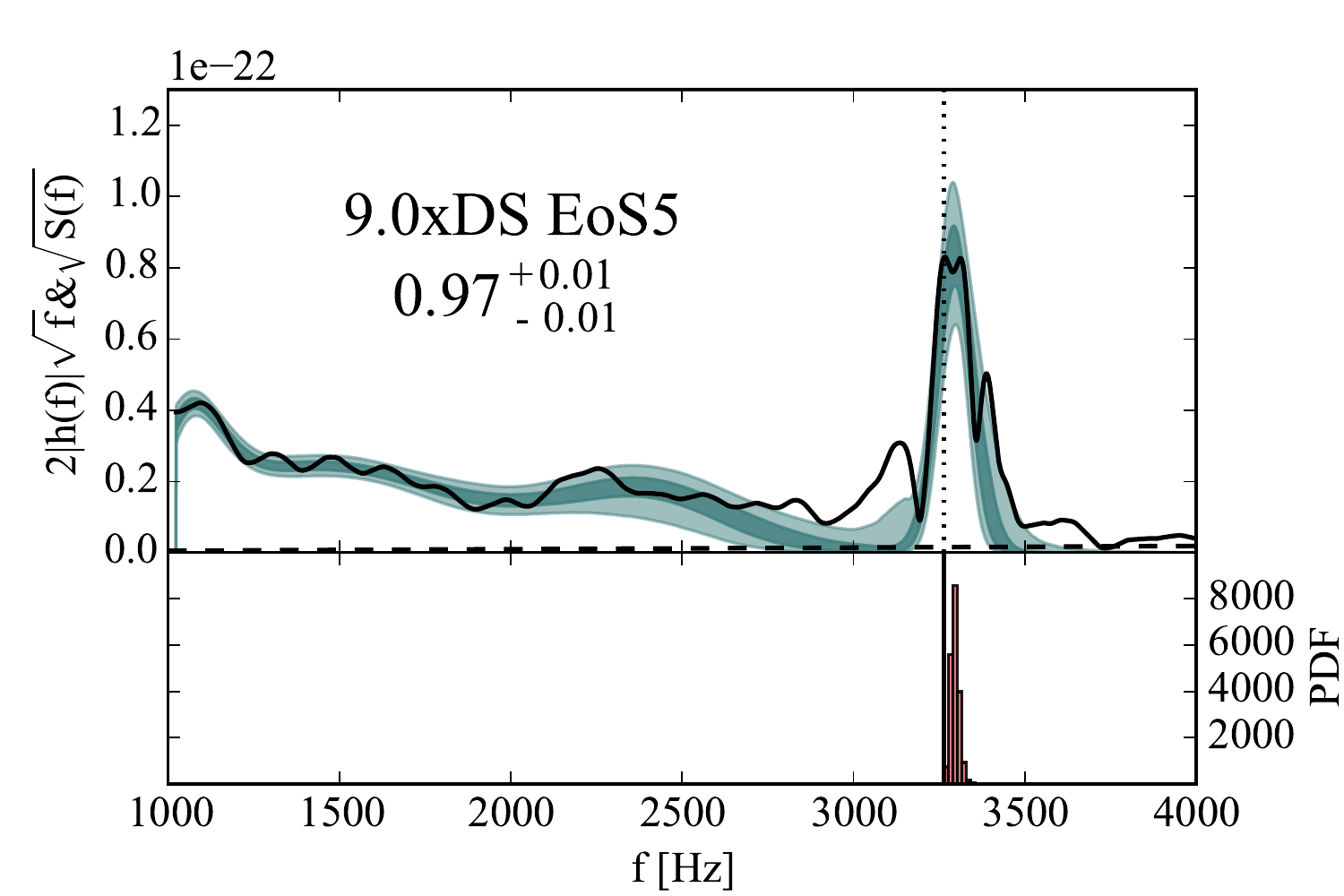}\\
\includegraphics[width=0.45\textwidth]{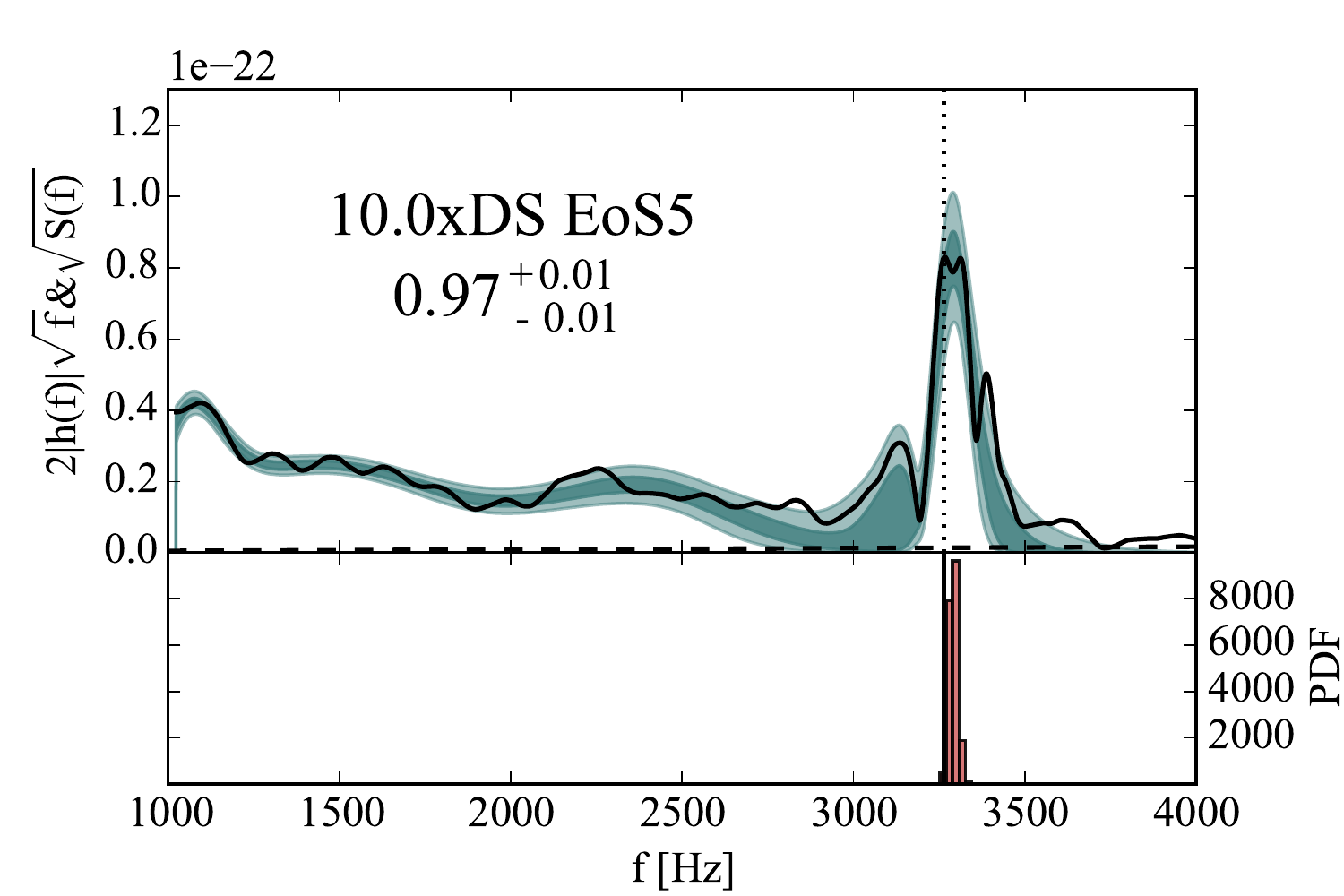}
\includegraphics[width=0.45\textwidth]{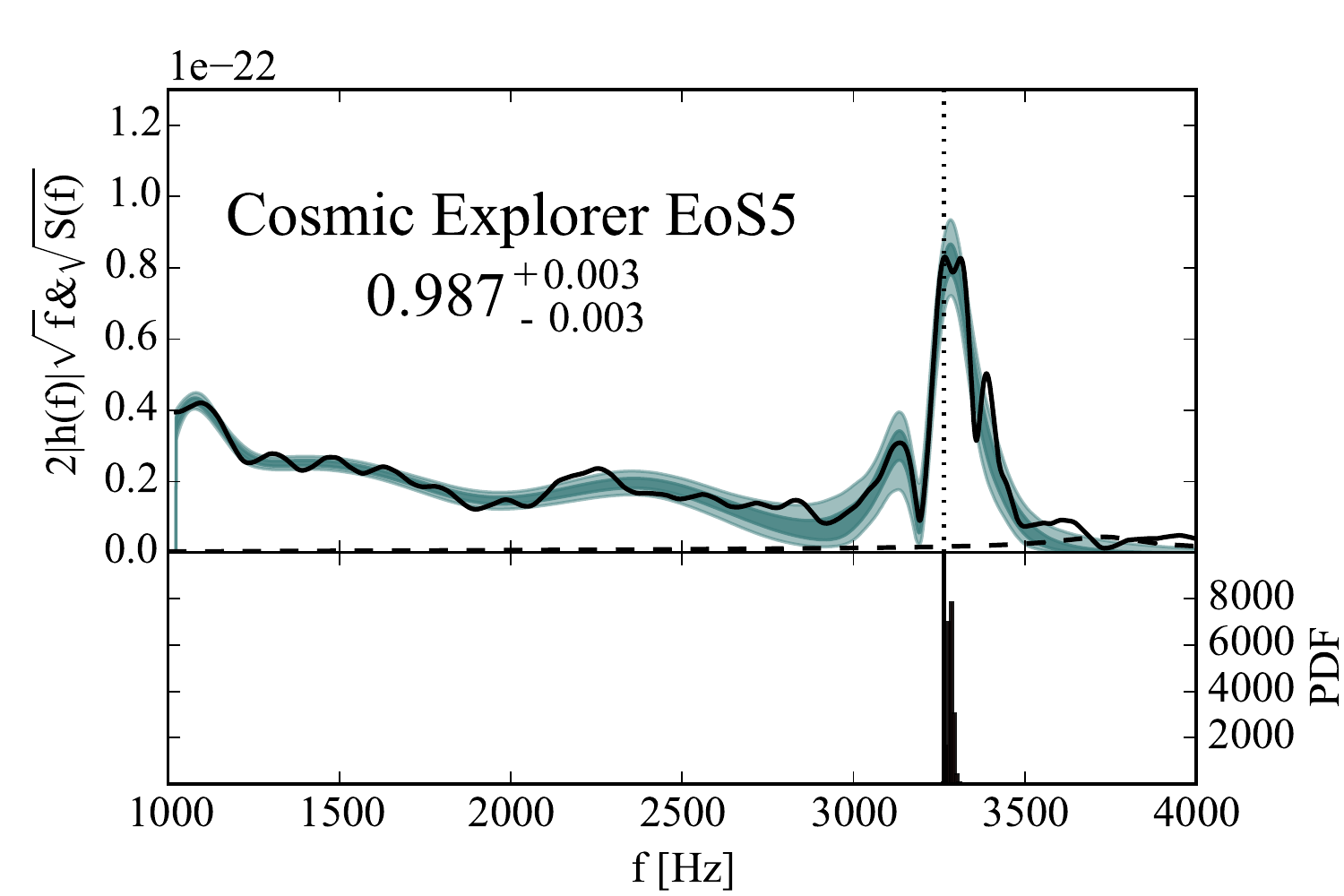}\\
\caption{Reconstruction of the post-merger signal emitted during the coalescence of an equal-mass binary with EoS5 at various network sensitivities characteristic of third-generation detectors. In each panel the top plot shows the $50\%$ and $90\%$ credible interval for the signal spectrum with dark and light shaded regions respectively. The bottom plot shows the posterior density for $\fpeak$ and $\fsub$.}
\label{fig: Design runs}
\end{figure*}

Figure~\ref{fig: Design runs} shows the post-merger reconstruction for an equal-mass binary with EoS5 injected in such networks. As before the top part of the plot shows the spectrum, while the bottom part shows the $\fpeak$ and $\fsub$ posterior densities. As expected, all sensitivities typical of third-generation detectors and networks that lead up to them will result in unambiguous identification of the post-merger signal and excellent measurement of $\fpeak$ with an accuracy of $10-20$Hz. Moreover, third generation detectors will be able to extract subdominant features of the signal, including, but not limited to, $\fsub$. For example, with CE (bottom right panel) we are able to reconstruct substructure in the signal, such as the small peak at around $3.2$kHz.

%---------------------------------------------------------------
\subsection{Direct collapse}

\begin{figure*}[!htbp]
\centering
\includegraphics[width=0.45\textwidth]{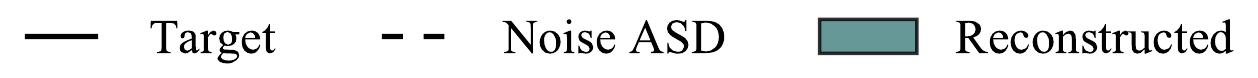}\\
\includegraphics[width=0.45\textwidth]{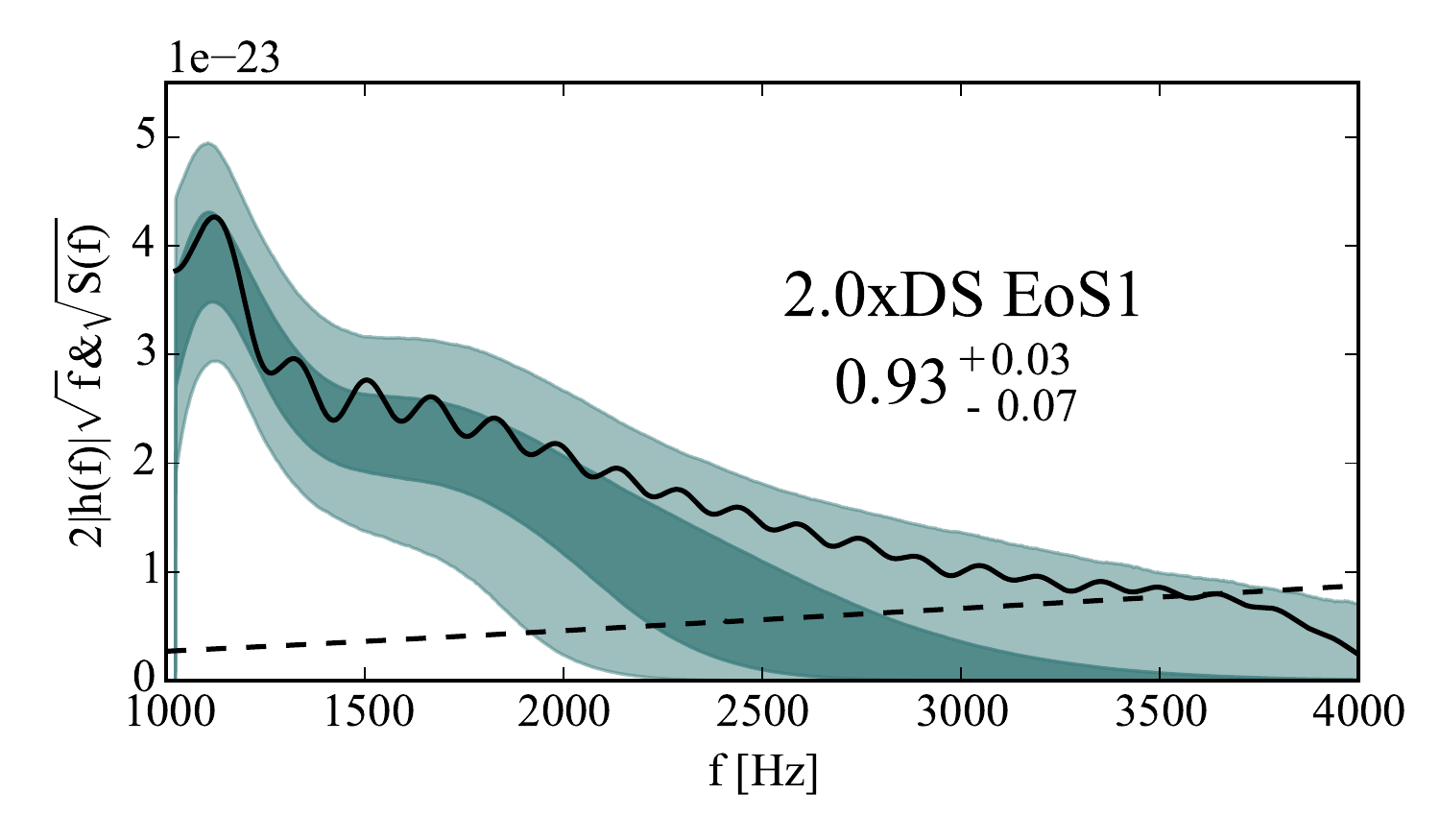}
\includegraphics[width=0.45\textwidth]{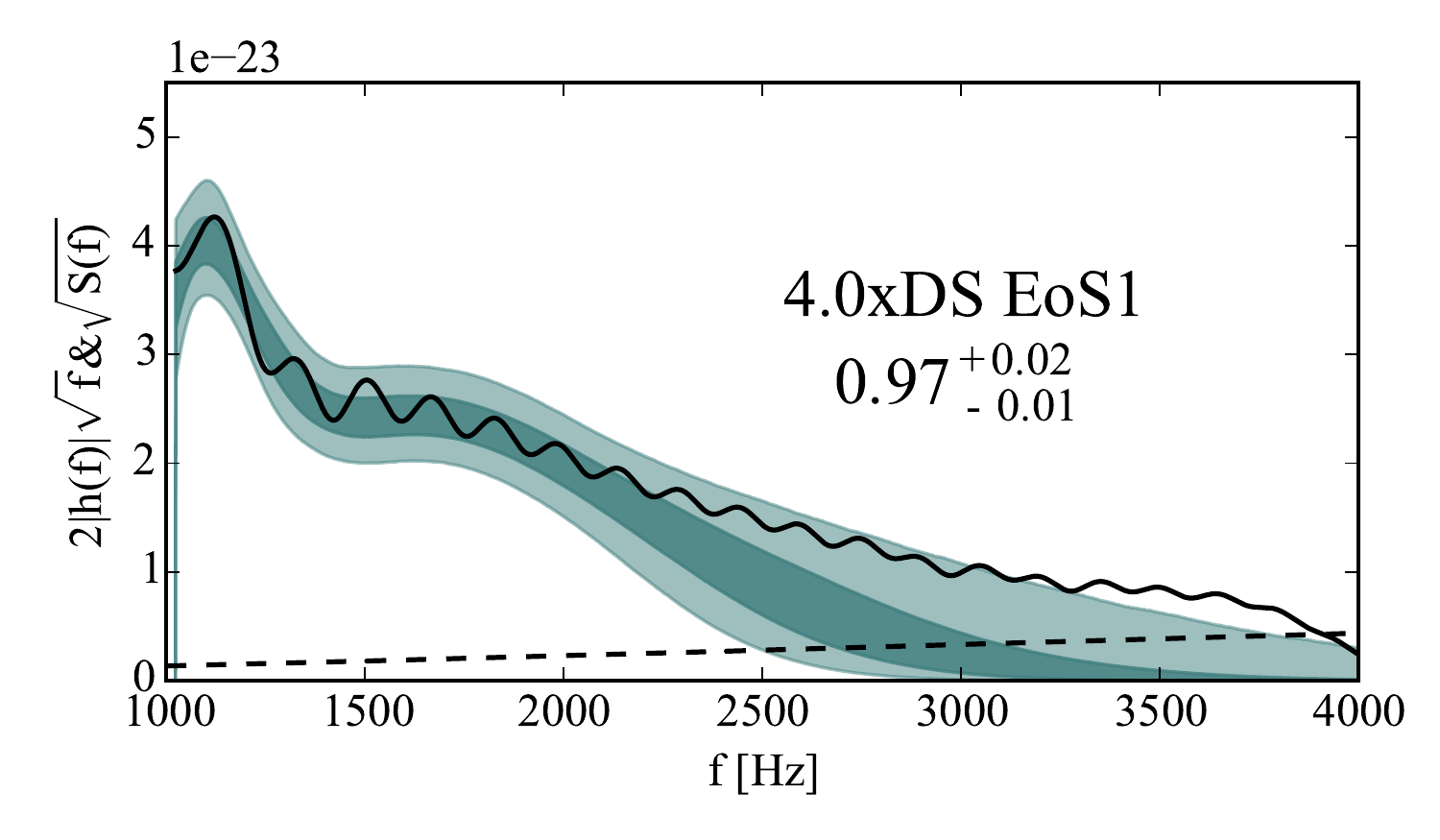}
\caption{Reconstruction of the post-merger spectrum for a case where the remnant collapses into a BH. We show EoS1 and a binary with equal masses injected in two different network sensitivities. The shape of the reconstructed spectrum allows us to determine whether the remnant collapsed promptly into a BH, placing constrains on the high-density EoS. }
    \label{fig: Equal Masses EoS1}
\end{figure*}

Besides the cases studied above, where the merger leads to a NSR and the signal spectrum exhibits a characteristic peak, another possibility is the direct gravitational collapse of the merger remnant into a BH on a dynamical time scale. This is indeed the case for the softest EoSs in our set, EoS1 and EoS8, for both mass ratios. Figure~\ref{fig: Equal Masses EoS1} shows the reconstructed spectrum for EoS1 and an equal-mass binary system for various network sensitivities, demonstrating the lack of a post-merger peak in the relevant frequency range. 

Comparing Figs.~\ref{fig:90c} and~\ref{fig: Equal Masses EoS1} suggests that at $\gtrapprox$2.0xDS {\tt BayesWave} can differentiate between the featureless (in the relevant frequency range) post-merger signal of a prompt-collapse event and an undetectable signal from a NSR. 
In that case and despite the absence of an $\fpeak$ measurement, the signal can still offer insight on the EoS of NSs.
In particular, if the post-merger signal is observed and identified as inconsistent with direct collapse, models such as EoS1 and EoS8 are ruled out, further constraining the soft end of the pressure posterior computed in~\cite{LigoEoS2018} from premerger data. Conversely, if a featureless post-merger signal is observed, we can conclude that the remnant collapsed after merger, suggesting that only soft EoSs such as EoS1 and EoS8 are viable.

Discrimination between prompt collapse and a NSR can be further used to study the high-density regime of the EoS. Specifically, determining the fate of the merger remnant can be used to estimate $M_{\mathrm{th}}$, the threshold mass above which the remnant collapses promptly into a BH. This can in turn be employed to determine $M_{\mathrm{max}}$, the maximum mass of non-rotating NSs, the value of which depends on the high-density EoS~\cite{2013PhRvL.111m1101B,2014arXiv1403.5301B}.
Arguments similar to this were employed in~\cite{2041-8205-850-2-L34} for the case of GW170817 already. There it was assumed that the electromagnetic observations suggest the presence of a NSR, leading to a lower limit on the GW170817 radius coming from the requirement that the EoS is not too soft, as it would have resulted in a prompt collapse. Interestingly, that radius lower limit agrees with the lower limit of~\cite{LigoEoS2018} which is the outcome of the requirement that the EoS supports NSs of at least $1.97 M_{\odot}$.

%---------------------------------------------------------------
\subsection{The case of EoS2}

\begin{figure*}[!htbp]
\centering
\includegraphics[width=0.55\textwidth]{legend_nofsub.pdf}\\
\includegraphics[width=0.32\textwidth]{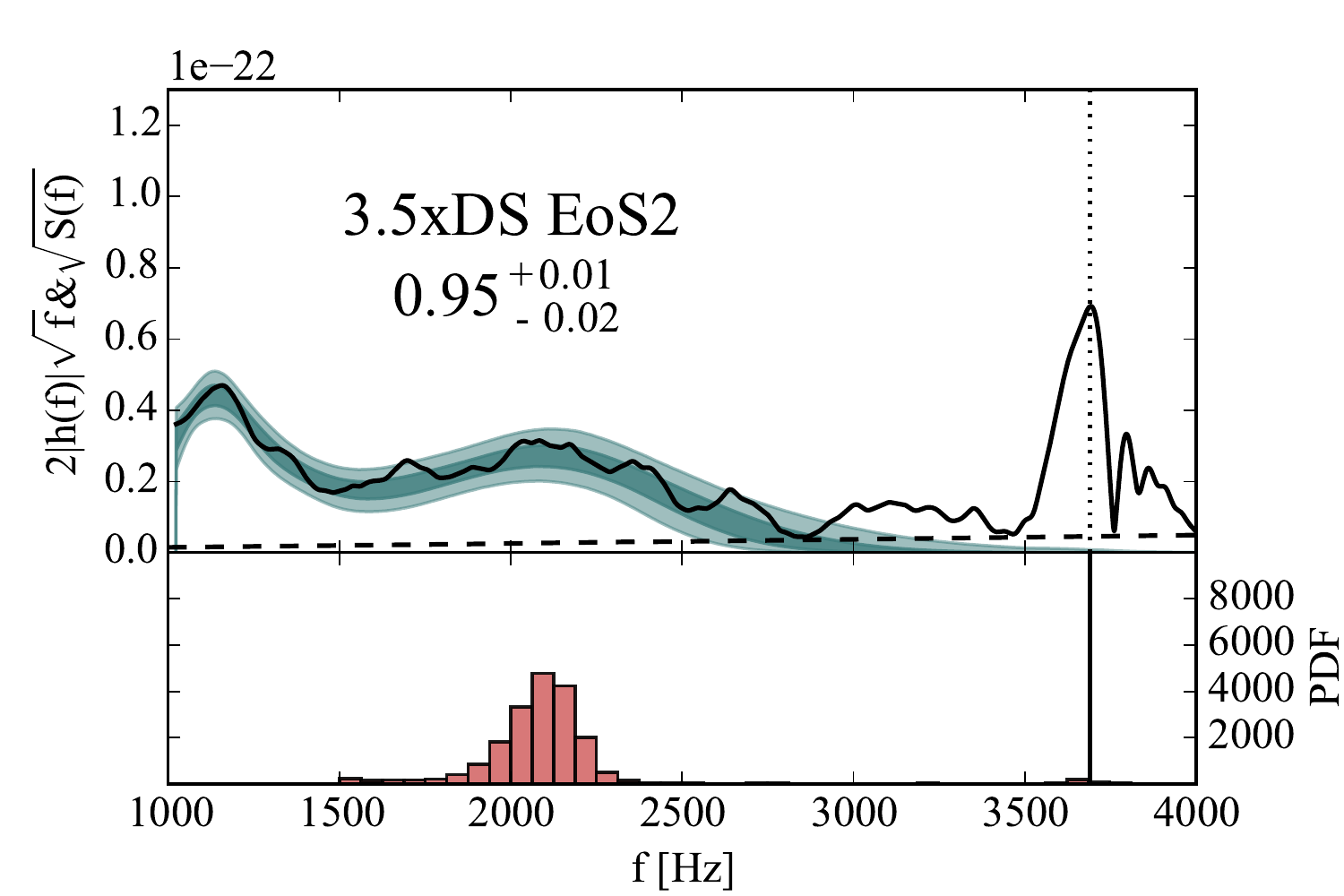}
\includegraphics[width=0.32\textwidth]{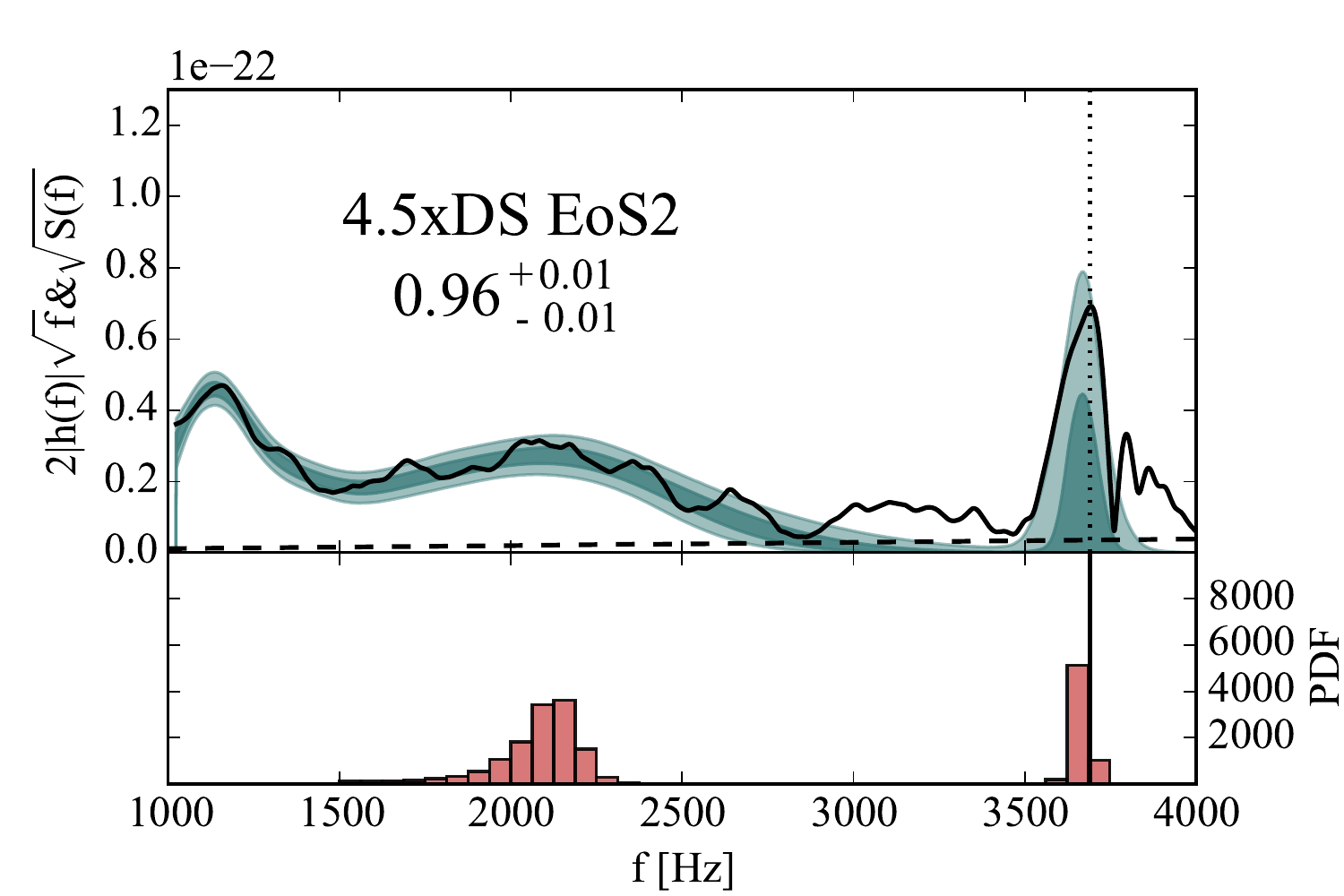}
\includegraphics[width=0.32\textwidth]{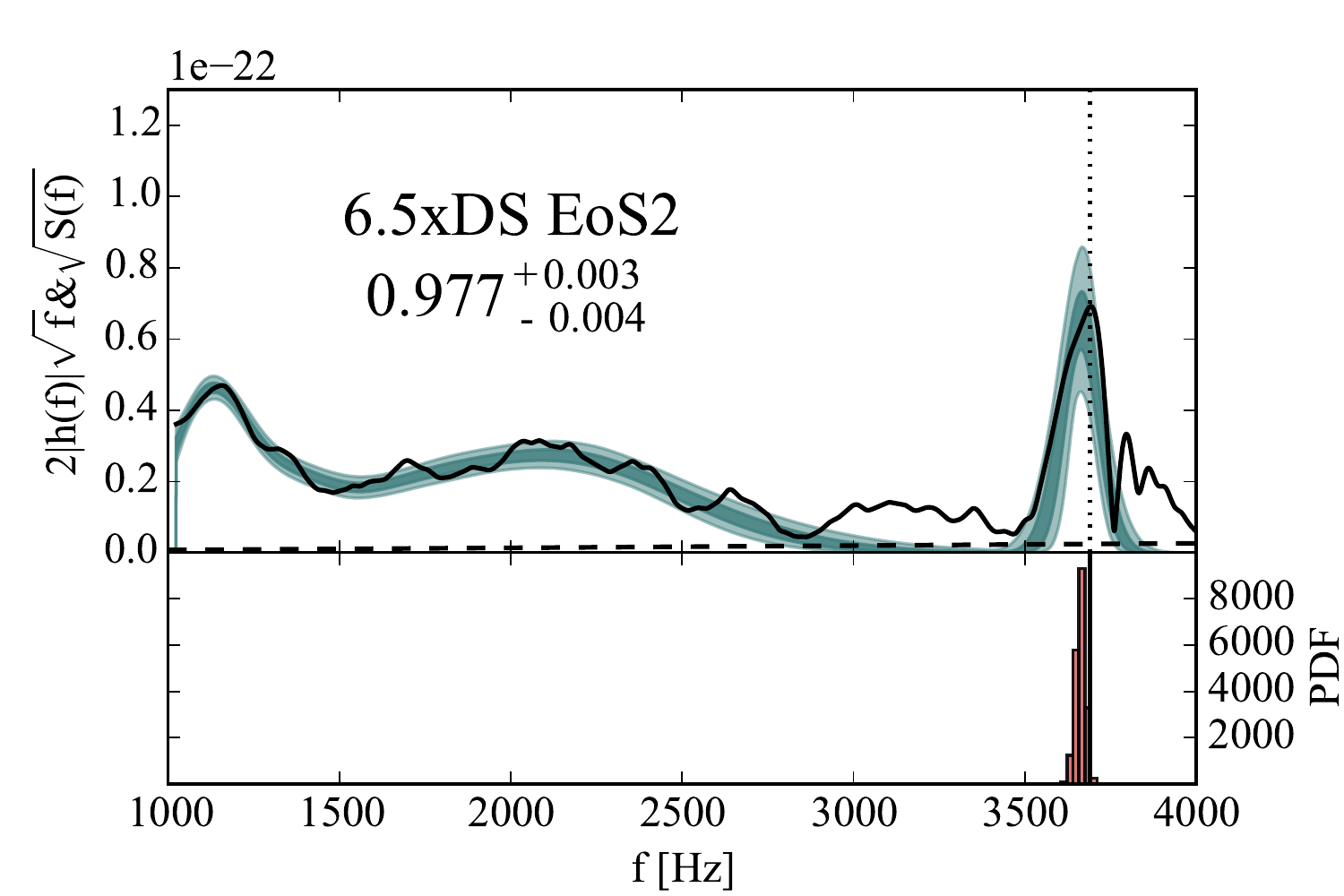}
\caption{Reconstruction of the post-merger signal emitted during the coalescence of an equal-mass binary with EoS2 at various network sensitivities. In each panel the top plot shows the $50\%$ and $90\%$ credible interval for the signal spectrum with dark and light shaded regions respectively. The bottom plot shows the posterior density for $\fpeak$, defined as the frequency of the highest peak of the post-merger spectrum. Due to the shape of the spectrum, a subdominant peak is reconstructed at lower sensitivity than the main peak.}
\label{fig:Eos2_Fpeak_1362}
\end{figure*}

Finally, we discuss the case of EoS2. Figure~\ref{fig:Eos2_Fpeak_1362} shows the reconstructed spectrum and $\fpeak$ and $\fsub$ posterior densities for this EoS and an equal-mass binary for various network sensitivities.
Since EoS2 is the softest EoS we study that does not lead to direct collapse, we expect it to result in a relatively high value $\fpeak \sim 3600$Hz. At the same time, the spectrum in Fig.~\ref{fig:Eos2_Fpeak_1362} exhibits a prominent and wide subdominant peak at around $\fsub \sim 2200$Hz. 

The combination of a large $\fpeak$ value -where the detector sensitivity is worse- and a wide $\fsub$ peak results in the subdominant peak being reconstructed at lower sensitivity than the dominant one. Indeed at 3.5xDS (left panel) the spectrum contains $\fsub$ only. At 4.5xDS (middle panel) there is a hint of another spectral peak of higher frequency, while at 6.5xDS (right panel) there is clear evidence of two peaks in the spectrum. Despite the reversal of which peak is measured first, we note that this case would not lead to a misidentification of the value of the dominant frequency mode for two reasons. First, the premerger data already suggest that for this system we should expect $\fpeak>2500$Hz. Second, the width of the subdominant peak is not typical of dominant peaks which are in general more narrow. Again, additional priors on the wavelets {\tt BayesWave} uses to reconstruct the spectrum would immediately be able to separate the two types of spectral peaks.

 %%%%%%%%%%%%%%%%%%%%%%%%%%%%%%%%%%%%%
 \section{Conclusions}
 \label{sec:conclusions}
 
 We studied the improvements in ground-based GW detectors required before the post-merger signal of a BNS coalescence can be extracted. We show that for a system similar to the recently-detected GW170817, the post-merger signal can be extracted once the two LIGO detectors operate at $\sim$ 2-3 times better than their design sensitivity. 
 This estimate is derived under the assumption that the numerical waveforms employed in this work approximately reflect the true strength of the emitted signal, which could be weaker if physical viscosity is very strong, or somewhat stronger if numerical damping leads to a significant underestimation.
 Since these upgrades are already under planning, we are optimistic about the prospects of observing post-merger signals and measuring their dominant frequency component with second-generation detectors.
 Moreover, we show that planned third-generation detectors will be able to extract even more information from post-merger signals. In particular we find that subdominant features of the signal will be measurable, enhancing the amount of EoS-related information we can extract from the signal.
 
 As a concluding remark, we again note that here we only focus on improvements on the GW detectors. In anticipation of these upcoming improvements we also plan to improve our analysis of these signals with {\tt BayesWave}. Possible improvements include using a different type of basis to reconstruct the signal, such as `chirplets'~\cite{Millhouse:2018dgi} that can account for a possible time-evolution of $\fpeak$, and additional priors that can facilitate reconstruction and extraction of the frequency components, such as priors on the width of the spectral peaks. Finally, this analysis assumes that the calibration of the detectors is known to large accuracy at high frequencies. Since this might not be expected to be the case, we plan in the future to study the calibration uncertainty requirements for these measurements to be feasible.

 %%%%%%%%%%%%%%%%%%%%%%%%%%%%%%%%%%%%%
 \acknowledgements
 We thank Matt Evans for useful discussions on GW detectors and instrumental upgrades and Jeff Kissel for discussions on detector calibration. We thank Meg Millhouse and Neil Cornish for helpful comments and discussions on the {\tt Bayeswave} results. AB acknowledges support by the European Research Council (ERC) under the European Union's Horizon 2020 research and innovation programme under grant agreement No. 759253 and the Klaus-Tschira Foundation. This research was done using resources provided by the Open Science Grid~\cite{pordes:2007,Sfiligoi:2009}, which is supported by the National Science Foundation award 1148698, and the U.S. Department of Energy's Office of Science. The authors are grateful for computational resources provided by the LIGO Laboratory and supported by National Science Foundation Grants PHY-0757058 and PHY-0823459. Figures were produced using {\tt matplotlib}~\cite{Hunter:2007}.

\bibliography{refs}

\begin{thebibliography}{110}%
\makeatletter
\providecommand \@ifxundefined [1]{%
 \@ifx{#1\undefined}
}%
\providecommand \@ifnum [1]{%
 \ifnum #1\expandafter \@firstoftwo
 \else \expandafter \@secondoftwo
 \fi
}%
\providecommand \@ifx [1]{%
 \ifx #1\expandafter \@firstoftwo
 \else \expandafter \@secondoftwo
 \fi
}%
\providecommand \natexlab [1]{#1}%
\providecommand \enquote  [1]{``#1''}%
\providecommand \bibnamefont  [1]{#1}%
\providecommand \bibfnamefont [1]{#1}%
\providecommand \citenamefont [1]{#1}%
\providecommand \href@noop [0]{\@secondoftwo}%
\providecommand \href [0]{\begingroup \@sanitize@url \@href}%
\providecommand \@href[1]{\@@startlink{#1}\@@href}%
\providecommand \@@href[1]{\endgroup#1\@@endlink}%
\providecommand \@sanitize@url [0]{\catcode `\\12\catcode `\$12\catcode
  `\&12\catcode `\#12\catcode `\^12\catcode `\_12\catcode `\%12\relax}%
\providecommand \@@startlink[1]{}%
\providecommand \@@endlink[0]{}%
\providecommand \url  [0]{\begingroup\@sanitize@url \@url }%
\providecommand \@url [1]{\endgroup\@href {#1}{\urlprefix }}%
\providecommand \urlprefix  [0]{URL }%
\providecommand \Eprint [0]{\href }%
\providecommand \doibase [0]{http://dx.doi.org/}%
\providecommand \selectlanguage [0]{\@gobble}%
\providecommand \bibinfo  [0]{\@secondoftwo}%
\providecommand \bibfield  [0]{\@secondoftwo}%
\providecommand \translation [1]{[#1]}%
\providecommand \BibitemOpen [0]{}%
\providecommand \bibitemStop [0]{}%
\providecommand \bibitemNoStop [0]{.\EOS\space}%
\providecommand \EOS [0]{\spacefactor3000\relax}%
\providecommand \BibitemShut  [1]{\csname bibitem#1\endcsname}%
\let\auto@bib@innerbib\@empty
%</preamble>
\bibitem [{\citenamefont {{Abbott {\it et al.} (LIGO and Virgo Scientific
  Collaboration)}}(2016{\natexlab{a}})}]{2016PhRvL.116f1102A}%
  \BibitemOpen
  \bibfield  {author} {\bibinfo {author} {\bibfnamefont {B.~P.}\ \bibnamefont
  {{Abbott {\it et al.} (LIGO and Virgo Scientific Collaboration)}}},\ }\href
  {\doibase 10.1103/PhysRevLett.116.061102} {\bibfield  {journal} {\bibinfo
  {journal} {\prl}\ }\textbf {\bibinfo {volume} {116}},\ \bibinfo {eid}
  {061102} (\bibinfo {year} {2016}{\natexlab{a}})},\ \Eprint
  {http://arxiv.org/abs/1602.03837} {arXiv:1602.03837 [gr-qc]} \BibitemShut
  {NoStop}%
\bibitem [{\citenamefont {{Abbott {\it et al.} (LIGO and Virgo Scientific
  Collaboration)}}(2016{\natexlab{b}})}]{2016PhRvL.116x1103A}%
  \BibitemOpen
  \bibfield  {author} {\bibinfo {author} {\bibfnamefont {B.~P.}\ \bibnamefont
  {{Abbott {\it et al.} (LIGO and Virgo Scientific Collaboration)}}},\ }\href
  {\doibase 10.1103/PhysRevLett.116.241103} {\bibfield  {journal} {\bibinfo
  {journal} {\prl}\ }\textbf {\bibinfo {volume} {116}},\ \bibinfo {eid}
  {241103} (\bibinfo {year} {2016}{\natexlab{b}})},\ \Eprint
  {http://arxiv.org/abs/1606.04855} {arXiv:1606.04855 [gr-qc]} \BibitemShut
  {NoStop}%
\bibitem [{\citenamefont {{Abbott {\it et al.} (LIGO and Virgo Scientific
  Collaboration)}}(2017)}]{2017PhRvL.118v1101A}%
  \BibitemOpen
  \bibfield  {author} {\bibinfo {author} {\bibfnamefont {B.~P.}\ \bibnamefont
  {{Abbott {\it et al.} (LIGO and Virgo Scientific Collaboration)}}},\ }\href
  {\doibase 10.1103/PhysRevLett.118.221101} {\bibfield  {journal} {\bibinfo
  {journal} {\prl}\ }\textbf {\bibinfo {volume} {118}},\ \bibinfo {eid}
  {221101} (\bibinfo {year} {2017})},\ \Eprint
  {http://arxiv.org/abs/1706.01812} {arXiv:1706.01812 [gr-qc]} \BibitemShut
  {NoStop}%
\bibitem [{\citenamefont {Abbott}\ \emph
  {et~al.}(2017{\natexlab{a}})\citenamefont {Abbott} \emph
  {et~al.}}]{Abbott:2017oio}%
  \BibitemOpen
  \bibfield  {author} {\bibinfo {author} {\bibfnamefont {B.~P.}\ \bibnamefont
  {Abbott}} \emph {et~al.} (\bibinfo {collaboration} {Virgo, LIGO
  Scientific}),\ }\href {\doibase 10.1103/PhysRevLett.119.141101} {\bibfield
  {journal} {\bibinfo  {journal} {Phys. Rev. Lett.}\ }\textbf {\bibinfo
  {volume} {119}},\ \bibinfo {pages} {141101} (\bibinfo {year}
  {2017}{\natexlab{a}})},\ \Eprint {http://arxiv.org/abs/1709.09660}
  {arXiv:1709.09660 [gr-qc]} \BibitemShut {NoStop}%
%%CITATION = ARXIV:1709.09660;%%
\bibitem [{\citenamefont {Abbott}\ \emph
  {et~al.}(2017{\natexlab{b}})\citenamefont {Abbott} \emph
  {et~al.}}]{TheLIGOScientific:2017qsa}%
  \BibitemOpen
  \bibfield  {author} {\bibinfo {author} {\bibfnamefont {B.}~\bibnamefont
  {Abbott}} \emph {et~al.} (\bibinfo {collaboration} {Virgo, LIGO
  Scientific}),\ }\href {\doibase 10.1103/PhysRevLett.119.161101} {\bibfield
  {journal} {\bibinfo  {journal} {Phys. Rev. Lett.}\ }\textbf {\bibinfo
  {volume} {119}},\ \bibinfo {pages} {161101} (\bibinfo {year}
  {2017}{\natexlab{b}})},\ \Eprint {http://arxiv.org/abs/1710.05832}
  {arXiv:1710.05832 [gr-qc]} \BibitemShut {NoStop}%
%%CITATION = ARXIV:1710.05832;%%
\bibitem [{\citenamefont {Abbott}\ \emph
  {et~al.}(2017{\natexlab{c}})\citenamefont {Abbott} \emph
  {et~al.}}]{Abbott:2017gyy}%
  \BibitemOpen
  \bibfield  {author} {\bibinfo {author} {\bibfnamefont {B.~P.}\ \bibnamefont
  {Abbott}} \emph {et~al.} (\bibinfo {collaboration} {Virgo, LIGO
  Scientific}),\ }\href {\doibase 10.3847/2041-8213/aa9f0c} {\bibfield
  {journal} {\bibinfo  {journal} {Astrophys. J.}\ }\textbf {\bibinfo {volume}
  {851}},\ \bibinfo {pages} {L35} (\bibinfo {year} {2017}{\natexlab{c}})},\
  \Eprint {http://arxiv.org/abs/1711.05578} {arXiv:1711.05578 [astro-ph.HE]}
  \BibitemShut {NoStop}%
%%CITATION = ARXIV:1711.05578;%%
\bibitem [{\citenamefont {{The LIGO Scientific Collaboration}}\ \emph
  {et~al.}(2015)\citenamefont {{The LIGO Scientific Collaboration}},
  \citenamefont {{Aasi}}, \citenamefont {{Abbott}}, \citenamefont {{Abbott}},
  \citenamefont {{Abbott}}, \citenamefont {{Abernathy}}, \citenamefont
  {{Ackley}}, \citenamefont {{Adams}}, \citenamefont {{Adams}}, \citenamefont
  {{Addesso}},\ and\ \citenamefont {et~al.}}]{2015CQGra..32g4001T}%
  \BibitemOpen
  \bibfield  {author} {\bibinfo {author} {\bibnamefont {{The LIGO Scientific
  Collaboration}}}, \bibinfo {author} {\bibfnamefont {J.}~\bibnamefont
  {{Aasi}}}, \bibinfo {author} {\bibfnamefont {B.~P.}\ \bibnamefont
  {{Abbott}}}, \bibinfo {author} {\bibfnamefont {R.}~\bibnamefont {{Abbott}}},
  \bibinfo {author} {\bibfnamefont {T.}~\bibnamefont {{Abbott}}}, \bibinfo
  {author} {\bibfnamefont {M.~R.}\ \bibnamefont {{Abernathy}}}, \bibinfo
  {author} {\bibfnamefont {K.}~\bibnamefont {{Ackley}}}, \bibinfo {author}
  {\bibfnamefont {C.}~\bibnamefont {{Adams}}}, \bibinfo {author} {\bibfnamefont
  {T.}~\bibnamefont {{Adams}}}, \bibinfo {author} {\bibfnamefont
  {P.}~\bibnamefont {{Addesso}}}, \ and\ \bibinfo {author} {\bibnamefont
  {et~al.}},\ }\href {\doibase 10.1088/0264-9381/32/7/074001} {\bibfield
  {journal} {\bibinfo  {journal} {Classical and Quantum Gravity}\ }\textbf
  {\bibinfo {volume} {32}},\ \bibinfo {eid} {074001} (\bibinfo {year}
  {2015})},\ \Eprint {http://arxiv.org/abs/1411.4547} {arXiv:1411.4547 [gr-qc]}
  \BibitemShut {NoStop}%
\bibitem [{\citenamefont {Acernese}\ \emph {et~al.}(2015)\citenamefont
  {Acernese} \emph {et~al.}}]{TheVirgo:2014hva}%
  \BibitemOpen
  \bibfield  {author} {\bibinfo {author} {\bibfnamefont {F.}~\bibnamefont
  {Acernese}} \emph {et~al.} (\bibinfo {collaboration} {VIRGO}),\ }\href
  {\doibase 10.1088/0264-9381/32/2/024001} {\bibfield  {journal} {\bibinfo
  {journal} {Class. Quant. Grav.}\ }\textbf {\bibinfo {volume} {32}},\ \bibinfo
  {pages} {024001} (\bibinfo {year} {2015})},\ \Eprint
  {http://arxiv.org/abs/1408.3978} {arXiv:1408.3978 [gr-qc]} \BibitemShut
  {NoStop}%
%%CITATION = ARXIV:1408.3978;%%
\bibitem [{\citenamefont {{{\"O}zel}}\ and\ \citenamefont
  {{Freire}}(2016)}]{Ozel:2016oaf}%
  \BibitemOpen
  \bibfield  {author} {\bibinfo {author} {\bibfnamefont {F.}~\bibnamefont
  {{{\"O}zel}}}\ and\ \bibinfo {author} {\bibfnamefont {P.}~\bibnamefont
  {{Freire}}},\ }\href {\doibase 10.1146/annurev-astro-081915-023322}
  {\bibfield  {journal} {\bibinfo  {journal} {Ann. Rev. Astron. Astrophys.}\
  }\textbf {\bibinfo {volume} {54}},\ \bibinfo {pages} {401} (\bibinfo {year}
  {2016})},\ \Eprint {http://arxiv.org/abs/1603.02698} {arXiv:1603.02698
  [astro-ph.HE]} \BibitemShut {NoStop}%
%%CITATION = ARXIV:1603.02698;%%
\bibitem [{\citenamefont {Lattimer}\ and\ \citenamefont
  {Prakash}(2016)}]{Lattimer:2015nhk}%
  \BibitemOpen
  \bibfield  {author} {\bibinfo {author} {\bibfnamefont {J.~M.}\ \bibnamefont
  {Lattimer}}\ and\ \bibinfo {author} {\bibfnamefont {M.}~\bibnamefont
  {Prakash}},\ }\href {\doibase 10.1016/j.physrep.2015.12.005} {\bibfield
  {journal} {\bibinfo  {journal} {Phys. Rept.}\ }\textbf {\bibinfo {volume}
  {621}},\ \bibinfo {pages} {127} (\bibinfo {year} {2016})},\ \Eprint
  {http://arxiv.org/abs/1512.07820} {arXiv:1512.07820 [astro-ph.SR]}
  \BibitemShut {NoStop}%
%%CITATION = ARXIV:1512.07820;%%
\bibitem [{\citenamefont {Oertel}\ \emph {et~al.}(2017)\citenamefont {Oertel},
  \citenamefont {Hempel}, \citenamefont {Kl\:ahn},\ and\ \citenamefont
  {Typel}}]{Oertel:2016bki}%
  \BibitemOpen
  \bibfield  {author} {\bibinfo {author} {\bibfnamefont {M.}~\bibnamefont
  {Oertel}}, \bibinfo {author} {\bibfnamefont {M.}~\bibnamefont {Hempel}},
  \bibinfo {author} {\bibfnamefont {T.}~\bibnamefont {Kl\:ahn}}, \ and\
  \bibinfo {author} {\bibfnamefont {S.}~\bibnamefont {Typel}},\ }\href
  {\doibase 10.1103/RevModPhys.89.015007} {\bibfield  {journal} {\bibinfo
  {journal} {Rev. Mod. Phys.}\ }\textbf {\bibinfo {volume} {89}},\ \bibinfo
  {pages} {015007} (\bibinfo {year} {2017})},\ \Eprint
  {http://arxiv.org/abs/1610.03361} {arXiv:1610.03361 [astro-ph.HE]}
  \BibitemShut {NoStop}%
%%CITATION = ARXIV:1610.03361;%%
\bibitem [{\citenamefont {Blanchet}(2014)}]{lrr-2014-2}%
  \BibitemOpen
  \bibfield  {author} {\bibinfo {author} {\bibfnamefont {L.}~\bibnamefont
  {Blanchet}},\ }\href {\doibase 10.12942/lrr-2014-2} {\bibfield  {journal}
  {\bibinfo  {journal} {Living Reviews in Relativity}\ }\textbf {\bibinfo
  {volume} {17}} (\bibinfo {year} {2014}),\ 10.12942/lrr-2014-2}\BibitemShut
  {NoStop}%
\bibitem [{\citenamefont {Faber}\ and\ \citenamefont
  {Rasio}(2012)}]{lrr-2012-8}%
  \BibitemOpen
  \bibfield  {author} {\bibinfo {author} {\bibfnamefont {J.~A.}\ \bibnamefont
  {Faber}}\ and\ \bibinfo {author} {\bibfnamefont {F.~A.}\ \bibnamefont
  {Rasio}},\ }\href {\doibase 10.12942/lrr-2012-8} {\bibfield  {journal}
  {\bibinfo  {journal} {Living Reviews in Relativity}\ }\textbf {\bibinfo
  {volume} {15}} (\bibinfo {year} {2012}),\ 10.12942/lrr-2012-8}\BibitemShut
  {NoStop}%
\bibitem [{\citenamefont {Baiotti}\ and\ \citenamefont
  {Rezzolla}(2017)}]{Baiotti:2016qnr}%
  \BibitemOpen
  \bibfield  {author} {\bibinfo {author} {\bibfnamefont {L.}~\bibnamefont
  {Baiotti}}\ and\ \bibinfo {author} {\bibfnamefont {L.}~\bibnamefont
  {Rezzolla}},\ }\href {\doibase 10.1088/1361-6633/aa67bb} {\bibfield
  {journal} {\bibinfo  {journal} {Rept. Prog. Phys.}\ }\textbf {\bibinfo
  {volume} {80}},\ \bibinfo {pages} {096901} (\bibinfo {year} {2017})},\
  \Eprint {http://arxiv.org/abs/1607.03540} {arXiv:1607.03540 [gr-qc]}
  \BibitemShut {NoStop}%
%%CITATION = ARXIV:1607.03540;%%
\bibitem [{\citenamefont {{Flanagan}}\ and\ \citenamefont
  {{Hinderer}}(2008)}]{2008PhRvD..77b1502F}%
  \BibitemOpen
  \bibfield  {author} {\bibinfo {author} {\bibfnamefont {{\'E}.~{\'E}.}\
  \bibnamefont {{Flanagan}}}\ and\ \bibinfo {author} {\bibfnamefont
  {T.}~\bibnamefont {{Hinderer}}},\ }\href {\doibase
  10.1103/PhysRevD.77.021502} {\bibfield  {journal} {\bibinfo  {journal}
  {\prd}\ }\textbf {\bibinfo {volume} {77}},\ \bibinfo {eid} {021502} (\bibinfo
  {year} {2008})},\ \Eprint {http://arxiv.org/abs/0709.1915} {arXiv:0709.1915}
  \BibitemShut {NoStop}%
\bibitem [{\citenamefont {{Hinderer}}\ \emph {et~al.}(2010)\citenamefont
  {{Hinderer}}, \citenamefont {{Lackey}}, \citenamefont {{Lang}},\ and\
  \citenamefont {{Read}}}]{2010PhRvD..81l3016H}%
  \BibitemOpen
  \bibfield  {author} {\bibinfo {author} {\bibfnamefont {T.}~\bibnamefont
  {{Hinderer}}}, \bibinfo {author} {\bibfnamefont {B.~D.}\ \bibnamefont
  {{Lackey}}}, \bibinfo {author} {\bibfnamefont {R.~N.}\ \bibnamefont
  {{Lang}}}, \ and\ \bibinfo {author} {\bibfnamefont {J.~S.}\ \bibnamefont
  {{Read}}},\ }\href {\doibase 10.1103/PhysRevD.81.123016} {\bibfield
  {journal} {\bibinfo  {journal} {\prd}\ }\textbf {\bibinfo {volume} {81}},\
  \bibinfo {eid} {123016} (\bibinfo {year} {2010})}\BibitemShut {NoStop}%
\bibitem [{\citenamefont {{Read}}\ \emph {et~al.}(2009)\citenamefont {{Read}},
  \citenamefont {{Markakis}}, \citenamefont {{Shibata}}, \citenamefont {{Ury{\=
  u}}}, \citenamefont {{Creighton}},\ and\ \citenamefont
  {{Friedman}}}]{2009PhRvD..79l4033R}%
  \BibitemOpen
  \bibfield  {author} {\bibinfo {author} {\bibfnamefont {J.~S.}\ \bibnamefont
  {{Read}}}, \bibinfo {author} {\bibfnamefont {C.}~\bibnamefont {{Markakis}}},
  \bibinfo {author} {\bibfnamefont {M.}~\bibnamefont {{Shibata}}}, \bibinfo
  {author} {\bibfnamefont {K.}~\bibnamefont {{Ury{\= u}}}}, \bibinfo {author}
  {\bibfnamefont {J.~D.~E.}\ \bibnamefont {{Creighton}}}, \ and\ \bibinfo
  {author} {\bibfnamefont {J.~L.}\ \bibnamefont {{Friedman}}},\ }\href
  {\doibase 10.1103/PhysRevD.79.124033} {\bibfield  {journal} {\bibinfo
  {journal} {\prd}\ }\textbf {\bibinfo {volume} {79}},\ \bibinfo {eid} {124033}
  (\bibinfo {year} {2009})},\ \Eprint {http://arxiv.org/abs/0901.3258}
  {arXiv:0901.3258 [gr-qc]} \BibitemShut {NoStop}%
\bibitem [{\citenamefont {{Read}}\ \emph {et~al.}(2013)\citenamefont {{Read}},
  \citenamefont {{Baiotti}}, \citenamefont {{Creighton}}, \citenamefont
  {{Friedman}}, \citenamefont {{Giacomazzo}}, \citenamefont {{Kyutoku}},
  \citenamefont {{Markakis}}, \citenamefont {{Rezzolla}}, \citenamefont
  {{Shibata}},\ and\ \citenamefont {{Taniguchi}}}]{2013PhRvD..88d4042R}%
  \BibitemOpen
  \bibfield  {author} {\bibinfo {author} {\bibfnamefont {J.~S.}\ \bibnamefont
  {{Read}}}, \bibinfo {author} {\bibfnamefont {L.}~\bibnamefont {{Baiotti}}},
  \bibinfo {author} {\bibfnamefont {J.~D.~E.}\ \bibnamefont {{Creighton}}},
  \bibinfo {author} {\bibfnamefont {J.~L.}\ \bibnamefont {{Friedman}}},
  \bibinfo {author} {\bibfnamefont {B.}~\bibnamefont {{Giacomazzo}}}, \bibinfo
  {author} {\bibfnamefont {K.}~\bibnamefont {{Kyutoku}}}, \bibinfo {author}
  {\bibfnamefont {C.}~\bibnamefont {{Markakis}}}, \bibinfo {author}
  {\bibfnamefont {L.}~\bibnamefont {{Rezzolla}}}, \bibinfo {author}
  {\bibfnamefont {M.}~\bibnamefont {{Shibata}}}, \ and\ \bibinfo {author}
  {\bibfnamefont {K.}~\bibnamefont {{Taniguchi}}},\ }\href {\doibase
  10.1103/PhysRevD.88.044042} {\bibfield  {journal} {\bibinfo  {journal}
  {\prd}\ }\textbf {\bibinfo {volume} {88}},\ \bibinfo {eid} {044042} (\bibinfo
  {year} {2013})},\ \Eprint {http://arxiv.org/abs/1306.4065} {arXiv:1306.4065
  [gr-qc]} \BibitemShut {NoStop}%
\bibitem [{\citenamefont {Del~Pozzo}\ \emph {et~al.}(2013)\citenamefont
  {Del~Pozzo}, \citenamefont {Li}, \citenamefont {Agathos}, \citenamefont {Van
  Den~Broeck},\ and\ \citenamefont {Vitale}}]{PhysRevLett.111.071101}%
  \BibitemOpen
  \bibfield  {author} {\bibinfo {author} {\bibfnamefont {W.}~\bibnamefont
  {Del~Pozzo}}, \bibinfo {author} {\bibfnamefont {T.~G.~F.}\ \bibnamefont
  {Li}}, \bibinfo {author} {\bibfnamefont {M.}~\bibnamefont {Agathos}},
  \bibinfo {author} {\bibfnamefont {C.}~\bibnamefont {Van Den~Broeck}}, \ and\
  \bibinfo {author} {\bibfnamefont {S.}~\bibnamefont {Vitale}},\ }\href
  {\doibase 10.1103/PhysRevLett.111.071101} {\bibfield  {journal} {\bibinfo
  {journal} {Phys. Rev. Lett.}\ }\textbf {\bibinfo {volume} {111}},\ \bibinfo
  {pages} {071101} (\bibinfo {year} {2013})}\BibitemShut {NoStop}%
\bibitem [{\citenamefont {{Wade}}\ \emph {et~al.}(2014)\citenamefont {{Wade}},
  \citenamefont {{Creighton}}, \citenamefont {{Ochsner}}, \citenamefont
  {{Lackey}}, \citenamefont {{Farr}}, \citenamefont {{Littenberg}},\ and\
  \citenamefont {{Raymond}}}]{2014PhRvD..89j3012W}%
  \BibitemOpen
  \bibfield  {author} {\bibinfo {author} {\bibfnamefont {L.}~\bibnamefont
  {{Wade}}}, \bibinfo {author} {\bibfnamefont {J.~D.~E.}\ \bibnamefont
  {{Creighton}}}, \bibinfo {author} {\bibfnamefont {E.}~\bibnamefont
  {{Ochsner}}}, \bibinfo {author} {\bibfnamefont {B.~D.}\ \bibnamefont
  {{Lackey}}}, \bibinfo {author} {\bibfnamefont {B.~F.}\ \bibnamefont
  {{Farr}}}, \bibinfo {author} {\bibfnamefont {T.~B.}\ \bibnamefont
  {{Littenberg}}}, \ and\ \bibinfo {author} {\bibfnamefont {V.}~\bibnamefont
  {{Raymond}}},\ }\href {\doibase 10.1103/PhysRevD.89.103012} {\bibfield
  {journal} {\bibinfo  {journal} {\prd}\ }\textbf {\bibinfo {volume} {89}},\
  \bibinfo {eid} {103012} (\bibinfo {year} {2014})},\ \Eprint
  {http://arxiv.org/abs/1402.5156} {arXiv:1402.5156 [gr-qc]} \BibitemShut
  {NoStop}%
\bibitem [{\citenamefont {{Lackey}}\ and\ \citenamefont
  {{Wade}}(2015)}]{2015PhRvD..91d3002L}%
  \BibitemOpen
  \bibfield  {author} {\bibinfo {author} {\bibfnamefont {B.~D.}\ \bibnamefont
  {{Lackey}}}\ and\ \bibinfo {author} {\bibfnamefont {L.}~\bibnamefont
  {{Wade}}},\ }\href {\doibase 10.1103/PhysRevD.91.043002} {\bibfield
  {journal} {\bibinfo  {journal} {\prd}\ }\textbf {\bibinfo {volume} {91}},\
  \bibinfo {eid} {043002} (\bibinfo {year} {2015})},\ \Eprint
  {http://arxiv.org/abs/1410.8866} {arXiv:1410.8866 [gr-qc]} \BibitemShut
  {NoStop}%
\bibitem [{\citenamefont {{Agathos}}\ \emph {et~al.}(2015)\citenamefont
  {{Agathos}}, \citenamefont {{Meidam}}, \citenamefont {{Del Pozzo}},
  \citenamefont {{Li}}, \citenamefont {{Tompitak}}, \citenamefont {{Veitch}},
  \citenamefont {{Vitale}},\ and\ \citenamefont {{Van Den
  Broeck}}}]{2015arXiv150305405A}%
  \BibitemOpen
  \bibfield  {author} {\bibinfo {author} {\bibfnamefont {M.}~\bibnamefont
  {{Agathos}}}, \bibinfo {author} {\bibfnamefont {J.}~\bibnamefont {{Meidam}}},
  \bibinfo {author} {\bibfnamefont {W.}~\bibnamefont {{Del Pozzo}}}, \bibinfo
  {author} {\bibfnamefont {T.~G.~F.}\ \bibnamefont {{Li}}}, \bibinfo {author}
  {\bibfnamefont {M.}~\bibnamefont {{Tompitak}}}, \bibinfo {author}
  {\bibfnamefont {J.}~\bibnamefont {{Veitch}}}, \bibinfo {author}
  {\bibfnamefont {S.}~\bibnamefont {{Vitale}}}, \ and\ \bibinfo {author}
  {\bibfnamefont {C.}~\bibnamefont {{Van Den Broeck}}},\ }\href {\doibase
  10.1103/PhysRevD.92.023012} {\bibfield  {journal} {\bibinfo  {journal}
  {\prd}\ }\textbf {\bibinfo {volume} {92}},\ \bibinfo {eid} {023012} (\bibinfo
  {year} {2015})},\ \Eprint {http://arxiv.org/abs/1503.05405} {arXiv:1503.05405
  [gr-qc]} \BibitemShut {NoStop}%
\bibitem [{\citenamefont {Chatziioannou}\ \emph {et~al.}(2015)\citenamefont
  {Chatziioannou}, \citenamefont {Yagi}, \citenamefont {Klein}, \citenamefont
  {Cornish},\ and\ \citenamefont {Yunes}}]{Chatziioannou:2015uea}%
  \BibitemOpen
  \bibfield  {author} {\bibinfo {author} {\bibfnamefont {K.}~\bibnamefont
  {Chatziioannou}}, \bibinfo {author} {\bibfnamefont {K.}~\bibnamefont {Yagi}},
  \bibinfo {author} {\bibfnamefont {A.}~\bibnamefont {Klein}}, \bibinfo
  {author} {\bibfnamefont {N.}~\bibnamefont {Cornish}}, \ and\ \bibinfo
  {author} {\bibfnamefont {N.}~\bibnamefont {Yunes}},\ }\href {\doibase
  10.1103/PhysRevD.92.104008} {\bibfield  {journal} {\bibinfo  {journal} {Phys.
  Rev.}\ }\textbf {\bibinfo {volume} {D92}},\ \bibinfo {pages} {104008}
  (\bibinfo {year} {2015})},\ \Eprint {http://arxiv.org/abs/1508.02062}
  {arXiv:1508.02062 [gr-qc]} \BibitemShut {NoStop}%
%%CITATION = ARXIV:1508.02062;%%
\bibitem [{\citenamefont {Hotokezaka}\ \emph {et~al.}(2016)\citenamefont
  {Hotokezaka}, \citenamefont {Kyutoku}, \citenamefont {Sekiguchi},\ and\
  \citenamefont {Shibata}}]{Hotokezaka:2016bzh}%
  \BibitemOpen
  \bibfield  {author} {\bibinfo {author} {\bibfnamefont {K.}~\bibnamefont
  {Hotokezaka}}, \bibinfo {author} {\bibfnamefont {K.}~\bibnamefont {Kyutoku}},
  \bibinfo {author} {\bibfnamefont {Y.-i.}\ \bibnamefont {Sekiguchi}}, \ and\
  \bibinfo {author} {\bibfnamefont {M.}~\bibnamefont {Shibata}},\ }\href
  {\doibase 10.1103/PhysRevD.93.064082} {\bibfield  {journal} {\bibinfo
  {journal} {Phys. Rev.}\ }\textbf {\bibinfo {volume} {D93}},\ \bibinfo {pages}
  {064082} (\bibinfo {year} {2016})},\ \Eprint
  {http://arxiv.org/abs/1603.01286} {arXiv:1603.01286 [gr-qc]} \BibitemShut
  {NoStop}%
%%CITATION = ARXIV:1603.01286;%%
\bibitem [{\citenamefont {Chatziioannou}\ \emph {et~al.}(2018)\citenamefont
  {Chatziioannou}, \citenamefont {Haster},\ and\ \citenamefont
  {Zimmerman}}]{Chatziioannou:2018vzf}%
  \BibitemOpen
  \bibfield  {author} {\bibinfo {author} {\bibfnamefont {K.}~\bibnamefont
  {Chatziioannou}}, \bibinfo {author} {\bibfnamefont {C.-J.}\ \bibnamefont
  {Haster}}, \ and\ \bibinfo {author} {\bibfnamefont {A.}~\bibnamefont
  {Zimmerman}},\ }\href {\doibase 10.1103/PhysRevD.97.104036} {\bibfield
  {journal} {\bibinfo  {journal} {Phys. Rev.}\ }\textbf {\bibinfo {volume}
  {D97}},\ \bibinfo {pages} {104036} (\bibinfo {year} {2018})},\ \Eprint
  {http://arxiv.org/abs/1804.03221} {arXiv:1804.03221 [gr-qc]} \BibitemShut
  {NoStop}%
%%CITATION = ARXIV:1804.03221;%%
\bibitem [{\citenamefont {Carney}\ \emph {et~al.}(2018)\citenamefont {Carney},
  \citenamefont {Wade},\ and\ \citenamefont {Irwin}}]{Carney:2018sdv}%
  \BibitemOpen
  \bibfield  {author} {\bibinfo {author} {\bibfnamefont {M.~F.}\ \bibnamefont
  {Carney}}, \bibinfo {author} {\bibfnamefont {L.~E.}\ \bibnamefont {Wade}}, \
  and\ \bibinfo {author} {\bibfnamefont {B.~S.}\ \bibnamefont {Irwin}},\ }\href
  {\doibase 10.1103/PhysRevD.98.063004} {\bibfield  {journal} {\bibinfo
  {journal} {Phys. Rev.}\ }\textbf {\bibinfo {volume} {D98}},\ \bibinfo {pages}
  {063004} (\bibinfo {year} {2018})},\ \Eprint
  {http://arxiv.org/abs/1805.11217} {arXiv:1805.11217 [gr-qc]} \BibitemShut
  {NoStop}%
%%CITATION = ARXIV:1805.11217;%%
\bibitem [{\citenamefont {{Zhuge}}\ \emph {et~al.}(1994)\citenamefont
  {{Zhuge}}, \citenamefont {{Centrella}},\ and\ \citenamefont
  {{McMillan}}}]{1994PhRvD..50.6247Z}%
  \BibitemOpen
  \bibfield  {author} {\bibinfo {author} {\bibfnamefont {X.}~\bibnamefont
  {{Zhuge}}}, \bibinfo {author} {\bibfnamefont {J.~M.}\ \bibnamefont
  {{Centrella}}}, \ and\ \bibinfo {author} {\bibfnamefont {S.~L.~W.}\
  \bibnamefont {{McMillan}}},\ }\href {\doibase 10.1103/PhysRevD.50.6247}
  {\bibfield  {journal} {\bibinfo  {journal} {\prd}\ }\textbf {\bibinfo
  {volume} {50}},\ \bibinfo {pages} {6247} (\bibinfo {year} {1994})},\ \Eprint
  {http://arxiv.org/abs/gr-qc/9411029} {gr-qc/9411029} \BibitemShut {NoStop}%
\bibitem [{\citenamefont {{Ruffert}}\ \emph {et~al.}(1996)\citenamefont
  {{Ruffert}}, \citenamefont {{Janka}},\ and\ \citenamefont
  {{Schaefer}}}]{1996A&A...311..532R}%
  \BibitemOpen
  \bibfield  {author} {\bibinfo {author} {\bibfnamefont {M.}~\bibnamefont
  {{Ruffert}}}, \bibinfo {author} {\bibfnamefont {H.-T.}\ \bibnamefont
  {{Janka}}}, \ and\ \bibinfo {author} {\bibfnamefont {G.}~\bibnamefont
  {{Schaefer}}},\ }\href@noop {} {\bibfield  {journal} {\bibinfo  {journal}
  {Astron. Astrophys.}\ }\textbf {\bibinfo {volume} {311}},\ \bibinfo {pages}
  {532} (\bibinfo {year} {1996})},\ \Eprint
  {http://arxiv.org/abs/astro-ph/9509006} {astro-ph/9509006} \BibitemShut
  {NoStop}%
\bibitem [{\citenamefont {{Shibata}}(2005)}]{2005PhRvL..94t1101S}%
  \BibitemOpen
  \bibfield  {author} {\bibinfo {author} {\bibfnamefont {M.}~\bibnamefont
  {{Shibata}}},\ }\href {\doibase 10.1103/PhysRevLett.94.201101} {\bibfield
  {journal} {\bibinfo  {journal} {Phys. Rev. Lett.}\ }\textbf {\bibinfo
  {volume} {94}},\ \bibinfo {eid} {201101} (\bibinfo {year} {2005})},\ \Eprint
  {http://arxiv.org/abs/gr-qc/0504082} {gr-qc/0504082} \BibitemShut {NoStop}%
\bibitem [{\citenamefont {{Shibata}}\ \emph {et~al.}(2005)\citenamefont
  {{Shibata}}, \citenamefont {{Taniguchi}},\ and\ \citenamefont {{Ury{\=
  u}}}}]{2005PhRvD..71h4021S}%
  \BibitemOpen
  \bibfield  {author} {\bibinfo {author} {\bibfnamefont {M.}~\bibnamefont
  {{Shibata}}}, \bibinfo {author} {\bibfnamefont {K.}~\bibnamefont
  {{Taniguchi}}}, \ and\ \bibinfo {author} {\bibfnamefont {K.}~\bibnamefont
  {{Ury{\= u}}}},\ }\href {\doibase 10.1103/PhysRevD.71.084021} {\bibfield
  {journal} {\bibinfo  {journal} {\prd}\ }\textbf {\bibinfo {volume} {71}},\
  \bibinfo {eid} {084021} (\bibinfo {year} {2005})},\ \Eprint
  {http://arxiv.org/abs/gr-qc/0503119} {gr-qc/0503119} \BibitemShut {NoStop}%
\bibitem [{\citenamefont {{Shibata}}\ and\ \citenamefont
  {{Taniguchi}}(2006)}]{shibata:06bns}%
  \BibitemOpen
  \bibfield  {author} {\bibinfo {author} {\bibfnamefont {M.}~\bibnamefont
  {{Shibata}}}\ and\ \bibinfo {author} {\bibfnamefont {K.}~\bibnamefont
  {{Taniguchi}}},\ }\href@noop {} {\bibfield  {journal} {\bibinfo  {journal}
  {\prd}\ }\textbf {\bibinfo {volume} {73}},\ \bibinfo {pages} {064027}
  (\bibinfo {year} {2006})}\BibitemShut {NoStop}%
\bibitem [{\citenamefont {{Oechslin}}\ and\ \citenamefont
  {{Janka}}(2007)}]{2007PhRvL..99l1102O}%
  \BibitemOpen
  \bibfield  {author} {\bibinfo {author} {\bibfnamefont {R.}~\bibnamefont
  {{Oechslin}}}\ and\ \bibinfo {author} {\bibfnamefont {H.-T.}\ \bibnamefont
  {{Janka}}},\ }\href {\doibase 10.1103/PhysRevLett.99.121102} {\bibfield
  {journal} {\bibinfo  {journal} {Phys. Rev. Lett.}\ }\textbf {\bibinfo
  {volume} {99}},\ \bibinfo {eid} {121102} (\bibinfo {year} {2007})},\ \Eprint
  {http://arxiv.org/abs/astro-ph/0702228} {astro-ph/0702228} \BibitemShut
  {NoStop}%
\bibitem [{\citenamefont {{Stergioulas}}\ \emph {et~al.}(2011)\citenamefont
  {{Stergioulas}}, \citenamefont {{Bauswein}}, \citenamefont {{Zagkouris}},\
  and\ \citenamefont {{Janka}}}]{2011MNRAS.418..427S}%
  \BibitemOpen
  \bibfield  {author} {\bibinfo {author} {\bibfnamefont {N.}~\bibnamefont
  {{Stergioulas}}}, \bibinfo {author} {\bibfnamefont {A.}~\bibnamefont
  {{Bauswein}}}, \bibinfo {author} {\bibfnamefont {K.}~\bibnamefont
  {{Zagkouris}}}, \ and\ \bibinfo {author} {\bibfnamefont {H.-T.}\ \bibnamefont
  {{Janka}}},\ }\href {\doibase 10.1111/j.1365-2966.2011.19493.x} {\bibfield
  {journal} {\bibinfo  {journal} {Mon. Not. Roy. Astron. Soc.}\ }\textbf
  {\bibinfo {volume} {418}},\ \bibinfo {pages} {427} (\bibinfo {year}
  {2011})}\BibitemShut {NoStop}%
\bibitem [{\citenamefont {{Hotokezaka}}\ \emph {et~al.}(2011)\citenamefont
  {{Hotokezaka}}, \citenamefont {{Kyutoku}}, \citenamefont {{Okawa}},
  \citenamefont {{Shibata}},\ and\ \citenamefont
  {{Kiuchi}}}]{2011PhRvD..83l4008H}%
  \BibitemOpen
  \bibfield  {author} {\bibinfo {author} {\bibfnamefont {K.}~\bibnamefont
  {{Hotokezaka}}}, \bibinfo {author} {\bibfnamefont {K.}~\bibnamefont
  {{Kyutoku}}}, \bibinfo {author} {\bibfnamefont {H.}~\bibnamefont {{Okawa}}},
  \bibinfo {author} {\bibfnamefont {M.}~\bibnamefont {{Shibata}}}, \ and\
  \bibinfo {author} {\bibfnamefont {K.}~\bibnamefont {{Kiuchi}}},\ }\href
  {\doibase 10.1103/PhysRevD.83.124008} {\bibfield  {journal} {\bibinfo
  {journal} {\prd}\ }\textbf {\bibinfo {volume} {83}},\ \bibinfo {eid} {124008}
  (\bibinfo {year} {2011})},\ \Eprint {http://arxiv.org/abs/1105.4370}
  {arXiv:1105.4370 [astro-ph.HE]} \BibitemShut {NoStop}%
\bibitem [{\citenamefont {{Bauswein}}\ and\ \citenamefont
  {{Janka}}(2012)}]{2012PhRvL.108a1101B}%
  \BibitemOpen
  \bibfield  {author} {\bibinfo {author} {\bibfnamefont {A.}~\bibnamefont
  {{Bauswein}}}\ and\ \bibinfo {author} {\bibfnamefont {H.-T.}\ \bibnamefont
  {{Janka}}},\ }\href {\doibase 10.1103/PhysRevLett.108.011101} {\bibfield
  {journal} {\bibinfo  {journal} {Physical Review Letters}\ }\textbf {\bibinfo
  {volume} {108}},\ \bibinfo {eid} {011101} (\bibinfo {year} {2012})},\ \Eprint
  {http://arxiv.org/abs/1106.1616} {arXiv:1106.1616 [astro-ph.SR]} \BibitemShut
  {NoStop}%
\bibitem [{\citenamefont {{Bauswein}}\ \emph
  {et~al.}(2012{\natexlab{a}})\citenamefont {{Bauswein}}, \citenamefont
  {{Janka}}, \citenamefont {{Hebeler}},\ and\ \citenamefont
  {{Schwenk}}}]{bauswein:12}%
  \BibitemOpen
  \bibfield  {author} {\bibinfo {author} {\bibfnamefont {A.}~\bibnamefont
  {{Bauswein}}}, \bibinfo {author} {\bibfnamefont {H.-T.}\ \bibnamefont
  {{Janka}}}, \bibinfo {author} {\bibfnamefont {K.}~\bibnamefont {{Hebeler}}},
  \ and\ \bibinfo {author} {\bibfnamefont {A.}~\bibnamefont {{Schwenk}}},\
  }\href {\doibase 10.1103/PhysRevD.86.063001} {\bibfield  {journal} {\bibinfo
  {journal} {\prd}\ }\textbf {\bibinfo {volume} {86}},\ \bibinfo {eid} {063001}
  (\bibinfo {year} {2012}{\natexlab{a}})},\ \Eprint
  {http://arxiv.org/abs/1204.1888} {arXiv:1204.1888 [astro-ph.SR]} \BibitemShut
  {NoStop}%
\bibitem [{\citenamefont {{Bauswein}}\ \emph
  {et~al.}(2013{\natexlab{a}})\citenamefont {{Bauswein}}, \citenamefont
  {{Baumgarte}},\ and\ \citenamefont {{Janka}}}]{2013PhRvL.111m1101B}%
  \BibitemOpen
  \bibfield  {author} {\bibinfo {author} {\bibfnamefont {A.}~\bibnamefont
  {{Bauswein}}}, \bibinfo {author} {\bibfnamefont {T.~W.}\ \bibnamefont
  {{Baumgarte}}}, \ and\ \bibinfo {author} {\bibfnamefont {H.-T.}\ \bibnamefont
  {{Janka}}},\ }\href {\doibase 10.1103/PhysRevLett.111.131101} {\bibfield
  {journal} {\bibinfo  {journal} {\prl}\ }\textbf {\bibinfo {volume} {111}},\
  \bibinfo {eid} {131101} (\bibinfo {year} {2013}{\natexlab{a}})}\BibitemShut
  {NoStop}%
\bibitem [{\citenamefont {{Baiotti}}\ \emph {et~al.}(2008)\citenamefont
  {{Baiotti}}, \citenamefont {{Giacomazzo}},\ and\ \citenamefont
  {{Rezzolla}}}]{PhysRevD.78.084033}%
  \BibitemOpen
  \bibfield  {author} {\bibinfo {author} {\bibfnamefont {L.}~\bibnamefont
  {{Baiotti}}}, \bibinfo {author} {\bibfnamefont {B.}~\bibnamefont
  {{Giacomazzo}}}, \ and\ \bibinfo {author} {\bibfnamefont {L.}~\bibnamefont
  {{Rezzolla}}},\ }\href {\doibase 10.1103/PhysRevD.78.084033} {\bibfield
  {journal} {\bibinfo  {journal} {Phys. Rev. D}\ }\textbf {\bibinfo {volume}
  {78}},\ \bibinfo {pages} {084033} (\bibinfo {year} {2008})}\BibitemShut
  {NoStop}%
\bibitem [{\citenamefont {{Sekiguchi}}\ \emph {et~al.}(2011)\citenamefont
  {{Sekiguchi}}, \citenamefont {{Kiuchi}}, \citenamefont {{Kyutoku}},\ and\
  \citenamefont {{Shibata}}}]{2011PhRvL.107e1102S}%
  \BibitemOpen
  \bibfield  {author} {\bibinfo {author} {\bibfnamefont {Y.}~\bibnamefont
  {{Sekiguchi}}}, \bibinfo {author} {\bibfnamefont {K.}~\bibnamefont
  {{Kiuchi}}}, \bibinfo {author} {\bibfnamefont {K.}~\bibnamefont {{Kyutoku}}},
  \ and\ \bibinfo {author} {\bibfnamefont {M.}~\bibnamefont {{Shibata}}},\
  }\href {\doibase 10.1103/PhysRevLett.107.051102} {\bibfield  {journal}
  {\bibinfo  {journal} {prl}\ }\textbf {\bibinfo {volume} {107}},\ \bibinfo
  {eid} {051102} (\bibinfo {year} {2011})},\ \Eprint
  {http://arxiv.org/abs/1105.2125} {arXiv:1105.2125 [gr-qc]} \BibitemShut
  {NoStop}%
\bibitem [{\citenamefont {{Hotokezaka}}\ \emph {et~al.}(2013)\citenamefont
  {{Hotokezaka}}, \citenamefont {{Kiuchi}}, \citenamefont {{Kyutoku}},
  \citenamefont {{Okawa}}, \citenamefont {{Sekiguchi}}, \citenamefont
  {{Shibata}},\ and\ \citenamefont {{Taniguchi}}}]{hotokezaka:13}%
  \BibitemOpen
  \bibfield  {author} {\bibinfo {author} {\bibfnamefont {K.}~\bibnamefont
  {{Hotokezaka}}}, \bibinfo {author} {\bibfnamefont {K.}~\bibnamefont
  {{Kiuchi}}}, \bibinfo {author} {\bibfnamefont {K.}~\bibnamefont {{Kyutoku}}},
  \bibinfo {author} {\bibfnamefont {H.}~\bibnamefont {{Okawa}}}, \bibinfo
  {author} {\bibfnamefont {Y.-i.}\ \bibnamefont {{Sekiguchi}}}, \bibinfo
  {author} {\bibfnamefont {M.}~\bibnamefont {{Shibata}}}, \ and\ \bibinfo
  {author} {\bibfnamefont {K.}~\bibnamefont {{Taniguchi}}},\ }\href {\doibase
  10.1103/PhysRevD.87.024001} {\bibfield  {journal} {\bibinfo  {journal}
  {\prd}\ }\textbf {\bibinfo {volume} {87}},\ \bibinfo {eid} {024001} (\bibinfo
  {year} {2013})},\ \Eprint {http://arxiv.org/abs/1212.0905} {arXiv:1212.0905
  [astro-ph.HE]} \BibitemShut {NoStop}%
\bibitem [{\citenamefont {{Takami}}\ \emph {et~al.}(2014)\citenamefont
  {{Takami}}, \citenamefont {{Rezzolla}},\ and\ \citenamefont
  {{Baiotti}}}]{2014PhRvL.113i1104T}%
  \BibitemOpen
  \bibfield  {author} {\bibinfo {author} {\bibfnamefont {K.}~\bibnamefont
  {{Takami}}}, \bibinfo {author} {\bibfnamefont {L.}~\bibnamefont
  {{Rezzolla}}}, \ and\ \bibinfo {author} {\bibfnamefont {L.}~\bibnamefont
  {{Baiotti}}},\ }\href {\doibase 10.1103/PhysRevLett.113.091104} {\bibfield
  {journal} {\bibinfo  {journal} {Physical Review Letters}\ }\textbf {\bibinfo
  {volume} {113}},\ \bibinfo {eid} {091104} (\bibinfo {year} {2014})},\ \Eprint
  {http://arxiv.org/abs/1403.5672} {arXiv:1403.5672 [gr-qc]} \BibitemShut
  {NoStop}%
\bibitem [{\citenamefont {{Takami}}\ \emph {et~al.}(2015)\citenamefont
  {{Takami}}, \citenamefont {{Rezzolla}},\ and\ \citenamefont
  {{Baiotti}}}]{2014arXiv1412.3240T}%
  \BibitemOpen
  \bibfield  {author} {\bibinfo {author} {\bibfnamefont {K.}~\bibnamefont
  {{Takami}}}, \bibinfo {author} {\bibfnamefont {L.}~\bibnamefont
  {{Rezzolla}}}, \ and\ \bibinfo {author} {\bibfnamefont {L.}~\bibnamefont
  {{Baiotti}}},\ }\href {\doibase 10.1103/PhysRevD.91.064001} {\bibfield
  {journal} {\bibinfo  {journal} {\prd}\ }\textbf {\bibinfo {volume} {91}},\
  \bibinfo {eid} {064001} (\bibinfo {year} {2015})}\BibitemShut {NoStop}%
\bibitem [{\citenamefont {{Bernuzzi}}\ \emph {et~al.}(2015)\citenamefont
  {{Bernuzzi}}, \citenamefont {{Dietrich}},\ and\ \citenamefont
  {{Nagar}}}]{2015arXiv150401764B}%
  \BibitemOpen
  \bibfield  {author} {\bibinfo {author} {\bibfnamefont {S.}~\bibnamefont
  {{Bernuzzi}}}, \bibinfo {author} {\bibfnamefont {T.}~\bibnamefont
  {{Dietrich}}}, \ and\ \bibinfo {author} {\bibfnamefont {A.}~\bibnamefont
  {{Nagar}}},\ }\href {\doibase 10.1103/PhysRevLett.115.091101} {\bibfield
  {journal} {\bibinfo  {journal} {Physical Review Letters}\ }\textbf {\bibinfo
  {volume} {115}},\ \bibinfo {eid} {091101} (\bibinfo {year} {2015})},\ \Eprint
  {http://arxiv.org/abs/1504.01764} {arXiv:1504.01764 [gr-qc]} \BibitemShut
  {NoStop}%
\bibitem [{\citenamefont {Bauswein}\ and\ \citenamefont
  {Stergioulas}(2015)}]{bauswein:15}%
  \BibitemOpen
  \bibfield  {author} {\bibinfo {author} {\bibfnamefont {A.}~\bibnamefont
  {Bauswein}}\ and\ \bibinfo {author} {\bibfnamefont {N.}~\bibnamefont
  {Stergioulas}},\ }\href {\doibase 10.1103/PhysRevD.91.124056} {\bibfield
  {journal} {\bibinfo  {journal} {Phys. Rev. D}\ }\textbf {\bibinfo {volume}
  {91}},\ \bibinfo {pages} {124056} (\bibinfo {year} {2015})}\BibitemShut
  {NoStop}%
\bibitem [{\citenamefont {{Foucart}}\ \emph {et~al.}(2016)\citenamefont
  {{Foucart}}, \citenamefont {{Haas}}, \citenamefont {{Duez}}, \citenamefont
  {{O'Connor}}, \citenamefont {{Ott}}, \citenamefont {{Roberts}}, \citenamefont
  {{Kidder}}, \citenamefont {{Lippuner}}, \citenamefont {{Pfeiffer}},\ and\
  \citenamefont {{Scheel}}}]{Foucart2016}%
  \BibitemOpen
  \bibfield  {author} {\bibinfo {author} {\bibfnamefont {F.}~\bibnamefont
  {{Foucart}}}, \bibinfo {author} {\bibfnamefont {R.}~\bibnamefont {{Haas}}},
  \bibinfo {author} {\bibfnamefont {M.~D.}\ \bibnamefont {{Duez}}}, \bibinfo
  {author} {\bibfnamefont {E.}~\bibnamefont {{O'Connor}}}, \bibinfo {author}
  {\bibfnamefont {C.~D.}\ \bibnamefont {{Ott}}}, \bibinfo {author}
  {\bibfnamefont {L.}~\bibnamefont {{Roberts}}}, \bibinfo {author}
  {\bibfnamefont {L.~E.}\ \bibnamefont {{Kidder}}}, \bibinfo {author}
  {\bibfnamefont {J.}~\bibnamefont {{Lippuner}}}, \bibinfo {author}
  {\bibfnamefont {H.~P.}\ \bibnamefont {{Pfeiffer}}}, \ and\ \bibinfo {author}
  {\bibfnamefont {M.~A.}\ \bibnamefont {{Scheel}}},\ }\href {\doibase
  10.1103/PhysRevD.93.044019} {\bibfield  {journal} {\bibinfo  {journal}
  {\prd}\ }\textbf {\bibinfo {volume} {93}},\ \bibinfo {eid} {044019} (\bibinfo
  {year} {2016})},\ \Eprint {http://arxiv.org/abs/1510.06398} {arXiv:1510.06398
  [astro-ph.HE]} \BibitemShut {NoStop}%
\bibitem [{\citenamefont {Lehner}\ \emph {et~al.}(2016)\citenamefont {Lehner},
  \citenamefont {Liebling}, \citenamefont {Palenzuela}, \citenamefont
  {Caballero}, \citenamefont {O'Connor}, \citenamefont {Anderson},\ and\
  \citenamefont {Neilsen}}]{Lehner:2016lxy}%
  \BibitemOpen
  \bibfield  {author} {\bibinfo {author} {\bibfnamefont {L.}~\bibnamefont
  {Lehner}}, \bibinfo {author} {\bibfnamefont {S.~L.}\ \bibnamefont
  {Liebling}}, \bibinfo {author} {\bibfnamefont {C.}~\bibnamefont
  {Palenzuela}}, \bibinfo {author} {\bibfnamefont {O.~L.}\ \bibnamefont
  {Caballero}}, \bibinfo {author} {\bibfnamefont {E.}~\bibnamefont {O'Connor}},
  \bibinfo {author} {\bibfnamefont {M.}~\bibnamefont {Anderson}}, \ and\
  \bibinfo {author} {\bibfnamefont {D.}~\bibnamefont {Neilsen}},\ }\href
  {\doibase 10.1088/0264-9381/33/18/184002} {\bibfield  {journal} {\bibinfo
  {journal} {Class. Quant. Grav.}\ }\textbf {\bibinfo {volume} {33}},\ \bibinfo
  {pages} {184002} (\bibinfo {year} {2016})},\ \Eprint
  {http://arxiv.org/abs/1603.00501} {arXiv:1603.00501 [gr-qc]} \BibitemShut
  {NoStop}%
%%CITATION = ARXIV:1603.00501;%%
\bibitem [{\citenamefont {{Kawamura}}\ \emph {et~al.}(2016)\citenamefont
  {{Kawamura}}, \citenamefont {{Giacomazzo}}, \citenamefont {{Kastaun}},
  \citenamefont {{Ciolfi}}, \citenamefont {{Endrizzi}}, \citenamefont
  {{Baiotti}},\ and\ \citenamefont {{Perna}}}]{Kawamura2016}%
  \BibitemOpen
  \bibfield  {author} {\bibinfo {author} {\bibfnamefont {T.}~\bibnamefont
  {{Kawamura}}}, \bibinfo {author} {\bibfnamefont {B.}~\bibnamefont
  {{Giacomazzo}}}, \bibinfo {author} {\bibfnamefont {W.}~\bibnamefont
  {{Kastaun}}}, \bibinfo {author} {\bibfnamefont {R.}~\bibnamefont {{Ciolfi}}},
  \bibinfo {author} {\bibfnamefont {A.}~\bibnamefont {{Endrizzi}}}, \bibinfo
  {author} {\bibfnamefont {L.}~\bibnamefont {{Baiotti}}}, \ and\ \bibinfo
  {author} {\bibfnamefont {R.}~\bibnamefont {{Perna}}},\ }\href {\doibase
  10.1103/PhysRevD.94.064012} {\bibfield  {journal} {\bibinfo  {journal}
  {\prd}\ }\textbf {\bibinfo {volume} {94}},\ \bibinfo {eid} {064012} (\bibinfo
  {year} {2016})},\ \Eprint {http://arxiv.org/abs/1607.01791} {arXiv:1607.01791
  [astro-ph.HE]} \BibitemShut {NoStop}%
\bibitem [{\citenamefont {{Maione}}\ \emph {et~al.}(2017)\citenamefont
  {{Maione}}, \citenamefont {{De Pietri}}, \citenamefont {{Feo}},\ and\
  \citenamefont {{L{\"o}ffler}}}]{Maione2017}%
  \BibitemOpen
  \bibfield  {author} {\bibinfo {author} {\bibfnamefont {F.}~\bibnamefont
  {{Maione}}}, \bibinfo {author} {\bibfnamefont {R.}~\bibnamefont {{De
  Pietri}}}, \bibinfo {author} {\bibfnamefont {A.}~\bibnamefont {{Feo}}}, \
  and\ \bibinfo {author} {\bibfnamefont {F.}~\bibnamefont {{L{\"o}ffler}}},\
  }\href {\doibase 10.1103/PhysRevD.96.063011} {\bibfield  {journal} {\bibinfo
  {journal} {\prd}\ }\textbf {\bibinfo {volume} {96}},\ \bibinfo {eid} {063011}
  (\bibinfo {year} {2017})},\ \Eprint {http://arxiv.org/abs/1707.03368}
  {arXiv:1707.03368 [gr-qc]} \BibitemShut {NoStop}%
\bibitem [{\citenamefont {Bauswein}\ \emph {et~al.}(2018)\citenamefont
  {Bauswein}, \citenamefont {Bastian}, \citenamefont {Blaschke}, \citenamefont
  {Chatziioannou}, \citenamefont {Clark}, \citenamefont {Fischer},\ and\
  \citenamefont {Oertel}}]{Bauswein:2018bma}%
  \BibitemOpen
  \bibfield  {author} {\bibinfo {author} {\bibfnamefont {A.}~\bibnamefont
  {Bauswein}}, \bibinfo {author} {\bibfnamefont {N.-U.~F.}\ \bibnamefont
  {Bastian}}, \bibinfo {author} {\bibfnamefont {D.~B.}\ \bibnamefont
  {Blaschke}}, \bibinfo {author} {\bibfnamefont {K.}~\bibnamefont
  {Chatziioannou}}, \bibinfo {author} {\bibfnamefont {J.~A.}\ \bibnamefont
  {Clark}}, \bibinfo {author} {\bibfnamefont {T.}~\bibnamefont {Fischer}}, \
  and\ \bibinfo {author} {\bibfnamefont {M.}~\bibnamefont {Oertel}},\
  }\href@noop {} {\  (\bibinfo {year} {2018})},\ \Eprint
  {http://arxiv.org/abs/1809.01116} {arXiv:1809.01116 [astro-ph.HE]}
  \BibitemShut {NoStop}%
%%CITATION = ARXIV:1809.01116;%%
\bibitem [{\citenamefont {Most}\ \emph
  {et~al.}(2018{\natexlab{a}})\citenamefont {Most}, \citenamefont {Papenfort},
  \citenamefont {Dexheimer}, \citenamefont {Hanauske}, \citenamefont {Schramm},
  \citenamefont {St{\"o}cker},\ and\ \citenamefont {Rezzolla}}]{Most2018a}%
  \BibitemOpen
  \bibfield  {author} {\bibinfo {author} {\bibfnamefont {E.~R.}\ \bibnamefont
  {Most}}, \bibinfo {author} {\bibfnamefont {L.~J.}\ \bibnamefont {Papenfort}},
  \bibinfo {author} {\bibfnamefont {V.}~\bibnamefont {Dexheimer}}, \bibinfo
  {author} {\bibfnamefont {M.}~\bibnamefont {Hanauske}}, \bibinfo {author}
  {\bibfnamefont {S.}~\bibnamefont {Schramm}}, \bibinfo {author} {\bibfnamefont
  {H.}~\bibnamefont {St{\"o}cker}}, \ and\ \bibinfo {author} {\bibfnamefont
  {L.}~\bibnamefont {Rezzolla}},\ }\href@noop {} {\bibfield  {journal}
  {\bibinfo  {journal} {ArXiv e-prints}\ } (\bibinfo {year}
  {2018}{\natexlab{a}})},\ \Eprint {http://arxiv.org/abs/1807.03684}
  {arXiv:1807.03684 [astro-ph.HE]} \BibitemShut {NoStop}%
\bibitem [{\citenamefont {Margalit}\ and\ \citenamefont
  {Metzger}(2017)}]{2041-8205-850-2-L19}%
  \BibitemOpen
  \bibfield  {author} {\bibinfo {author} {\bibfnamefont {B.}~\bibnamefont
  {Margalit}}\ and\ \bibinfo {author} {\bibfnamefont {B.~D.}\ \bibnamefont
  {Metzger}},\ }\href {http://stacks.iop.org/2041-8205/850/i=2/a=L19}
  {\bibfield  {journal} {\bibinfo  {journal} {The Astrophysical Journal
  Letters}\ }\textbf {\bibinfo {volume} {850}},\ \bibinfo {pages} {L19}
  (\bibinfo {year} {2017})}\BibitemShut {NoStop}%
\bibitem [{\citenamefont {Bauswein}\ \emph {et~al.}(2017)\citenamefont
  {Bauswein}, \citenamefont {Just}, \citenamefont {Janka},\ and\ \citenamefont
  {Stergioulas}}]{2041-8205-850-2-L34}%
  \BibitemOpen
  \bibfield  {author} {\bibinfo {author} {\bibfnamefont {A.}~\bibnamefont
  {Bauswein}}, \bibinfo {author} {\bibfnamefont {O.}~\bibnamefont {Just}},
  \bibinfo {author} {\bibfnamefont {H.-T.}\ \bibnamefont {Janka}}, \ and\
  \bibinfo {author} {\bibfnamefont {N.}~\bibnamefont {Stergioulas}},\ }\href
  {http://stacks.iop.org/2041-8205/850/i=2/a=L34} {\bibfield  {journal}
  {\bibinfo  {journal} {The Astrophysical Journal Letters}\ }\textbf {\bibinfo
  {volume} {850}},\ \bibinfo {pages} {L34} (\bibinfo {year}
  {2017})}\BibitemShut {NoStop}%
\bibitem [{\citenamefont {Zhou}\ \emph {et~al.}(2018)\citenamefont {Zhou},
  \citenamefont {Zhou},\ and\ \citenamefont {Li}}]{Zhou:2017pha}%
  \BibitemOpen
  \bibfield  {author} {\bibinfo {author} {\bibfnamefont {E.-P.}\ \bibnamefont
  {Zhou}}, \bibinfo {author} {\bibfnamefont {X.}~\bibnamefont {Zhou}}, \ and\
  \bibinfo {author} {\bibfnamefont {A.}~\bibnamefont {Li}},\ }\href {\doibase
  10.1103/PhysRevD.97.083015} {\bibfield  {journal} {\bibinfo  {journal}
  {\prd}\ }\textbf {\bibinfo {volume} {97}},\ \bibinfo {pages} {083015}
  (\bibinfo {year} {2018})},\ \Eprint {http://arxiv.org/abs/1711.04312}
  {arXiv:1711.04312 [astro-ph.HE]} \BibitemShut {NoStop}%
%%CITATION = ARXIV:1711.04312;%%
\bibitem [{\citenamefont {Rezzolla}\ \emph {et~al.}(2018)\citenamefont
  {Rezzolla}, \citenamefont {Most},\ and\ \citenamefont
  {Weih}}]{2041-8205-852-2-L25}%
  \BibitemOpen
  \bibfield  {author} {\bibinfo {author} {\bibfnamefont {L.}~\bibnamefont
  {Rezzolla}}, \bibinfo {author} {\bibfnamefont {E.~R.}\ \bibnamefont {Most}},
  \ and\ \bibinfo {author} {\bibfnamefont {L.~R.}\ \bibnamefont {Weih}},\
  }\href {http://stacks.iop.org/2041-8205/852/i=2/a=L25} {\bibfield  {journal}
  {\bibinfo  {journal} {The Astrophysical Journal Letters}\ }\textbf {\bibinfo
  {volume} {852}},\ \bibinfo {pages} {L25} (\bibinfo {year}
  {2018})}\BibitemShut {NoStop}%
\bibitem [{\citenamefont {Fattoyev}\ \emph {et~al.}(2018)\citenamefont
  {Fattoyev}, \citenamefont {Piekarewicz},\ and\ \citenamefont
  {Horowitz}}]{PhysRevLett.120.172702}%
  \BibitemOpen
  \bibfield  {author} {\bibinfo {author} {\bibfnamefont {F.~J.}\ \bibnamefont
  {Fattoyev}}, \bibinfo {author} {\bibfnamefont {J.}~\bibnamefont
  {Piekarewicz}}, \ and\ \bibinfo {author} {\bibfnamefont {C.~J.}\ \bibnamefont
  {Horowitz}},\ }\href {\doibase 10.1103/PhysRevLett.120.172702} {\bibfield
  {journal} {\bibinfo  {journal} {Phys. Rev. Lett.}\ }\textbf {\bibinfo
  {volume} {120}},\ \bibinfo {pages} {172702} (\bibinfo {year}
  {2018})}\BibitemShut {NoStop}%
\bibitem [{\citenamefont {Nandi}\ and\ \citenamefont
  {Char}(2018)}]{Nandi:2017rhy}%
  \BibitemOpen
  \bibfield  {author} {\bibinfo {author} {\bibfnamefont {R.}~\bibnamefont
  {Nandi}}\ and\ \bibinfo {author} {\bibfnamefont {P.}~\bibnamefont {Char}},\
  }\href {\doibase 10.3847/1538-4357/aab78c} {\bibfield  {journal} {\bibinfo
  {journal} {Astrophys. J.}\ }\textbf {\bibinfo {volume} {857}},\ \bibinfo
  {pages} {12} (\bibinfo {year} {2018})},\ \Eprint
  {http://arxiv.org/abs/1712.08094} {arXiv:1712.08094 [astro-ph.HE]}
  \BibitemShut {NoStop}%
%%CITATION = ARXIV:1712.08094;%%
\bibitem [{\citenamefont {Ruiz}\ \emph {et~al.}(2018)\citenamefont {Ruiz},
  \citenamefont {Shapiro},\ and\ \citenamefont
  {Tsokaros}}]{PhysRevD.97.021501}%
  \BibitemOpen
  \bibfield  {author} {\bibinfo {author} {\bibfnamefont {M.}~\bibnamefont
  {Ruiz}}, \bibinfo {author} {\bibfnamefont {S.~L.}\ \bibnamefont {Shapiro}}, \
  and\ \bibinfo {author} {\bibfnamefont {A.}~\bibnamefont {Tsokaros}},\ }\href
  {\doibase 10.1103/PhysRevD.97.021501} {\bibfield  {journal} {\bibinfo
  {journal} {Phys. Rev. D}\ }\textbf {\bibinfo {volume} {97}},\ \bibinfo
  {pages} {021501} (\bibinfo {year} {2018})}\BibitemShut {NoStop}%
\bibitem [{\citenamefont {Annala}\ \emph {et~al.}(2018)\citenamefont {Annala},
  \citenamefont {Gorda}, \citenamefont {Kurkela},\ and\ \citenamefont
  {Vuorinen}}]{PhysRevLett.120.172703}%
  \BibitemOpen
  \bibfield  {author} {\bibinfo {author} {\bibfnamefont {E.}~\bibnamefont
  {Annala}}, \bibinfo {author} {\bibfnamefont {T.}~\bibnamefont {Gorda}},
  \bibinfo {author} {\bibfnamefont {A.}~\bibnamefont {Kurkela}}, \ and\
  \bibinfo {author} {\bibfnamefont {A.}~\bibnamefont {Vuorinen}},\ }\href
  {\doibase 10.1103/PhysRevLett.120.172703} {\bibfield  {journal} {\bibinfo
  {journal} {Phys. Rev. Lett.}\ }\textbf {\bibinfo {volume} {120}},\ \bibinfo
  {pages} {172703} (\bibinfo {year} {2018})}\BibitemShut {NoStop}%
\bibitem [{\citenamefont {Raithel}\ \emph {et~al.}(2018)\citenamefont
  {Raithel}, \citenamefont {{\"{O}}zel},\ and\ \citenamefont
  {Psaltis}}]{Raithel:2018ncd}%
  \BibitemOpen
  \bibfield  {author} {\bibinfo {author} {\bibfnamefont {C.}~\bibnamefont
  {Raithel}}, \bibinfo {author} {\bibfnamefont {F.}~\bibnamefont {{\"{O}}zel}},
  \ and\ \bibinfo {author} {\bibfnamefont {D.}~\bibnamefont {Psaltis}},\
  }\href@noop {} {\  (\bibinfo {year} {2018})},\ \Eprint
  {http://arxiv.org/abs/1803.07687} {arXiv:1803.07687 [astro-ph.HE]}
  \BibitemShut {NoStop}%
%%CITATION = ARXIV:1803.07687;%%
\bibitem [{\citenamefont {Most}\ \emph
  {et~al.}(2018{\natexlab{b}})\citenamefont {Most}, \citenamefont {Weih},
  \citenamefont {Rezzolla},\ and\ \citenamefont
  {Schaffner-Bielich}}]{Most:2018hfd}%
  \BibitemOpen
  \bibfield  {author} {\bibinfo {author} {\bibfnamefont {E.~R.}\ \bibnamefont
  {Most}}, \bibinfo {author} {\bibfnamefont {L.~R.}\ \bibnamefont {Weih}},
  \bibinfo {author} {\bibfnamefont {L.}~\bibnamefont {Rezzolla}}, \ and\
  \bibinfo {author} {\bibfnamefont {J.}~\bibnamefont {Schaffner-Bielich}},\
  }\href@noop {} {\  (\bibinfo {year} {2018}{\natexlab{b}})},\ \Eprint
  {http://arxiv.org/abs/1803.00549} {arXiv:1803.00549 [gr-qc]} \BibitemShut
  {NoStop}%
%%CITATION = ARXIV:1803.00549;%%
\bibitem [{\citenamefont {Radice}\ \emph {et~al.}(2018)\citenamefont {Radice},
  \citenamefont {Perego}, \citenamefont {Zappa},\ and\ \citenamefont
  {Bernuzzi}}]{Radice2018}%
  \BibitemOpen
  \bibfield  {author} {\bibinfo {author} {\bibfnamefont {D.}~\bibnamefont
  {Radice}}, \bibinfo {author} {\bibfnamefont {A.}~\bibnamefont {Perego}},
  \bibinfo {author} {\bibfnamefont {F.}~\bibnamefont {Zappa}}, \ and\ \bibinfo
  {author} {\bibfnamefont {S.}~\bibnamefont {Bernuzzi}},\ }\href {\doibase
  10.3847/2041-8213/aaa402} {\bibfield  {journal} {\bibinfo  {journal}
  {Astrophys. J. L.}\ }\textbf {\bibinfo {volume} {852}},\ \bibinfo {eid} {L29}
  (\bibinfo {year} {2018})},\ \Eprint {http://arxiv.org/abs/1711.03647}
  {arXiv:1711.03647 [astro-ph.HE]} \BibitemShut {NoStop}%
\bibitem [{\citenamefont {De}\ \emph {et~al.}(2018)\citenamefont {De},
  \citenamefont {Finstad}, \citenamefont {Lattimer}, \citenamefont {Brown},
  \citenamefont {Berger},\ and\ \citenamefont {Biwer}}]{De:2018uhw}%
  \BibitemOpen
  \bibfield  {author} {\bibinfo {author} {\bibfnamefont {S.}~\bibnamefont
  {De}}, \bibinfo {author} {\bibfnamefont {D.}~\bibnamefont {Finstad}},
  \bibinfo {author} {\bibfnamefont {J.~M.}\ \bibnamefont {Lattimer}}, \bibinfo
  {author} {\bibfnamefont {D.~A.}\ \bibnamefont {Brown}}, \bibinfo {author}
  {\bibfnamefont {E.}~\bibnamefont {Berger}}, \ and\ \bibinfo {author}
  {\bibfnamefont {C.~M.}\ \bibnamefont {Biwer}},\ }\href@noop {} {\  (\bibinfo
  {year} {2018})},\ \Eprint {http://arxiv.org/abs/1804.08583} {arXiv:1804.08583
  [astro-ph.HE]} \BibitemShut {NoStop}%
%%CITATION = ARXIV:1804.08583;%%
\bibitem [{\citenamefont {Abbott}\ \emph {et~al.}(2018)\citenamefont {Abbott}
  \emph {et~al.}}]{Abbott:2018wiz}%
  \BibitemOpen
  \bibfield  {author} {\bibinfo {author} {\bibfnamefont {B.~P.}\ \bibnamefont
  {Abbott}} \emph {et~al.} (\bibinfo {collaboration} {Virgo, LIGO
  Scientific}),\ }\href@noop {} {\  (\bibinfo {year} {2018})},\ \Eprint
  {http://arxiv.org/abs/1805.11579} {arXiv:1805.11579 [gr-qc]} \BibitemShut
  {NoStop}%
%%CITATION = ARXIV:1805.11579;%%
\bibitem [{\citenamefont {{The LIGO Scientific Collaboration}}\ \emph
  {et~al.}(2018)\citenamefont {{The LIGO Scientific Collaboration}},
  \citenamefont {{the Virgo Collaboration}}, \citenamefont {{Abbott}},
  \citenamefont {{Abbott}}, \citenamefont {{Abbott}}, \citenamefont
  {{Acernese}}, \citenamefont {{Ackley}}, \citenamefont {{Adams}},
  \citenamefont {{Adams}}, \citenamefont {{Addesso}},\ and\ \citenamefont
  {et~al.}}]{LigoEoS2018}%
  \BibitemOpen
  \bibfield  {author} {\bibinfo {author} {\bibnamefont {{The LIGO Scientific
  Collaboration}}}, \bibinfo {author} {\bibnamefont {{the Virgo
  Collaboration}}}, \bibinfo {author} {\bibfnamefont {B.~P.}\ \bibnamefont
  {{Abbott}}}, \bibinfo {author} {\bibfnamefont {R.}~\bibnamefont {{Abbott}}},
  \bibinfo {author} {\bibfnamefont {T.~D.}\ \bibnamefont {{Abbott}}}, \bibinfo
  {author} {\bibfnamefont {F.}~\bibnamefont {{Acernese}}}, \bibinfo {author}
  {\bibfnamefont {K.}~\bibnamefont {{Ackley}}}, \bibinfo {author}
  {\bibfnamefont {C.}~\bibnamefont {{Adams}}}, \bibinfo {author} {\bibfnamefont
  {T.}~\bibnamefont {{Adams}}}, \bibinfo {author} {\bibfnamefont
  {P.}~\bibnamefont {{Addesso}}}, \ and\ \bibinfo {author} {\bibnamefont
  {et~al.}},\ }\href@noop {} {\bibfield  {journal} {\bibinfo  {journal} {ArXiv
  e-prints}\ } (\bibinfo {year} {2018})},\ \Eprint
  {http://arxiv.org/abs/1805.11581} {arXiv:1805.11581 [gr-qc]} \BibitemShut
  {NoStop}%
\bibitem [{\citenamefont {Tsang}\ \emph {et~al.}(2018)\citenamefont {Tsang},
  \citenamefont {Tsang}, \citenamefont {Danielewicz}, \citenamefont {Lynch},\
  and\ \citenamefont {Fattoyev}}]{Tsang:2018kqj}%
  \BibitemOpen
  \bibfield  {author} {\bibinfo {author} {\bibfnamefont {C.~Y.}\ \bibnamefont
  {Tsang}}, \bibinfo {author} {\bibfnamefont {M.~B.}\ \bibnamefont {Tsang}},
  \bibinfo {author} {\bibfnamefont {P.}~\bibnamefont {Danielewicz}}, \bibinfo
  {author} {\bibfnamefont {W.~G.}\ \bibnamefont {Lynch}}, \ and\ \bibinfo
  {author} {\bibfnamefont {F.~J.}\ \bibnamefont {Fattoyev}},\ }\href@noop {} {\
   (\bibinfo {year} {2018})},\ \Eprint {http://arxiv.org/abs/1807.06571}
  {arXiv:1807.06571 [nucl-ex]} \BibitemShut {NoStop}%
%%CITATION = ARXIV:1807.06571;%%
\bibitem [{\citenamefont {Abbott}\ \emph
  {et~al.}(2017{\natexlab{d}})\citenamefont {Abbott} \emph
  {et~al.}}]{Abbott:2017dke}%
  \BibitemOpen
  \bibfield  {author} {\bibinfo {author} {\bibfnamefont {B.~P.}\ \bibnamefont
  {Abbott}} \emph {et~al.} (\bibinfo {collaboration} {Virgo, LIGO
  Scientific}),\ }\href {\doibase 10.3847/2041-8213/aa9a35} {\bibfield
  {journal} {\bibinfo  {journal} {Astrophys. J.}\ }\textbf {\bibinfo {volume}
  {851}},\ \bibinfo {pages} {L16} (\bibinfo {year} {2017}{\natexlab{d}})},\
  \Eprint {http://arxiv.org/abs/1710.09320} {arXiv:1710.09320 [astro-ph.HE]}
  \BibitemShut {NoStop}%
%%CITATION = ARXIV:1710.09320;%%
\bibitem [{\citenamefont {Clark}\ \emph {et~al.}(2016)\citenamefont {Clark},
  \citenamefont {Bauswein}, \citenamefont {Stergioulas},\ and\ \citenamefont
  {Shoemaker}}]{Clark:2015zxa}%
  \BibitemOpen
  \bibfield  {author} {\bibinfo {author} {\bibfnamefont {J.~A.}\ \bibnamefont
  {Clark}}, \bibinfo {author} {\bibfnamefont {A.}~\bibnamefont {Bauswein}},
  \bibinfo {author} {\bibfnamefont {N.}~\bibnamefont {Stergioulas}}, \ and\
  \bibinfo {author} {\bibfnamefont {D.}~\bibnamefont {Shoemaker}},\ }\href
  {\doibase 10.1088/0264-9381/33/8/085003} {\bibfield  {journal} {\bibinfo
  {journal} {Class. Quant. Grav.}\ }\textbf {\bibinfo {volume} {33}},\ \bibinfo
  {pages} {085003} (\bibinfo {year} {2016})},\ \Eprint
  {http://arxiv.org/abs/1509.08522} {arXiv:1509.08522 [astro-ph.HE]}
  \BibitemShut {NoStop}%
%%CITATION = ARXIV:1509.08522;%%
\bibitem [{\citenamefont {{Bauswein}}\ \emph {et~al.}(2016)\citenamefont
  {{Bauswein}}, \citenamefont {{Stergioulas}},\ and\ \citenamefont
  {{Janka}}}]{bauswein:july15}%
  \BibitemOpen
  \bibfield  {author} {\bibinfo {author} {\bibfnamefont {A.}~\bibnamefont
  {{Bauswein}}}, \bibinfo {author} {\bibfnamefont {N.}~\bibnamefont
  {{Stergioulas}}}, \ and\ \bibinfo {author} {\bibfnamefont {H.-T.}\
  \bibnamefont {{Janka}}},\ }\href {\doibase 10.1140/epja/i2016-16056-7}
  {\bibfield  {journal} {\bibinfo  {journal} {European Physical Journal A}\
  }\textbf {\bibinfo {volume} {52}},\ \bibinfo {eid} {56} (\bibinfo {year}
  {2016})},\ \Eprint {http://arxiv.org/abs/1508.05493} {arXiv:1508.05493
  [astro-ph.HE]} \BibitemShut {NoStop}%
\bibitem [{\citenamefont {{Clark}}\ \emph {et~al.}(2014)\citenamefont
  {{Clark}}, \citenamefont {{Bauswein}}, \citenamefont {{Cadonati}},
  \citenamefont {{Janka}}, \citenamefont {{Pankow}},\ and\ \citenamefont
  {{Stergioulas}}}]{2014PhRvD..90f2004C}%
  \BibitemOpen
  \bibfield  {author} {\bibinfo {author} {\bibfnamefont {J.}~\bibnamefont
  {{Clark}}}, \bibinfo {author} {\bibfnamefont {A.}~\bibnamefont {{Bauswein}}},
  \bibinfo {author} {\bibfnamefont {L.}~\bibnamefont {{Cadonati}}}, \bibinfo
  {author} {\bibfnamefont {H.-T.}\ \bibnamefont {{Janka}}}, \bibinfo {author}
  {\bibfnamefont {C.}~\bibnamefont {{Pankow}}}, \ and\ \bibinfo {author}
  {\bibfnamefont {N.}~\bibnamefont {{Stergioulas}}},\ }\href {\doibase
  10.1103/PhysRevD.90.062004} {\bibfield  {journal} {\bibinfo  {journal} {Phys.
  Rev. D}\ }\textbf {\bibinfo {volume} {90}},\ \bibinfo {eid} {062004}
  (\bibinfo {year} {2014})},\ \Eprint {http://arxiv.org/abs/1406.5444}
  {arXiv:1406.5444 [astro-ph.HE]} \BibitemShut {NoStop}%
\bibitem [{\citenamefont {Chatziioannou}\ \emph {et~al.}(2017)\citenamefont
  {Chatziioannou}, \citenamefont {Clark}, \citenamefont {Bauswein},
  \citenamefont {Millhouse}, \citenamefont {Littenberg},\ and\ \citenamefont
  {Cornish}}]{Chatziioannou:2017ixj}%
  \BibitemOpen
  \bibfield  {author} {\bibinfo {author} {\bibfnamefont {K.}~\bibnamefont
  {Chatziioannou}}, \bibinfo {author} {\bibfnamefont {J.~A.}\ \bibnamefont
  {Clark}}, \bibinfo {author} {\bibfnamefont {A.}~\bibnamefont {Bauswein}},
  \bibinfo {author} {\bibfnamefont {M.}~\bibnamefont {Millhouse}}, \bibinfo
  {author} {\bibfnamefont {T.~B.}\ \bibnamefont {Littenberg}}, \ and\ \bibinfo
  {author} {\bibfnamefont {N.}~\bibnamefont {Cornish}},\ }\href {\doibase
  10.1103/PhysRevD.96.124035} {\bibfield  {journal} {\bibinfo  {journal} {Phys.
  Rev.}\ }\textbf {\bibinfo {volume} {D96}},\ \bibinfo {pages} {124035}
  (\bibinfo {year} {2017})},\ \Eprint {http://arxiv.org/abs/1711.00040}
  {arXiv:1711.00040 [gr-qc]} \BibitemShut {NoStop}%
%%CITATION = ARXIV:1711.00040;%%
\bibitem [{\citenamefont {Cornish}\ and\ \citenamefont
  {Littenberg}(2015)}]{Cornish:2014kda}%
  \BibitemOpen
  \bibfield  {author} {\bibinfo {author} {\bibfnamefont {N.~J.}\ \bibnamefont
  {Cornish}}\ and\ \bibinfo {author} {\bibfnamefont {T.~B.}\ \bibnamefont
  {Littenberg}},\ }\href {\doibase 10.1088/0264-9381/32/13/135012} {\bibfield
  {journal} {\bibinfo  {journal} {Class. Quant. Grav.}\ }\textbf {\bibinfo
  {volume} {32}},\ \bibinfo {pages} {135012} (\bibinfo {year} {2015})},\
  \Eprint {http://arxiv.org/abs/1410.3835} {arXiv:1410.3835 [gr-qc]}
  \BibitemShut {NoStop}%
%%CITATION = ARXIV:1410.3835;%%
\bibitem [{\citenamefont {Littenberg}\ and\ \citenamefont
  {Cornish}(2015)}]{Littenberg:2014oda}%
  \BibitemOpen
  \bibfield  {author} {\bibinfo {author} {\bibfnamefont {T.~B.}\ \bibnamefont
  {Littenberg}}\ and\ \bibinfo {author} {\bibfnamefont {N.~J.}\ \bibnamefont
  {Cornish}},\ }\href {\doibase 10.1103/PhysRevD.91.084034} {\bibfield
  {journal} {\bibinfo  {journal} {Phys. Rev.}\ }\textbf {\bibinfo {volume}
  {D91}},\ \bibinfo {pages} {084034} (\bibinfo {year} {2015})},\ \Eprint
  {http://arxiv.org/abs/1410.3852} {arXiv:1410.3852 [gr-qc]} \BibitemShut
  {NoStop}%
%%CITATION = ARXIV:1410.3852;%%
\bibitem [{\citenamefont {Abbott}\ \emph {et~al.}(2013)\citenamefont {Abbott}
  \emph {et~al.}}]{Aasi:2013wya}%
  \BibitemOpen
  \bibfield  {author} {\bibinfo {author} {\bibfnamefont {B.~P.}\ \bibnamefont
  {Abbott}} \emph {et~al.} (\bibinfo {collaboration} {VIRGO, LIGO
  Scientific}),\ }\href {\doibase 10.1007/lrr-2016-1} {\  (\bibinfo {year}
  {2013}),\ 10.1007/lrr-2016-1},\ \bibinfo {note} {[Living Rev.
  Rel.19,1(2016)]},\ \Eprint {http://arxiv.org/abs/1304.0670} {arXiv:1304.0670
  [gr-qc]} \BibitemShut {NoStop}%
%%CITATION = ARXIV:1304.0670;%%
\bibitem [{\citenamefont {{Punturo}}\ \emph {et~al.}(2010)\citenamefont
  {{Punturo}} \emph {et~al.}}]{2010CQGra..27h4007P}%
  \BibitemOpen
  \bibfield  {author} {\bibinfo {author} {\bibfnamefont {M.}~\bibnamefont
  {{Punturo}}} \emph {et~al.},\ }\href {\doibase 10.1088/0264-9381/27/8/084007}
  {\bibfield  {journal} {\bibinfo  {journal} {Classical and Quantum Gravity}\
  }\textbf {\bibinfo {volume} {27}},\ \bibinfo {eid} {084007} (\bibinfo {year}
  {2010})}\BibitemShut {NoStop}%
\bibitem [{\citenamefont {{Hild}}\ \emph {et~al.}(2011)\citenamefont {{Hild}}
  \emph {et~al.}}]{2011CQGra..28i4013H}%
  \BibitemOpen
  \bibfield  {author} {\bibinfo {author} {\bibfnamefont {S.}~\bibnamefont
  {{Hild}}} \emph {et~al.},\ }\href {\doibase 10.1088/0264-9381/28/9/094013}
  {\bibfield  {journal} {\bibinfo  {journal} {Classical and Quantum Gravity}\
  }\textbf {\bibinfo {volume} {28}},\ \bibinfo {eid} {094013} (\bibinfo {year}
  {2011})},\ \Eprint {http://arxiv.org/abs/1012.0908} {arXiv:1012.0908 [gr-qc]}
  \BibitemShut {NoStop}%
\bibitem [{\citenamefont {Collaboration}(2017)}]{ISwhitePaper}%
  \BibitemOpen
  \bibfield  {author} {\bibinfo {author} {\bibfnamefont {L.~S.}\ \bibnamefont
  {Collaboration}},\ }\href {https://dcc.ligo.org/LIGO-T1700231} {\bibfield
  {journal} {\bibinfo  {journal} {https://dcc.ligo.org/LIGO-T1700231}\ }
  (\bibinfo {year} {2017})}\BibitemShut {NoStop}%
\bibitem [{\citenamefont {Miao}\ \emph {et~al.}(2018)\citenamefont {Miao},
  \citenamefont {Yang},\ and\ \citenamefont {Martynov}}]{Miao:2017qot}%
  \BibitemOpen
  \bibfield  {author} {\bibinfo {author} {\bibfnamefont {H.}~\bibnamefont
  {Miao}}, \bibinfo {author} {\bibfnamefont {H.}~\bibnamefont {Yang}}, \ and\
  \bibinfo {author} {\bibfnamefont {D.}~\bibnamefont {Martynov}},\ }\href
  {\doibase 10.1103/PhysRevD.98.044044} {\bibfield  {journal} {\bibinfo
  {journal} {Phys. Rev.}\ }\textbf {\bibinfo {volume} {D98}},\ \bibinfo {pages}
  {044044} (\bibinfo {year} {2018})},\ \Eprint
  {http://arxiv.org/abs/1712.07345} {arXiv:1712.07345 [gr-qc]} \BibitemShut
  {NoStop}%
%%CITATION = ARXIV:1712.07345;%%
\bibitem [{\citenamefont {{Miller}}\ \emph {et~al.}(2015)\citenamefont
  {{Miller}}, \citenamefont {{Barsotti}}, \citenamefont {{Vitale}},
  \citenamefont {{Fritschel}}, \citenamefont {{Evans}},\ and\ \citenamefont
  {{Sigg}}}]{Miller2015}%
  \BibitemOpen
  \bibfield  {author} {\bibinfo {author} {\bibfnamefont {J.}~\bibnamefont
  {{Miller}}}, \bibinfo {author} {\bibfnamefont {L.}~\bibnamefont
  {{Barsotti}}}, \bibinfo {author} {\bibfnamefont {S.}~\bibnamefont
  {{Vitale}}}, \bibinfo {author} {\bibfnamefont {P.}~\bibnamefont
  {{Fritschel}}}, \bibinfo {author} {\bibfnamefont {M.}~\bibnamefont
  {{Evans}}}, \ and\ \bibinfo {author} {\bibfnamefont {D.}~\bibnamefont
  {{Sigg}}},\ }\href {\doibase 10.1103/PhysRevD.91.062005} {\bibfield
  {journal} {\bibinfo  {journal} {\prd}\ }\textbf {\bibinfo {volume} {91}},\
  \bibinfo {eid} {062005} (\bibinfo {year} {2015})},\ \Eprint
  {http://arxiv.org/abs/1410.5882} {arXiv:1410.5882 [gr-qc]} \BibitemShut
  {NoStop}%
\bibitem [{\citenamefont {Vines}\ \emph {et~al.}(2011)\citenamefont {Vines},
  \citenamefont {Flanagan},\ and\ \citenamefont {Hinderer}}]{Vines:2011ud}%
  \BibitemOpen
  \bibfield  {author} {\bibinfo {author} {\bibfnamefont {J.}~\bibnamefont
  {Vines}}, \bibinfo {author} {\bibfnamefont {E.~E.}\ \bibnamefont {Flanagan}},
  \ and\ \bibinfo {author} {\bibfnamefont {T.}~\bibnamefont {Hinderer}},\
  }\href {\doibase 10.1103/PhysRevD.83.084051} {\bibfield  {journal} {\bibinfo
  {journal} {Phys. Rev.}\ }\textbf {\bibinfo {volume} {D83}},\ \bibinfo {pages}
  {084051} (\bibinfo {year} {2011})},\ \Eprint {http://arxiv.org/abs/1101.1673}
  {arXiv:1101.1673 [gr-qc]} \BibitemShut {NoStop}%
%%CITATION = ARXIV:1101.1673;%%
\bibitem [{\citenamefont {Yagi}\ and\ \citenamefont
  {Yunes}(2016)}]{Yagi:2015pkc}%
  \BibitemOpen
  \bibfield  {author} {\bibinfo {author} {\bibfnamefont {K.}~\bibnamefont
  {Yagi}}\ and\ \bibinfo {author} {\bibfnamefont {N.}~\bibnamefont {Yunes}},\
  }\href {\doibase 10.1088/0264-9381/33/13/13LT01} {\bibfield  {journal}
  {\bibinfo  {journal} {Class. Quant. Grav.}\ }\textbf {\bibinfo {volume}
  {33}},\ \bibinfo {pages} {13LT01} (\bibinfo {year} {2016})},\ \Eprint
  {http://arxiv.org/abs/1512.02639} {arXiv:1512.02639 [gr-qc]} \BibitemShut
  {NoStop}%
%%CITATION = ARXIV:1512.02639;%%
\bibitem [{\citenamefont {Lindblom}(2010)}]{PhysRevD.82.103011}%
  \BibitemOpen
  \bibfield  {author} {\bibinfo {author} {\bibfnamefont {L.}~\bibnamefont
  {Lindblom}},\ }\href {\doibase 10.1103/PhysRevD.82.103011} {\bibfield
  {journal} {\bibinfo  {journal} {Phys. Rev. D}\ }\textbf {\bibinfo {volume}
  {82}},\ \bibinfo {pages} {103011} (\bibinfo {year} {2010})}\BibitemShut
  {NoStop}%
\bibitem [{\citenamefont {Antoniadis}\ \emph {et~al.}(2013)\citenamefont
  {Antoniadis}, \citenamefont {Freire}, \citenamefont {Wex}, \citenamefont
  {Tauris}, \citenamefont {Lynch}, \citenamefont {van Kerkwijk}, \citenamefont
  {Kramer}, \citenamefont {Bassa}, \citenamefont {Dhillon}, \citenamefont
  {Driebe}, \citenamefont {Hessels}, \citenamefont {Kaspi}, \citenamefont
  {Kondratiev}, \citenamefont {Langer}, \citenamefont {Marsh}, \citenamefont
  {McLaughlin}, \citenamefont {Pennucci}, \citenamefont {Ransom}, \citenamefont
  {Stairs}, \citenamefont {van Leeuwen}, \citenamefont {Verbiest},\ and\
  \citenamefont {Whelan}}]{Antoniadis26042013}%
  \BibitemOpen
  \bibfield  {author} {\bibinfo {author} {\bibfnamefont {J.}~\bibnamefont
  {Antoniadis}}, \bibinfo {author} {\bibfnamefont {P.~C.~C.}\ \bibnamefont
  {Freire}}, \bibinfo {author} {\bibfnamefont {N.}~\bibnamefont {Wex}},
  \bibinfo {author} {\bibfnamefont {T.~M.}\ \bibnamefont {Tauris}}, \bibinfo
  {author} {\bibfnamefont {R.~S.}\ \bibnamefont {Lynch}}, \bibinfo {author}
  {\bibfnamefont {M.~H.}\ \bibnamefont {van Kerkwijk}}, \bibinfo {author}
  {\bibfnamefont {M.}~\bibnamefont {Kramer}}, \bibinfo {author} {\bibfnamefont
  {C.}~\bibnamefont {Bassa}}, \bibinfo {author} {\bibfnamefont {V.~S.}\
  \bibnamefont {Dhillon}}, \bibinfo {author} {\bibfnamefont {T.}~\bibnamefont
  {Driebe}}, \bibinfo {author} {\bibfnamefont {J.~W.~T.}\ \bibnamefont
  {Hessels}}, \bibinfo {author} {\bibfnamefont {V.~M.}\ \bibnamefont {Kaspi}},
  \bibinfo {author} {\bibfnamefont {V.~I.}\ \bibnamefont {Kondratiev}},
  \bibinfo {author} {\bibfnamefont {N.}~\bibnamefont {Langer}}, \bibinfo
  {author} {\bibfnamefont {T.~R.}\ \bibnamefont {Marsh}}, \bibinfo {author}
  {\bibfnamefont {M.~A.}\ \bibnamefont {McLaughlin}}, \bibinfo {author}
  {\bibfnamefont {T.~T.}\ \bibnamefont {Pennucci}}, \bibinfo {author}
  {\bibfnamefont {S.~M.}\ \bibnamefont {Ransom}}, \bibinfo {author}
  {\bibfnamefont {I.~H.}\ \bibnamefont {Stairs}}, \bibinfo {author}
  {\bibfnamefont {J.}~\bibnamefont {van Leeuwen}}, \bibinfo {author}
  {\bibfnamefont {J.~P.~W.}\ \bibnamefont {Verbiest}}, \ and\ \bibinfo {author}
  {\bibfnamefont {D.~G.}\ \bibnamefont {Whelan}},\ }\href@noop {} {\bibfield
  {journal} {\bibinfo  {journal} {Science}\ }\textbf {\bibinfo {volume} {340}}
  (\bibinfo {year} {2013})}\BibitemShut {NoStop}%
\bibitem [{\citenamefont {Damour}\ and\ \citenamefont
  {Nagar}(2010)}]{PhysRevD.81.084016}%
  \BibitemOpen
  \bibfield  {author} {\bibinfo {author} {\bibfnamefont {T.}~\bibnamefont
  {Damour}}\ and\ \bibinfo {author} {\bibfnamefont {A.}~\bibnamefont {Nagar}},\
  }\href {\doibase 10.1103/PhysRevD.81.084016} {\bibfield  {journal} {\bibinfo
  {journal} {Phys. Rev. D}\ }\textbf {\bibinfo {volume} {81}},\ \bibinfo
  {pages} {084016} (\bibinfo {year} {2010})}\BibitemShut {NoStop}%
\bibitem [{\citenamefont {Baiotti}\ \emph {et~al.}(2011)\citenamefont
  {Baiotti}, \citenamefont {Damour}, \citenamefont {Giacomazzo}, \citenamefont
  {Nagar},\ and\ \citenamefont {Rezzolla}}]{Baiotti:2011am}%
  \BibitemOpen
  \bibfield  {author} {\bibinfo {author} {\bibfnamefont {L.}~\bibnamefont
  {Baiotti}}, \bibinfo {author} {\bibfnamefont {T.}~\bibnamefont {Damour}},
  \bibinfo {author} {\bibfnamefont {B.}~\bibnamefont {Giacomazzo}}, \bibinfo
  {author} {\bibfnamefont {A.}~\bibnamefont {Nagar}}, \ and\ \bibinfo {author}
  {\bibfnamefont {L.}~\bibnamefont {Rezzolla}},\ }\href {\doibase
  10.1103/PhysRevD.84.024017} {\bibfield  {journal} {\bibinfo  {journal} {Phys.
  Rev.}\ }\textbf {\bibinfo {volume} {D84}},\ \bibinfo {pages} {024017}
  (\bibinfo {year} {2011})},\ \Eprint {http://arxiv.org/abs/1103.3874}
  {arXiv:1103.3874 [gr-qc]} \BibitemShut {NoStop}%
%%CITATION = ARXIV:1103.3874;%%
\bibitem [{\citenamefont {Bernuzzi}\ \emph {et~al.}(2012)\citenamefont
  {Bernuzzi}, \citenamefont {Nagar}, \citenamefont {Thierfelder},\ and\
  \citenamefont {Brugmann}}]{Bernuzzi:2012ci}%
  \BibitemOpen
  \bibfield  {author} {\bibinfo {author} {\bibfnamefont {S.}~\bibnamefont
  {Bernuzzi}}, \bibinfo {author} {\bibfnamefont {A.}~\bibnamefont {Nagar}},
  \bibinfo {author} {\bibfnamefont {M.}~\bibnamefont {Thierfelder}}, \ and\
  \bibinfo {author} {\bibfnamefont {B.}~\bibnamefont {Brugmann}},\ }\href
  {\doibase 10.1103/PhysRevD.86.044030} {\bibfield  {journal} {\bibinfo
  {journal} {Phys. Rev.}\ }\textbf {\bibinfo {volume} {D86}},\ \bibinfo {pages}
  {044030} (\bibinfo {year} {2012})},\ \Eprint {http://arxiv.org/abs/1205.3403}
  {arXiv:1205.3403 [gr-qc]} \BibitemShut {NoStop}%
%%CITATION = ARXIV:1205.3403;%%
\bibitem [{\citenamefont {Bernuzzi}\ \emph {et~al.}(2014)\citenamefont
  {Bernuzzi}, \citenamefont {Nagar}, \citenamefont {Balmelli}, \citenamefont
  {Dietrich},\ and\ \citenamefont {Ujevic}}]{PhysRevLett.112.201101}%
  \BibitemOpen
  \bibfield  {author} {\bibinfo {author} {\bibfnamefont {S.}~\bibnamefont
  {Bernuzzi}}, \bibinfo {author} {\bibfnamefont {A.}~\bibnamefont {Nagar}},
  \bibinfo {author} {\bibfnamefont {S.}~\bibnamefont {Balmelli}}, \bibinfo
  {author} {\bibfnamefont {T.}~\bibnamefont {Dietrich}}, \ and\ \bibinfo
  {author} {\bibfnamefont {M.}~\bibnamefont {Ujevic}},\ }\href {\doibase
  10.1103/PhysRevLett.112.201101} {\bibfield  {journal} {\bibinfo  {journal}
  {Phys. Rev. Lett.}\ }\textbf {\bibinfo {volume} {112}},\ \bibinfo {pages}
  {201101} (\bibinfo {year} {2014})}\BibitemShut {NoStop}%
\bibitem [{\citenamefont {{Dietrich}}\ \emph {et~al.}(2018)\citenamefont
  {{Dietrich}}, \citenamefont {{Khan}}, \citenamefont {{Dudi}}, \citenamefont
  {{Kapadia}}, \citenamefont {{Kumar}}, \citenamefont {{Nagar}}, \citenamefont
  {{Ohme}}, \citenamefont {{Pannarale}}, \citenamefont {{Samajdar}},
  \citenamefont {{Bernuzzi}}, \citenamefont {{Carullo}}, \citenamefont {{Del
  Pozzo}}, \citenamefont {{Haney}}, \citenamefont {{Markakis}}, \citenamefont
  {{Puerrer}}, \citenamefont {{Riemenschneider}}, \citenamefont {{Eka
  Setyawati}}, \citenamefont {{Tsang}},\ and\ \citenamefont {{Van Den
  Broeck}}}]{TDietrich2018}%
  \BibitemOpen
  \bibfield  {author} {\bibinfo {author} {\bibfnamefont {T.}~\bibnamefont
  {{Dietrich}}}, \bibinfo {author} {\bibfnamefont {S.}~\bibnamefont {{Khan}}},
  \bibinfo {author} {\bibfnamefont {R.}~\bibnamefont {{Dudi}}}, \bibinfo
  {author} {\bibfnamefont {S.~J.}\ \bibnamefont {{Kapadia}}}, \bibinfo {author}
  {\bibfnamefont {P.}~\bibnamefont {{Kumar}}}, \bibinfo {author} {\bibfnamefont
  {A.}~\bibnamefont {{Nagar}}}, \bibinfo {author} {\bibfnamefont
  {F.}~\bibnamefont {{Ohme}}}, \bibinfo {author} {\bibfnamefont
  {F.}~\bibnamefont {{Pannarale}}}, \bibinfo {author} {\bibfnamefont
  {A.}~\bibnamefont {{Samajdar}}}, \bibinfo {author} {\bibfnamefont
  {S.}~\bibnamefont {{Bernuzzi}}}, \bibinfo {author} {\bibfnamefont
  {G.}~\bibnamefont {{Carullo}}}, \bibinfo {author} {\bibfnamefont
  {W.}~\bibnamefont {{Del Pozzo}}}, \bibinfo {author} {\bibfnamefont
  {M.}~\bibnamefont {{Haney}}}, \bibinfo {author} {\bibfnamefont
  {C.}~\bibnamefont {{Markakis}}}, \bibinfo {author} {\bibfnamefont
  {M.}~\bibnamefont {{Puerrer}}}, \bibinfo {author} {\bibfnamefont
  {G.}~\bibnamefont {{Riemenschneider}}}, \bibinfo {author} {\bibfnamefont
  {Y.}~\bibnamefont {{Eka Setyawati}}}, \bibinfo {author} {\bibfnamefont
  {K.~W.}\ \bibnamefont {{Tsang}}}, \ and\ \bibinfo {author} {\bibfnamefont
  {C.}~\bibnamefont {{Van Den Broeck}}},\ }\href@noop {} {\bibfield  {journal}
  {\bibinfo  {journal} {ArXiv e-prints}\ } (\bibinfo {year} {2018})},\ \Eprint
  {http://arxiv.org/abs/1804.02235} {arXiv:1804.02235 [gr-qc]} \BibitemShut
  {NoStop}%
\bibitem [{\citenamefont {{Oechslin}}\ \emph {et~al.}(2002)\citenamefont
  {{Oechslin}}, \citenamefont {{Rosswog}},\ and\ \citenamefont
  {{Thielemann}}}]{2002PhRvD..65j3005O}%
  \BibitemOpen
  \bibfield  {author} {\bibinfo {author} {\bibfnamefont {R.}~\bibnamefont
  {{Oechslin}}}, \bibinfo {author} {\bibfnamefont {S.}~\bibnamefont
  {{Rosswog}}}, \ and\ \bibinfo {author} {\bibfnamefont {F.-K.}\ \bibnamefont
  {{Thielemann}}},\ }\href@noop {} {\bibfield  {journal} {\bibinfo  {journal}
  {\prd}\ }\textbf {\bibinfo {volume} {65}},\ \bibinfo {pages} {103005}
  (\bibinfo {year} {2002})}\BibitemShut {NoStop}%
\bibitem [{\citenamefont {{Oechslin}}\ \emph {et~al.}(2007)\citenamefont
  {{Oechslin}}, \citenamefont {{Janka}},\ and\ \citenamefont
  {{Marek}}}]{2007A&A...467..395O}%
  \BibitemOpen
  \bibfield  {author} {\bibinfo {author} {\bibfnamefont {R.}~\bibnamefont
  {{Oechslin}}}, \bibinfo {author} {\bibfnamefont {H.-T.}\ \bibnamefont
  {{Janka}}}, \ and\ \bibinfo {author} {\bibfnamefont {A.}~\bibnamefont
  {{Marek}}},\ }\href {\doibase 10.1051/0004-6361:20066682} {\bibfield
  {journal} {\bibinfo  {journal} {Astron. Astrophys.}\ }\textbf {\bibinfo
  {volume} {467}},\ \bibinfo {pages} {395} (\bibinfo {year}
  {2007})}\BibitemShut {NoStop}%
\bibitem [{\citenamefont {{Bauswein}}\ \emph {et~al.}(2010)\citenamefont
  {{Bauswein}}, \citenamefont {{Janka}},\ and\ \citenamefont
  {{Oechslin}}}]{2010PhRvD..82h4043B}%
  \BibitemOpen
  \bibfield  {author} {\bibinfo {author} {\bibfnamefont {A.}~\bibnamefont
  {{Bauswein}}}, \bibinfo {author} {\bibfnamefont {H.-T.}\ \bibnamefont
  {{Janka}}}, \ and\ \bibinfo {author} {\bibfnamefont {R.}~\bibnamefont
  {{Oechslin}}},\ }\href {\doibase 10.1103/PhysRevD.82.084043} {\bibfield
  {journal} {\bibinfo  {journal} {\prd}\ }\textbf {\bibinfo {volume} {82}},\
  \bibinfo {pages} {084043} (\bibinfo {year} {2010})}\BibitemShut {NoStop}%
\bibitem [{\citenamefont {{Bauswein}}\ \emph
  {et~al.}(2012{\natexlab{b}})\citenamefont {{Bauswein}}, \citenamefont
  {{Janka}}, \citenamefont {{Hebeler}},\ and\ \citenamefont
  {{Schwenk}}}]{2012PhRvD..86f3001B}%
  \BibitemOpen
  \bibfield  {author} {\bibinfo {author} {\bibfnamefont {A.}~\bibnamefont
  {{Bauswein}}}, \bibinfo {author} {\bibfnamefont {H.-T.}\ \bibnamefont
  {{Janka}}}, \bibinfo {author} {\bibfnamefont {K.}~\bibnamefont {{Hebeler}}},
  \ and\ \bibinfo {author} {\bibfnamefont {A.}~\bibnamefont {{Schwenk}}},\
  }\href {\doibase 10.1103/PhysRevD.86.063001} {\bibfield  {journal} {\bibinfo
  {journal} {\prd}\ }\textbf {\bibinfo {volume} {86}},\ \bibinfo {eid} {063001}
  (\bibinfo {year} {2012}{\natexlab{b}})}\BibitemShut {NoStop}%
\bibitem [{\citenamefont {{Bauswein}}\ \emph
  {et~al.}(2013{\natexlab{b}})\citenamefont {{Bauswein}}, \citenamefont
  {{Goriely}},\ and\ \citenamefont {{Janka}}}]{2013ApJ...773...78B}%
  \BibitemOpen
  \bibfield  {author} {\bibinfo {author} {\bibfnamefont {A.}~\bibnamefont
  {{Bauswein}}}, \bibinfo {author} {\bibfnamefont {S.}~\bibnamefont
  {{Goriely}}}, \ and\ \bibinfo {author} {\bibfnamefont {H.-T.}\ \bibnamefont
  {{Janka}}},\ }\href {\doibase 10.1088/0004-637X/773/1/78} {\bibfield
  {journal} {\bibinfo  {journal} {Astrophys. J.}\ }\textbf {\bibinfo {volume}
  {773}},\ \bibinfo {eid} {78} (\bibinfo {year}
  {2013}{\natexlab{b}})}\BibitemShut {NoStop}%
\bibitem [{\citenamefont {{Isenberg}}\ and\ \citenamefont
  {{Nester}}(1980)}]{1980grg..conf...23I}%
  \BibitemOpen
  \bibfield  {author} {\bibinfo {author} {\bibfnamefont {J.}~\bibnamefont
  {{Isenberg}}}\ and\ \bibinfo {author} {\bibfnamefont {J.}~\bibnamefont
  {{Nester}}},\ }in\ \href@noop {} {\emph {\bibinfo {booktitle} {General
  Relativity and Gravitation}}}\ (\bibinfo  {publisher} {Plenum Press, New
  York},\ \bibinfo {year} {1980})\ p.~\bibinfo {pages} {23}\BibitemShut
  {NoStop}%
\bibitem [{\citenamefont {{Wilson}}\ \emph {et~al.}(1996)\citenamefont
  {{Wilson}}, \citenamefont {{Mathews}},\ and\ \citenamefont
  {{Marronetti}}}]{1996PhRvD..54.1317W}%
  \BibitemOpen
  \bibfield  {author} {\bibinfo {author} {\bibfnamefont {J.~R.}\ \bibnamefont
  {{Wilson}}}, \bibinfo {author} {\bibfnamefont {G.~J.}\ \bibnamefont
  {{Mathews}}}, \ and\ \bibinfo {author} {\bibfnamefont {P.}~\bibnamefont
  {{Marronetti}}},\ }\href {\doibase 10.1103/PhysRevD.54.1317} {\bibfield
  {journal} {\bibinfo  {journal} {\prd}\ }\textbf {\bibinfo {volume} {54}},\
  \bibinfo {pages} {1317} (\bibinfo {year} {1996})}\BibitemShut {NoStop}%
\bibitem [{\citenamefont {{Rezzolla}}\ and\ \citenamefont
  {{Takami}}(2016)}]{Rezzolla2016}%
  \BibitemOpen
  \bibfield  {author} {\bibinfo {author} {\bibfnamefont {L.}~\bibnamefont
  {{Rezzolla}}}\ and\ \bibinfo {author} {\bibfnamefont {K.}~\bibnamefont
  {{Takami}}},\ }\href {\doibase 10.1103/PhysRevD.93.124051} {\bibfield
  {journal} {\bibinfo  {journal} {\prd}\ }\textbf {\bibinfo {volume} {93}},\
  \bibinfo {eid} {124051} (\bibinfo {year} {2016})},\ \Eprint
  {http://arxiv.org/abs/1604.00246} {arXiv:1604.00246 [gr-qc]} \BibitemShut
  {NoStop}%
\bibitem [{\citenamefont {Schmidt}\ \emph {et~al.}(2017)\citenamefont
  {Schmidt}, \citenamefont {Harry},\ and\ \citenamefont
  {Pfeiffer}}]{Schmidt:2017btt}%
  \BibitemOpen
  \bibfield  {author} {\bibinfo {author} {\bibfnamefont {P.}~\bibnamefont
  {Schmidt}}, \bibinfo {author} {\bibfnamefont {I.~W.}\ \bibnamefont {Harry}},
  \ and\ \bibinfo {author} {\bibfnamefont {H.~P.}\ \bibnamefont {Pfeiffer}},\
  }\href@noop {} {\  (\bibinfo {year} {2017})},\ \Eprint
  {http://arxiv.org/abs/1703.01076} {arXiv:1703.01076 [gr-qc]} \BibitemShut
  {NoStop}%
%%CITATION = ARXIV:1703.01076;%%
\bibitem [{\citenamefont {Nissanke}\ \emph {et~al.}(2010)\citenamefont
  {Nissanke}, \citenamefont {Holz}, \citenamefont {Hughes}, \citenamefont
  {Dalal},\ and\ \citenamefont {Sievers}}]{Nissanke:2009kt}%
  \BibitemOpen
  \bibfield  {author} {\bibinfo {author} {\bibfnamefont {S.}~\bibnamefont
  {Nissanke}}, \bibinfo {author} {\bibfnamefont {D.~E.}\ \bibnamefont {Holz}},
  \bibinfo {author} {\bibfnamefont {S.~A.}\ \bibnamefont {Hughes}}, \bibinfo
  {author} {\bibfnamefont {N.}~\bibnamefont {Dalal}}, \ and\ \bibinfo {author}
  {\bibfnamefont {J.~L.}\ \bibnamefont {Sievers}},\ }\href {\doibase
  10.1088/0004-637X/725/1/496} {\bibfield  {journal} {\bibinfo  {journal}
  {Astrophys. J.}\ }\textbf {\bibinfo {volume} {725}},\ \bibinfo {pages} {496}
  (\bibinfo {year} {2010})},\ \Eprint {http://arxiv.org/abs/0904.1017}
  {arXiv:0904.1017 [astro-ph.CO]} \BibitemShut {NoStop}%
%%CITATION = ARXIV:0904.1017;%%
\bibitem [{\citenamefont {Aso}\ \emph {et~al.}(2013)\citenamefont {Aso},
  \citenamefont {Michimura}, \citenamefont {Somiya}, \citenamefont {Ando},
  \citenamefont {Miyakawa}, \citenamefont {Sekiguchi}, \citenamefont
  {Tatsumi},\ and\ \citenamefont {Yamamoto}}]{PhysRevD.88.043007}%
  \BibitemOpen
  \bibfield  {author} {\bibinfo {author} {\bibfnamefont {Y.}~\bibnamefont
  {Aso}}, \bibinfo {author} {\bibfnamefont {Y.}~\bibnamefont {Michimura}},
  \bibinfo {author} {\bibfnamefont {K.}~\bibnamefont {Somiya}}, \bibinfo
  {author} {\bibfnamefont {M.}~\bibnamefont {Ando}}, \bibinfo {author}
  {\bibfnamefont {O.}~\bibnamefont {Miyakawa}}, \bibinfo {author}
  {\bibfnamefont {T.}~\bibnamefont {Sekiguchi}}, \bibinfo {author}
  {\bibfnamefont {D.}~\bibnamefont {Tatsumi}}, \ and\ \bibinfo {author}
  {\bibfnamefont {H.}~\bibnamefont {Yamamoto}} (\bibinfo {collaboration} {The
  KAGRA Collaboration}),\ }\href {\doibase 10.1103/PhysRevD.88.043007}
  {\bibfield  {journal} {\bibinfo  {journal} {Phys. Rev. D}\ }\textbf {\bibinfo
  {volume} {88}},\ \bibinfo {pages} {043007} (\bibinfo {year}
  {2013})}\BibitemShut {NoStop}%
\bibitem [{\citenamefont {Iyer}\ \emph {et~al.}(2011)\citenamefont {Iyer} \emph
  {et~al.}}]{LIGOINDIA}%
  \BibitemOpen
  \bibfield  {author} {\bibinfo {author} {\bibfnamefont {B.}~\bibnamefont
  {Iyer}} \emph {et~al.},\ }\href {https://dcc.ligo.org/LIGO-M1100296/public}
  {\emph {\bibinfo {title} {{LIGO-India}}}}\ (\bibinfo  {publisher}
  {LIGO-India},\ \bibinfo {year} {2011})\BibitemShut {NoStop}%
\bibitem [{\citenamefont {{Shoemaker}}(2010)}]{AdvLIGO-noise}%
  \BibitemOpen
  \bibfield  {author} {\bibinfo {author} {\bibfnamefont {D.}~\bibnamefont
  {{Shoemaker}}},\ }\href@noop {} {\emph {\bibinfo {title} {Advanced LIGO
  anticipated sensitivity curves}}}\ (\bibinfo  {publisher} {Tech. Rep.
  LIGO-T0900288-v3},\ \bibinfo {year} {2010})\BibitemShut {NoStop}%
\bibitem [{\citenamefont {{Fritschel}}\ \emph {et~al.}(2017)\citenamefont
  {{Fritschel}}, \citenamefont {{Coyne}} \emph {et~al.}}]{GWINC}%
  \BibitemOpen
  \bibfield  {author} {\bibinfo {author} {\bibfnamefont {D.}~\bibnamefont
  {{Fritschel}}}, \bibinfo {author} {\bibfnamefont {R.}~\bibnamefont
  {{Coyne}}},  \emph {et~al.},\ }\href {https://dcc.ligo.org/T010075/public}
  {\enquote {\bibinfo {title} {{Advanced LIGO Systems Design}},}\ } (\bibinfo
  {year} {2017}),\ \bibinfo {note}
  {\url{https://git.ligo.org/gwinc/pygwinc}}\BibitemShut {NoStop}%
\bibitem [{\citenamefont {Becsy}\ \emph {et~al.}(2017)\citenamefont {Becsy},
  \citenamefont {Raffai}, \citenamefont {Cornish}, \citenamefont {Essick},
  \citenamefont {Kanner}, \citenamefont {Katsavounidis}, \citenamefont
  {Littenberg}, \citenamefont {Millhouse},\ and\ \citenamefont
  {Vitale}}]{Becsy:2016ofp}%
  \BibitemOpen
  \bibfield  {author} {\bibinfo {author} {\bibfnamefont {B.}~\bibnamefont
  {Becsy}}, \bibinfo {author} {\bibfnamefont {P.}~\bibnamefont {Raffai}},
  \bibinfo {author} {\bibfnamefont {N.~J.}\ \bibnamefont {Cornish}}, \bibinfo
  {author} {\bibfnamefont {R.}~\bibnamefont {Essick}}, \bibinfo {author}
  {\bibfnamefont {J.}~\bibnamefont {Kanner}}, \bibinfo {author} {\bibfnamefont
  {E.}~\bibnamefont {Katsavounidis}}, \bibinfo {author} {\bibfnamefont {T.~B.}\
  \bibnamefont {Littenberg}}, \bibinfo {author} {\bibfnamefont
  {M.}~\bibnamefont {Millhouse}}, \ and\ \bibinfo {author} {\bibfnamefont
  {S.}~\bibnamefont {Vitale}},\ }\href {\doibase 10.3847/1538-4357/aa63ef}
  {\bibfield  {journal} {\bibinfo  {journal} {Astrophys. J.}\ }\textbf
  {\bibinfo {volume} {839}},\ \bibinfo {pages} {15} (\bibinfo {year} {2017})},\
  \bibinfo {note} {[Astrophys. J.839,15(2017)]},\ \Eprint
  {http://arxiv.org/abs/1612.02003} {arXiv:1612.02003 [astro-ph.HE]}
  \BibitemShut {NoStop}%
%%CITATION = ARXIV:1612.02003;%%
\bibitem [{\citenamefont {Pannarale}\ \emph {et~al.}(2018)\citenamefont
  {Pannarale}, \citenamefont {Macas},\ and\ \citenamefont
  {Sutton}}]{Pannarale:2018cct}%
  \BibitemOpen
  \bibfield  {author} {\bibinfo {author} {\bibfnamefont {F.}~\bibnamefont
  {Pannarale}}, \bibinfo {author} {\bibfnamefont {R.}~\bibnamefont {Macas}}, \
  and\ \bibinfo {author} {\bibfnamefont {P.~J.}\ \bibnamefont {Sutton}},\
  }\href@noop {} {\  (\bibinfo {year} {2018})},\ \Eprint
  {http://arxiv.org/abs/1807.01939} {arXiv:1807.01939 [gr-qc]} \BibitemShut
  {NoStop}%
%%CITATION = ARXIV:1807.01939;%%
\bibitem [{\citenamefont {Littenberg}\ \emph {et~al.}(2016)\citenamefont
  {Littenberg}, \citenamefont {Kanner}, \citenamefont {Cornish},\ and\
  \citenamefont {Millhouse}}]{Littenberg:2015kpb}%
  \BibitemOpen
  \bibfield  {author} {\bibinfo {author} {\bibfnamefont {T.~B.}\ \bibnamefont
  {Littenberg}}, \bibinfo {author} {\bibfnamefont {J.~B.}\ \bibnamefont
  {Kanner}}, \bibinfo {author} {\bibfnamefont {N.~J.}\ \bibnamefont {Cornish}},
  \ and\ \bibinfo {author} {\bibfnamefont {M.}~\bibnamefont {Millhouse}},\
  }\href {\doibase 10.1103/PhysRevD.94.044050} {\bibfield  {journal} {\bibinfo
  {journal} {Phys. Rev.}\ }\textbf {\bibinfo {volume} {D94}},\ \bibinfo {pages}
  {044050} (\bibinfo {year} {2016})},\ \Eprint
  {http://arxiv.org/abs/1511.08752} {arXiv:1511.08752 [gr-qc]} \BibitemShut
  {NoStop}%
%%CITATION = ARXIV:1511.08752;%%
\bibitem [{\citenamefont {{Kanner}}\ \emph {et~al.}(2016)\citenamefont
  {{Kanner}}, \citenamefont {{Littenberg}}, \citenamefont {{Cornish}},
  \citenamefont {{Millhouse}}, \citenamefont {{Xhakaj}}, \citenamefont
  {{Salemi}}, \citenamefont {{Drago}}, \citenamefont {{Vedovato}},\ and\
  \citenamefont {{Klimenko}}}]{2016PhRvD..93b2002K}%
  \BibitemOpen
  \bibfield  {author} {\bibinfo {author} {\bibfnamefont {J.~B.}\ \bibnamefont
  {{Kanner}}}, \bibinfo {author} {\bibfnamefont {T.~B.}\ \bibnamefont
  {{Littenberg}}}, \bibinfo {author} {\bibfnamefont {N.}~\bibnamefont
  {{Cornish}}}, \bibinfo {author} {\bibfnamefont {M.}~\bibnamefont
  {{Millhouse}}}, \bibinfo {author} {\bibfnamefont {E.}~\bibnamefont
  {{Xhakaj}}}, \bibinfo {author} {\bibfnamefont {F.}~\bibnamefont {{Salemi}}},
  \bibinfo {author} {\bibfnamefont {M.}~\bibnamefont {{Drago}}}, \bibinfo
  {author} {\bibfnamefont {G.}~\bibnamefont {{Vedovato}}}, \ and\ \bibinfo
  {author} {\bibfnamefont {S.}~\bibnamefont {{Klimenko}}},\ }\href {\doibase
  10.1103/PhysRevD.93.022002} {\bibfield  {journal} {\bibinfo  {journal}
  {\prd}\ }\textbf {\bibinfo {volume} {93}},\ \bibinfo {eid} {022002} (\bibinfo
  {year} {2016})},\ \Eprint {http://arxiv.org/abs/1509.06423} {arXiv:1509.06423
  [astro-ph.IM]} \BibitemShut {NoStop}%
\bibitem [{\citenamefont {{Bauswein}}\ \emph {et~al.}(2014)\citenamefont
  {{Bauswein}}, \citenamefont {{Stergioulas}},\ and\ \citenamefont
  {{Janka}}}]{2014arXiv1403.5301B}%
  \BibitemOpen
  \bibfield  {author} {\bibinfo {author} {\bibfnamefont {A.}~\bibnamefont
  {{Bauswein}}}, \bibinfo {author} {\bibfnamefont {N.}~\bibnamefont
  {{Stergioulas}}}, \ and\ \bibinfo {author} {\bibfnamefont {H.-T.}\
  \bibnamefont {{Janka}}},\ }\href {\doibase 10.1103/PhysRevD.90.023002}
  {\bibfield  {journal} {\bibinfo  {journal} {\prd}\ }\textbf {\bibinfo
  {volume} {90}},\ \bibinfo {eid} {023002} (\bibinfo {year} {2014})},\ \Eprint
  {http://arxiv.org/abs/1403.5301} {arXiv:1403.5301 [astro-ph.SR]} \BibitemShut
  {NoStop}%
\bibitem [{\citenamefont {Millhouse}\ \emph {et~al.}(2018)\citenamefont
  {Millhouse}, \citenamefont {Cornish},\ and\ \citenamefont
  {Littenberg}}]{Millhouse:2018dgi}%
  \BibitemOpen
  \bibfield  {author} {\bibinfo {author} {\bibfnamefont {M.}~\bibnamefont
  {Millhouse}}, \bibinfo {author} {\bibfnamefont {N.~J.}\ \bibnamefont
  {Cornish}}, \ and\ \bibinfo {author} {\bibfnamefont {T.}~\bibnamefont
  {Littenberg}},\ }\href {\doibase 10.1103/PhysRevD.97.104057} {\bibfield
  {journal} {\bibinfo  {journal} {Phys. Rev.}\ }\textbf {\bibinfo {volume}
  {D97}},\ \bibinfo {pages} {104057} (\bibinfo {year} {2018})},\ \Eprint
  {http://arxiv.org/abs/1804.03239} {arXiv:1804.03239 [gr-qc]} \BibitemShut
  {NoStop}%
%%CITATION = ARXIV:1804.03239;%%
\bibitem [{\citenamefont {{Pordes}}\ \emph {et~al.}(2007)\citenamefont
  {{Pordes}} \emph {et~al.}}]{pordes:2007}%
  \BibitemOpen
  \bibfield  {author} {\bibinfo {author} {\bibfnamefont {R.}~\bibnamefont
  {{Pordes}}} \emph {et~al.},\ }\href@noop {} {\bibfield  {journal} {\bibinfo
  {journal} {J. Phys. Conf. Ser.}\ }\textbf {\bibinfo {volume} {78}},\ \bibinfo
  {pages} {012057} (\bibinfo {year} {2007})}\BibitemShut {NoStop}%
\bibitem [{\citenamefont {Sfiligoi}\ \emph {et~al.}(2009)\citenamefont
  {Sfiligoi}, \citenamefont {Bradley}, \citenamefont {Holzman}, \citenamefont
  {Mhashilkar}, \citenamefont {Padhi},\ and\ \citenamefont
  {Wurthwein}}]{Sfiligoi:2009}%
  \BibitemOpen
  \bibfield  {author} {\bibinfo {author} {\bibfnamefont {I.}~\bibnamefont
  {Sfiligoi}}, \bibinfo {author} {\bibfnamefont {D.~C.}\ \bibnamefont
  {Bradley}}, \bibinfo {author} {\bibfnamefont {B.}~\bibnamefont {Holzman}},
  \bibinfo {author} {\bibfnamefont {P.}~\bibnamefont {Mhashilkar}}, \bibinfo
  {author} {\bibfnamefont {S.}~\bibnamefont {Padhi}}, \ and\ \bibinfo {author}
  {\bibfnamefont {F.}~\bibnamefont {Wurthwein}},\ }in\ \href {\doibase
  10.1109/CSIE.2009.950} {\emph {\bibinfo {booktitle} {Proceedings of the 2009
  WRI World Congress on Computer Science and Information Engineering - Volume
  02}}},\ \bibinfo {series and number} {CSIE '09}\ (\bibinfo  {publisher} {IEEE
  Computer Society},\ \bibinfo {address} {Washington, DC, USA},\ \bibinfo
  {year} {2009})\ pp.\ \bibinfo {pages} {428--432}\BibitemShut {NoStop}%
\bibitem [{\citenamefont {Hunter}(2007)}]{Hunter:2007}%
  \BibitemOpen
  \bibfield  {author} {\bibinfo {author} {\bibfnamefont {J.~D.}\ \bibnamefont
  {Hunter}},\ }\href {\doibase 10.1109/MCSE.2007.55} {\bibfield  {journal}
  {\bibinfo  {journal} {Computing In Science \& Engineering}\ }\textbf
  {\bibinfo {volume} {9}},\ \bibinfo {pages} {90} (\bibinfo {year}
  {2007})}\BibitemShut {NoStop}%
\end{thebibliography}%

\end{document}